%Paper: hep-th/9412117
%From: candelas@guinness.ias.edu (Philip Candelas)
%Date: Tue, 13 Dec 94 15:33:20 EST

%%%%%%%%%%%%%%%%%%%%%%%%%%%%%%%%%%%%%%%%%%%%%%%%%%%%%%%%%%%%%%%%%%%%%%%%%%%
%%
%%              Mirror Symmetry for Calabi-Yau Hypersurfaces
%%                         in Weighted P_4 and
%%                  Extensions of Landau Ginzburg Theory
%%
%%                                  by
%%
%%
%%            Philip   Candelas        candelas@dillo.ph.utexas.edu
%%            Xenia    de la Ossa      delaossa@guinness.ias.edu
%%            Sheldon  Katz            katz@math.okstate.edu
%%
%%%%%%%%%%%%%%%%%%%%%%%%%%%%%%%%%%%%%%%%%%%%%%%%%%%%%%%%%%%%%%%%%%%%%%%%%%%
%%
%%                             INSTRUCTIONS:
%%
%%  THERE ARE TWO POSSIBLE SOURCES OF DIFFICULTY IN PROCESSING THE FILE:
%%
%%  1) THERE ARE TWO FIGURES THAT HAVE BEEN SUBMITTED AS UUENCODED PS FILES.
%%  THE FIGURES REQUIRE THE MACRO epsf. INSTRUCTIONS FOR UNPACKING THE FIGURES
%%  ARE INCLUDED AT THE HEAD OF THE FIGURE FILE.
%%  IF YOU DO NOT WANT THE FIGURES, CHANGE THE STATEMENT \figmodetrue
%%  BELOW TO \figmodefalse .
%%
%%  2) THERE IS A PLOT AT THE END OF THE PAPER THAT WILL EXCEED THE LIMIT ON
%%  THE PAGE MEMORY UNLESS YOU HAVE AN EXTENDED MEMORY VERSION OF TEX. IF
%%  YOU DO NOT HAVE AN EXTENDED MEMORY VERSION OF TEX YOU WILL GET AN ERROR
%%  MESSAGE ABOUT HAVING EXCEEDED THE MEMORY. TO SUPPRESS THE PLOT USE YOUR
%%  EDITOR TO FIND THE LINE " %\listreferences\bye " AND REMOVE THE " % ".
%%
%%%%%%%%%%%%%%%%%%%%%%%%%%%%%%%%%%%%%%%%%%%%%%%%%%%%%%%%%%%%%%%%%%%%%%%%%%%

\newif\iffigmode
\figmodetrue
\iffigmode\input epsf\message{THIS SHOULD INCLUDE THE FIGURES}%
\else\message{THIS WILL NOT INCLUDE THE FIGURES}\fi

%%%%%%%%%%%%%%%%%%%%%%%%%%%%%%%labelfile%%%%%%%%%%%%%%%%%%%%%%%%%%%%%%%%%%%

\expandafter \def \csname CHAPLABELintro\endcsname {1}
\expandafter \def \csname CHAPLABELtoric\endcsname {2}
\expandafter \def \csname CHAPLABELBHrule\endcsname {3}
\expandafter \def \csname EQLABELnewwts\endcsname {3.1?}
\expandafter \def \csname EQLABELnewdeg\endcsname {3.2?}
\expandafter \def \csname EQLABELV'\endcsname {3.3?}
\expandafter \def \csname EQLABELreduct\endcsname {3.4?}
\expandafter \def \csname EQLABELcong\endcsname {3.5?}
\expandafter \def \csname EQLABELsimpcong\endcsname {3.6?}
\expandafter \def \csname EQLABELterms\endcsname {3.7?}
\expandafter \def \csname EQLABELwteqn\endcsname {3.8?}
\expandafter \def \csname EQLABELwtsoln\endcsname {3.9?}
\expandafter \def \csname CHAPLABELnonpoly\endcsname {4}
\expandafter \def \csname EQLABELqnumb\endcsname {4.1?}
\expandafter \def \csname FIGLABELphases\endcsname {4.1?}
\expandafter \def \csname CHAPLABELfrac\endcsname {5}
\expandafter \def \csname EQLABELtransf\endcsname {5.1?}
\expandafter \def \csname EQLABELtwisted\endcsname {5.2?}
\expandafter \def \csname CHAPLABELplot\endcsname {-2}

%%%%%%%%%%%%%%%%%%%%%%%%%%%%%% xenia.mac %%%%%%%%%%%%%%%%%%%%%%%%%%%%%%%%%%

%%%%%%%%%%%%%%%%%%%%%%%%%%%%%%%%%%%%%%%%%%%%%%%%%%%%%%%%%%%%%%%%%%%%%%%%%%%
%Fonts%%%%%%%%%%%%%%%%%%%%%%%%%%%%%%%%%%%%%%%%%%%%%%%%%%%%%%%%%%%%%%%%%%%%%
%%%%%%%%%%%%%%%%%%%%%%%%%%%%%%%%%%%%%%%%%%%%%%%%%%%%%%%%%%%%%%%%%%%%%%%%%%%

\font\eightrm=cmr8 at 8pt

\font\seventeenrm=cmr17 at 17pt
\font\twentyonerm=cmr17 at 21pt

\font\ss=cmss10

\font\csc=cmcsc10

\font\twelvecal=cmsy10 at 12pt

\font\twelvemath=cmmi12

\font\seventeenbold=cmbx7 at 17pt

\font\fively=lasy5
\font\sevenly=lasy7
\font\tenly=lasy10

\textfont10=\tenly
\scriptfont10=\sevenly
\scriptscriptfont10=\fively
%%%%%%%%%%%%%%%%%%%%%%%%%%%%%%%%%%%%%%%%%%%%%%%%%%%%%%%%%%%%%%%%%%%%%%%%%%%
%Formatting%%%%%%%%%%%%%%%%%%%%%%%%%%%%%%%%%%%%%%%%%%%%%%%%%%%%%%%%%%%%%%%%
%%%%%%%%%%%%%%%%%%%%%%%%%%%%%%%%%%%%%%%%%%%%%%%%%%%%%%%%%%%%%%%%%%%%%%%%%%%
\magnification=1200
\parskip=10pt
\parindent=20pt
\def\today{\ifcase\month\or January\or February\or March\or April\or May\or
June\or July\or August\or September\or October\or November\or December\fi
       \space\number\day, \number\year}

\def\title#1{\footline={\ifnum\pageno<2\hfil
       \else\hss\tenrm\folio\hss\fi}\vskip1truein\centerline{{#1}
       \footnote{\raise1ex\hbox{*}}{\eightrm Supported in part
       by the Robert A. Welch Foundation and N.S.F. Grants
       PHY-880637 and\break PHY-8605978.}}}

\def\newpage{\vfill\eject}
\def\abstract#1{\centerline{\bf ABSTRACT}\vskip.2truein{\narrower\noindent#1
       \smallskip}}
\def\acknowledgements{\noindent\line{\bf Acknowledgements\hfill}\nobreak
    \vskip.1truein\nobreak\noindent\ignorespaces}
\def\runninghead#1#2{\voffset=2\baselineskip\nopagenumbers
       \headline={\ifodd\pageno\rightheadline\else \leftheadline\fi}
       \def\rightheadline{{\sl#1}\hfill{\rm\folio}}
       \def\leftheadline{{\rm\folio}\hfill{\sl#2}}}

\newcount\footnoteno
\def\Footnote#1{\advance\footnoteno by 1
                \let\SF=\empty
                \ifhmode\edef\SF{\spacefactor=\the\spacefactor}\/\fi
                $^{\the\footnoteno}$\ignorespaces
                \SF\vfootnote{$^{\the\footnoteno}$}{#1}}

\def\figbox#1#2#3{\vbox{\vskip15pt
                   \vbox{\hrule
                    \hbox{\vrule
                     \vbox{\vskip12truept\centerline #1 \vskip6truept
                          {\hskip.4truein\vbox{\hsize=5truein\noindent
                          {\bf Figure\hskip5truept#2:}\hskip5truept#3}}
                     \vskip18truept}
                    \vrule}
                   \hrule}}}
\def\place#1#2#3{\vbox to0pt{\kern-\parskip\kern-7pt
                             \kern-#2truein\hbox{\kern#1truein #3}
                             \vss}\nointerlineskip}
\def\figurecaption#1#2{\kern.75truein\vbox{\hsize=5truein\noindent{\bf Figure
    \figlabel{#1}:} #2}}
\def\tablecaption#1#2{\kern.75truein\lower12truept\hbox{\vbox{\hsize=5truein
    \noindent{\bf Table\hskip5truept\tablabel{#1}:} #2}}}
\def\boxed#1{\lower3pt\hbox{
                       \vbox{\hrule\hbox{\vrule

\vbox{\kern2pt\hbox{\kern3pt#1\kern3pt}\kern3pt}\vrule}
                         \hrule}}}
%%%%%%%%%%%%%%%%%%%%%%%%%%%%%%%%%%%%%%%%%%%%%%%%%%%%%%%%%%%%%%%%%%%%%%%%
%Greek characters%%%%%%%%%%%%%%%%%%%%%%%%%%%%%%%%%%%%%%%%%%%%%%%%%%%%%%%
%%%%%%%%%%%%%%%%%%%%%%%%%%%%%%%%%%%%%%%%%%%%%%%%%%%%%%%%%%%%%%%%%%%%%%%%
\def\a{\alpha}
\def\b{\beta}
\def\g{\gamma}
\def\d{\delta}\def\D{\Delta}

\def\th{\theta}

\def\l{\lambda}\def\L{\Lambda}

%%%%%%%%%%%%%%%%%%%%%%%%%%%%%%%%%%%%%%%%%%%%%%%%%%%%%%%%%%%%%%%%%%%%%%%%
%Calligraphic capitals%%%%%%%%%%%%%%%%%%%%%%%%%%%%%%%%%%%%%%%%%%%%%%%%%%
%%%%%%%%%%%%%%%%%%%%%%%%%%%%%%%%%%%%%%%%%%%%%%%%%%%%%%%%%%%%%%%%%%%%%%%%
\def\ca#1{\relax\ifmmode {{\cal #1}}\else $\cal #1$\fi}

\def\calb{{\cal B}}

\def\calm{{\cal M}}

%%%%%%%%%%%%%%%%%%%%%%%%%%%%%%%%%%%%%%%%%%%%%%%%%%%%%%%%%%%%%%%%%%%%%%%%
% Poor man's Blackboard Bold%%%%%%%%%%%%%%%%%%%%%%%%%%%%%%%%%%%%%%%%%%%%
%%%%%%%%%%%%%%%%%%%%%%%%%%%%%%%%%%%%%%%%%%%%%%%%%%%%%%%%%%%%%%%%%%%%%%%%
\def\inbar{\vrule height1.5ex width.4pt depth0pt}
\def\IB{\relax{\rm I\kern-.18em B}}
\def\IC{\relax\hbox{\kern.25em$\inbar\kern-.3em{\rm C}$}}
\def\ID{\relax{\rm I\kern-.18em D}}
\def\IE{\relax{\rm I\kern-.18em E}}
\def\IF{\relax{\rm I\kern-.18em F}}
\def\IG{\relax\hbox{\kern.25em$\inbar\kern-.3em{\rm G}$}}
\def\IH{\relax{\rm I\kern-.18em H}}
\def\II{\relax{\rm I\kern-.18em I}}
\def\IK{\relax{\rm I\kern-.18em K}}
\def\IL{\relax{\rm I\kern-.18em L}}
\def\IM{\relax{\rm I\kern-.18em M}}
\def\IN{\relax{\rm I\kern-.18em N}}
\def\IO{\relax\hbox{\kern.25em$\inbar\kern-.3em{\rm O}$}}
\def\IP{\relax{\rm I\kern-.18em P}}
\def\IQ{\relax\hbox{\kern.25em$\inbar\kern-.3em{\rm Q}$}}
\def\IR{\relax{\rm I\kern-.18em R}}
\def\IZ{\relax\ifmmode\hbox{\ss Z\kern-.4em Z}\else{\ss Z\kern-.4em Z}\fi}
\def\IGa{\relax{\rm I}\kern-.18em\Gamma}
\def\IPi{\relax{\rm I}\kern-.18em\Pi}
\def\ITh{\relax\hbox{\kern.25em$\inbar\kern-.3em\Theta$}}
\def\IOm{\relax\thinspace\inbar\kern1.95pt\inbar\kern-5.525pt\Omega}

%Papers, Lecture Notes on Complex Manifolds etc.

\def\ie{{\it i.e.\ \/}}

\def\noblackboxes{\overfullrule=0pt}

\def\cy{Calabi--Yau}
\def\cym{Calabi--Yau manifold}
\def\cys{Calabi--Yau manifolds}

\def\H#1#2{\relax\ifmmode {H^{#1#2}}\else $H^{#1 #2}$\fi}
\def\M{\relax\ifmmode{\calm}\else $\calm$\fi}

\def\Bigcheck{\lower3.8pt\hbox{\smash{\hbox{{\twentyonerm \v{}}}}}}
\def\bigboldcheck{\smash{\hbox{{\seventeenbold\v{}}}}}

\def\Bighat{\lower3.8pt\hbox{\smash{\hbox{{\twentyonerm \^{}}}}}}

\def\Msharp{\relax\ifmmode{\calm^\sharp}\else $\smash{\calm^\sharp}$\fi}
\def\Mflat{\relax\ifmmode{\calm^\flat}\else $\smash{\calm^\flat}$\fi}
\def\preMcheck{\kern2pt\hbox{\Bigcheck\kern-12pt{$\cal M$}}}
\def\Mcheck{\relax\ifmmode\preMcheck\else $\preMcheck$\fi}
\def\preMhat{\kern2pt\hbox{\Bighat\kern-12pt{$\cal M$}}}
\def\Mhat{\relax\ifmmode\preMhat\else $\preMhat$\fi}

\def\Bsharp{\relax\ifmmode{\calb^\sharp}\else $\calb^\sharp$\fi}
\def\Bflat{\relax\ifmmode{\calb^\flat}\else $\calb^\flat$ \fi}
\def\preBcheck{\hbox{\Bigcheck\kern-9pt{$\cal B$}}}
\def\Bcheck{\relax\ifmmode\preBcheck\else $\preBcheck$\fi}
\def\preBhat{\hbox{\Bighat\kern-9pt{$\cal B$}}}
\def\Bhat{\relax\ifmmode\preBhat\else $\preBhat$\fi}

\def\figBcheck{\kern3pt\hbox{\raise1pt\hbox{\bigboldcheck}\kern-11pt
    {\twelvecal B}}}
\def\figBsharp{{\twelvecal B}\raise5pt\hbox{$\twelvemath\sharp$}}
\def\figBflat{{\twelvecal B}\raise5pt\hbox{$\twelvemath\flat$}}

\def\gcheck{\hbox{\lower2.5pt\hbox{\Bigcheck}\kern-8pt$\g$}}
\def\lhat{\hbox{\raise.5pt\hbox{\Bighat}\kern-8pt$\l$}}

\def\Fcheck{\kern2pt\hbox{\raise1pt\hbox{\Bigcheck}\kern-10pt{$\cal F$}}}
\def\Fhat{\kern2pt\hbox{\raise1pt\hbox{\Bighat}\kern-10pt{$\cal F$}}}

\def\cp#1{\relax\ifmmode {\IP\kern-2pt{}_{#1}}\else $\IP\kern-2pt{}_{#1}$\fi}
\def\h#1#2{\relax\ifmmode {b_{#1#2}}\else $b_{#1#2}$\fi}

\def\frac#1#2{{#1\over #2}}

\def\pd#1#2{{\partial #1\over\partial #2}}

\def\cone{\relax\thinspace\hbox{$<\kern-.8em{)}$}}
\mathchardef\mho"0A30

\def\-{\hphantom{-}}

%References

\def\npb#1{Nucl.\ Phys.\ {\bf B#1}}

\def\cmp#1{Commun. Math. Phys. {\bf #1}}
\def\plb#1{Phys. Lett. {\bf #1B}}

%%%%%%%%%%%%%%%%%%%%%%%%%%%%%%%%%%%%%%%%%%%%%%%%%%%%%%%%%%%%%%%%%%%%%%%%%%
%Equation macros%%%%%%%%%%%%%%%%%%%%%%%%%%%%%%%%%%%%%%%%%%%%%%%%%%%%%%%%%%
%%%%%%%%%%%%%%%%%%%%%%%%%%%%%%%%%%%%%%%%%%%%%%%%%%%%%%%%%%%%%%%%%%%%%%%%%%

\newif\ifproofmode
\proofmodefalse

\newif\ifforwardreference
\forwardreferencefalse

\newif\ifchapternumbers
\chapternumbersfalse

\newif\ifcontinuousnumbering
\continuousnumberingfalse

\newif\iffigurechapternumbers
\figurechapternumbersfalse

\newif\ifcontinuousfigurenumbering
\continuousfigurenumberingfalse

\newif\iftablechapternumbers
\tablechapternumbersfalse

\newif\ifcontinuoustablenumbering
\continuoustablenumberingfalse

\font\eqsixrm=cmr6

\def\marginstyle{\eqsixrm}

\newtoks\chapletter
\newcount\chapno
\newcount\eqlabelno
\newcount\figureno
\newcount\tableno

\chapno=0
\eqlabelno=0
\figureno=0
\tableno=0

\def\chapfolio{\ifnum\chapno>0 \the\chapno\else\the\chapletter\fi}

\def\bumpchapno{\ifnum\chapno>-1 \global\advance\chapno by 1
\else\global\advance\chapno by -1 \setletter\chapno\fi
\ifcontinuousnumbering\else\global\eqlabelno=0 \fi
\ifcontinuousfigurenumbering\else\global\figureno=0 \fi
\ifcontinuoustablenumbering\else\global\tableno=0 \fi}

\def\setletter#1{\ifcase-#1{}\or{}%
\or\global\chapletter={A}%
\or\global\chapletter={B}%
\or\global\chapletter={C}%
\or\global\chapletter={D}%
\or\global\chapletter={E}%
\or\global\chapletter={F}%
\or\global\chapletter={G}%
\or\global\chapletter={H}%
\or\global\chapletter={I}%
\or\global\chapletter={J}%
\or\global\chapletter={K}%
\or\global\chapletter={L}%
\or\global\chapletter={M}%
\or\global\chapletter={N}%
\or\global\chapletter={O}%
\or\global\chapletter={P}%
\or\global\chapletter={Q}%
\or\global\chapletter={R}%
\or\global\chapletter={S}%
\or\global\chapletter={T}%
\or\global\chapletter={U}%
\or\global\chapletter={V}%
\or\global\chapletter={W}%
\or\global\chapletter={X}%
\or\global\chapletter={Y}%
\or\global\chapletter={Z}\fi}

\def\tempsetletter#1{\ifcase-#1{}\or{}%
\or\global\chapletter={A}%
\or\global\chapletter={B}%
\or\global\chapletter={C}%
\or\global\chapletter={D}%
\or\global\chapletter={E}%
\or\global\chapletter={F}%
\or\global\chapletter={G}%
\or\global\chapletter={H}%
\or\global\chapletter={I}%
\or\global\chapletter={J}%
\or\global\chapletter={K}%
\or\global\chapletter={L}%
\or\global\chapletter={M}%
\or\global\chapletter={N}%
\or\global\chapletter={O}%
\or\global\chapletter={P}%
\or\global\chapletter={Q}%
\or\global\chapletter={R}%
\or\global\chapletter={S}%
\or\global\chapletter={T}%
\or\global\chapletter={U}%
\or\global\chapletter={V}%
\or\global\chapletter={W}%
\or\global\chapletter={X}%
\or\global\chapletter={Y}%
\or\global\chapletter={Z}\fi}

\def\chapshow#1{\ifnum#1>0 \relax#1%
\else{\tempsetletter{\number#1}\chapno=#1\chapfolio}\fi}

\def\chaplabel#1{\bumpchapno\ifproofmode\ifforwardreference
\immediate\write1{\noexpand\expandafter\noexpand\def
\noexpand\csname CHAPLABEL#1\endcsname{\the\chapno}}\fi\fi
\global\expandafter\edef\csname CHAPLABEL#1\endcsname
{\the\chapno}\ifproofmode\llap{\hbox{\marginstyle #1\ }}\fi\chapfolio}

\def\eqnum{\global\advance\eqlabelno by 1
\eqno(\ifchapternumbers\chapfolio.\fi\the\eqlabelno)}

\def\eqlabel#1{\global\advance\eqlabelno by 1 \ifproofmode\ifforwardreference
\immediate\write1{\noexpand\expandafter\noexpand\def
\noexpand\csname EQLABEL#1\endcsname{\the\chapno.\the\eqlabelno?}}\fi\fi
\global\expandafter\edef\csname EQLABEL#1\endcsname
{\the\chapno.\the\eqlabelno?}\eqno(\ifchapternumbers\chapfolio.\fi
\the\eqlabelno)\ifproofmode\rlap{\hbox{\marginstyle #1}}\fi}

\def\eqalignnum{\global\advance\eqlabelno by 1
&(\ifchapternumbers\chapfolio.\fi\the\eqlabelno)}

\def\eqalignlabel#1{\global\advance\eqlabelno by 1 \ifproofmode
\ifforwardreference\immediate\write1{\noexpand\expandafter\noexpand\def
\noexpand\csname EQLABEL#1\endcsname{\the\chapno.\the\eqlabelno?}}\fi\fi
\global\expandafter\edef\csname EQLABEL#1\endcsname
{\the\chapno.\the\eqlabelno?}&(\ifchapternumbers\chapfolio.\fi
\the\eqlabelno)\ifproofmode\rlap{\hbox{\marginstyle #1}}\fi}

\def\eqref#1{\hbox{(\ifundefined{EQLABEL#1}***)\ifproofmode\ifforwardreference%
\else\write16{ ***Undefined Equation Reference #1*** }\fi
\else\write16{ ***Undefined Equation Reference #1*** }\fi
\else\edef\LABxx{\getlabel{EQLABEL#1}}%
\def\LAByy{\expandafter\stripchap\LABxx}\ifchapternumbers%
\chapshow{\LAByy}.\expandafter\stripeq\LABxx%
\else\ifnum\number\LAByy=\chapno\relax\expandafter\stripeq\LABxx%
\else\chapshow{\LAByy}.\expandafter\stripeq\LABxx\fi\fi)\fi}%
\ifproofmode\write2{Equation #1}\fi}

\def\fignum{\global\advance\figureno by 1
\relax\iffigurechapternumbers\chapfolio.\fi\the\figureno}

\def\figlabel#1{\global\advance\figureno by 1
\relax\ifproofmode\ifforwardreference
\immediate\write1{\noexpand\expandafter\noexpand\def
\noexpand\csname FIGLABEL#1\endcsname{\the\chapno.\the\figureno?}}\fi\fi
\global\expandafter\edef\csname FIGLABEL#1\endcsname
{\the\chapno.\the\figureno?}\iffigurechapternumbers\chapfolio.\fi
\ifproofmode\llap{\hbox{\marginstyle#1
\kern1.2truein}}\relax\fi\the\figureno}

%% FOLLOWING LINE CANNOT BE BROKEN BEFORE 80 CHAR
\def\figref#1{\hbox{\ifundefined{FIGLABEL#1}!!!!\ifproofmode\ifforwardreference%
\else\write16{ ***Undefined Figure Reference #1*** }\fi
\else\write16{ ***Undefined Figure Reference #1*** }\fi
\else\edef\LABxx{\getlabel{FIGLABEL#1}}%
\def\LAByy{\expandafter\stripchap\LABxx}\iffigurechapternumbers%
\chapshow{\LAByy}.\expandafter\stripeq\LABxx%
\else\ifnum \number\LAByy=\chapno\relax\expandafter\stripeq\LABxx%
\else\chapshow{\LAByy}.\expandafter\stripeq\LABxx\fi\fi\fi}%
\ifproofmode\write2{Figure #1}\fi}

\def\tabnum{\global\advance\tableno by 1
\relax\iftablechapternumbers\chapfolio.\fi\the\tableno}

\def\tablabel#1{\global\advance\tableno by 1
\relax\ifproofmode\ifforwardreference
\immediate\write1{\noexpand\expandafter\noexpand\def
\noexpand\csname TABLABEL#1\endcsname{\the\chapno.\the\tableno?}}\fi\fi
\global\expandafter\edef\csname TABLABEL#1\endcsname
{\the\chapno.\the\tableno?}\iftablechapternumbers\chapfolio.\fi
\ifproofmode\llap{\hbox{\marginstyle#1
\kern1.2truein}}\relax\fi\the\tableno}

%% FOLLOWING LINE CANNOT BE BROKEN BEFORE 80 CHAR
\def\tabref#1{\hbox{\ifundefined{TABLABEL#1}!!!!\ifproofmode\ifforwardreference%
\else\write16{ ***Undefined Table Reference #1*** }\fi
\else\write16{ ***Undefined Table Reference #1*** }\fi
\else\edef\LABtt{\getlabel{TABLABEL#1}}%
\def\LABTT{\expandafter\stripchap\LABtt}\iftablechapternumbers%
\chapshow{\LABTT}.\expandafter\stripeq\LABtt%
\else\ifnum\number\LABTT=\chapno\relax\expandafter\stripeq\LABtt%
\else\chapshow{\LABTT}.\expandafter\stripeq\LABtt\fi\fi\fi}%
\ifproofmode\write2{Table#1}\fi}

\def\eq{Eq.~}

\newdimen\sectionskip     \sectionskip=20truept
\newcount\sectno
\def\section#1#2{\sectno=0 \null\vskip\sectionskip
    \centerline{\chaplabel{#1}.~~{\bf#2}}\nobreak\vskip.2truein
    \noindent\ignorespaces}

\def\advancesectno{\global\advance\sectno by 1}
\def\sectfolio{\number\sectno}
\def\subsection#1{\goodbreak\advancesectno\null\vskip10pt
                  \noindent\chapfolio.~\sectfolio.~{\bf #1}
                  \nobreak\vskip.05truein\noindent\ignorespaces}

\def\uttg#1{\null\vskip.1truein
    \ifproofmode \line{\hfill{\bf Draft}:
    UTTG--{#1}--\number\year}\line{\hfill\today}
    \else \line{\hfill UTTG--{#1}--\number\year}
    \line{\hfill\ifcase\month\or January\or February\or March\or April\or
May\or June
    \or July\or August\or September\or October\or November\or December\fi
    \space\number\year}\fi}

\def\contents{\noindent
   {\bf Contents\Z}\nobreak\vskip.05truein\noindent\ignorespaces}

\def\getlabel#1{\csname#1\endcsname}
\def\ifundefined#1{\expandafter\ifx\csname#1\endcsname\relax}
\def\stripchap#1.#2?{#1}
\def\stripeq#1.#2?{#2}

%%%%%%%%%%%%%%%%%%%%%%%%%%%%%%%%%%%%%%%%%%%%%%%%%%%%%%%%%%%%%%%%%%%%%%%
%Reference macros%%%%%%%%%%%%%%%%%%%%%%%%%%%%%%%%%%%%%%%%%%%%%%%%%%%%%%
%%%%%%%%%%%%%%%%%%%%%%%%%%%%%%%%%%%%%%%%%%%%%%%%%%%%%%%%%%%%%%%%%%%%%%%
%
\catcode`@=11 % This allows us to modify PLAIN macros.
\def\space@ver#1{\let\@sf=\empty\ifmmode#1\else\ifhmode%
\edef\@sf{\spacefactor=\the\spacefactor}\unskip${}#1$\relax\fi\fi}
\newcount\referencecount     \referencecount=0
\newif\ifreferenceopen       \newwrite\referencewrite
\newtoks\rw@toks
\def\refmark#1{\relax[#1]}
\def\refend{\refmark{\number\referencecount}}
\newcount\lastrefsbegincount \lastrefsbegincount=0
\def\refsend{\refmark{\count255=\referencecount%
\advance\count255 by -\lastrefsbegincount%
\ifcase\count255 \number\referencecount%
\or\number\lastrefsbegincount,\number\referencecount%
\else\number\lastrefsbegincount-\number\referencecount\fi}}
\def\refch@ck{\chardef\rw@write=\referencewrite
\ifreferenceopen\else\referenceopentrue
\immediate\openout\referencewrite=referenc.texauxil \fi}
%
% In \obeyendofline, we say `\let^^M=\relax
{\catcode`\^^M=\active % these lines must end with %
  \gdef\obeyendofline{\catcode`\^^M\active \let^^M\ }}%
%
% In \ignoreendofline, we say `\let^^M=\relax
{\catcode`\^^M=\active % these lines must end with %
  \gdef\ignoreendofline{\catcode`\^^M=5}}
{\obeyendofline\gdef\rw@start#1{\def\t@st{#1}\ifx\t@st\blankend%
\endgroup\@sf\relax\else\ifx\t@st\bl@nkend\endgroup\@sf\relax%
\else\rw@begin#1
\backtotext
\fi\fi}}
{\obeyendofline\gdef\rw@begin#1
{\def\n@xt{#1}\rw@toks={#1}\relax%
\rw@next}}
\def\blankend{}
{\obeylines\gdef\bl@nkend{
}}
\newif\iffirstrefline  \firstreflinetrue
\def\rwr@teswitch{\ifx\n@xt\blankend\let\n@xt=\rw@begin%
\else\iffirstrefline\global\firstreflinefalse%
\immediate\write\rw@write{\noexpand\obeyendofline\the\rw@toks}%
\let\n@xt=\rw@begin%
\else\ifx\n@xt\rw@@d \def\n@xt{\immediate\write\rw@write{%
\noexpand\ignoreendofline}\endgroup\@sf}%
\else\immediate\write\rw@write{\the\rw@toks}%
\let\n@xt=\rw@begin\fi\fi\fi}
\def\rw@next{\rwr@teswitch\n@xt}
\def\rw@@d{\backtotext} \let\rw@end=\relax
\let\backtotext=\relax

\newdimen\refindent     \refindent=30pt
\def\Textindent#1{\noindent\llap{#1\enspace}\ignorespaces}
\def\refitem#1{\par\hangafter=0 \hangindent=\refindent\Textindent{#1}}
\def\REFNUM#1{\space@ver{}\refch@ck\firstreflinetrue%
\global\advance\referencecount by 1 \xdef#1{\the\referencecount}}
\def\refnum#1{\space@ver{}\refch@ck\firstreflinetrue%
\global\advance\referencecount by 1\xdef#1{\the\referencecount}\refend}

\def\REF#1{\REFNUM#1%
\immediate\write\referencewrite{%
\noexpand\refitem{#1.}}%
\begingroup\obeyendofline\rw@start}
\def\ref{\refnum\?%
\immediate\write\referencewrite{\noexpand\refitem{\?.}}%
\begingroup\obeyendofline\rw@start}
\def\Ref#1{\refnum#1%
\immediate\write\referencewrite{\noexpand\refitem{#1.}}%
\begingroup\obeyendofline\rw@start}
\def\REFS#1{\REFNUM#1\global\lastrefsbegincount=\referencecount%
\immediate\write\referencewrite{\noexpand\refitem{#1.}}%
\begingroup\obeyendofline\rw@start}

\def\REFSCON#1{\REF#1}

\def\cite#1{\refmark#1}
\catcode`@=12 % at signs are no longer letters
%

%%%%%%%%%%%%%%%%%%%%%%%%%%%%%%%% other macros %%%%%%%%%%%%%%%%%%%%%%%%%%%%%%%%

\def\hourandminute{\count255=\time\divide\count255 by 60
\xdef\hour{\number\count255}
\multiply\count255 by -60\advance\count255 by\time
\hour:\ifnum\count255<10 0\fi\the\count255}

\def\subsection#1{\goodbreak\advancesectno\null\vskip10pt
                  \noindent{\it \chapfolio.\sectfolio.~#1}
                  \nobreak\vskip.05truein\noindent\ignorespaces}

\def\subsubsection#1{\goodbreak\null\vskip10pt
                     \noindent$\underline{\hbox{#1}}$
                     \nobreak\vskip.05truein\noindent\ignorespaces}

\def\cite#1{\refmark{#1}}

\def\:{\kern-1.5truept}

\def\\{\hfill\break}

\def\cropen#1{\crcr\noalign{\vskip #1}}%For use with \eqalign

\def\contents{\line{{\bf Contents}\hfill}\nobreak\vskip.05truein\noindent%
              \ignorespaces}

\def\bigcp{$
    \hbox{\seventeenrm I\kern-.18em P}_{\hbox{\seventeenrm\kern-.3em 4}}$}

\def\M{\ca{M}}

\def\W{\ca{W}}

\def\LG{Landau--Ginzburg}

\def\vone{{\bf 1}}

\def\vk{{\bf k}}

\def\khat{\hat{k}}

\def\dhat{\hat{d}}

\def\phat{\hat{p}}

\def\vkhat{{\bf \khat}}

\def\vm{{\bf m}}

\def\va{{\bf a}}

\def\vv{{\bf v}}

\def\vga{{\vec \a}}

\def\vgb{{\vec \b}}

\def\vgc{{\vec \g}}

\def\vw{{\bf w}}

\def\vzero{{\bf 0}}

\def\rV{\hbox{V}}

%%%%%%%%%%%%%%%%%%%%%%%%%%%%%%%%% set flags %%%%%%%%%%%%%%%%%%%%%%%%%%%%%%%%

\proofmodefalse
\baselineskip=13pt plus 2pt minus 1pt
\parskip=2pt
\chapternumberstrue
\figurechapternumberstrue
\tablechapternumberstrue
\forwardreferencefalse
\ifproofmode
\immediate\openout2=allcrossreferfile \fi
\ifforwardreference\input labelfile
\ifproofmode\immediate\openout1=labelfile \fi\fi
\noblackboxes
\hfuzz=1pt
\vfuzz=2pt

%%%%%%%%%%%%%%%%%%%%%%%%%%%%%%%%%% the paper %%%%%%%%%%%%%%%%%%%%%%%%%%%%%%%
\nopagenumbers\pageno=0
\null\vskip-30pt
\rightline{\eightrm IASSNS-HEP-94/100}\vskip-3pt
\rightline{\eightrm NEIP-94-009}\vskip-3pt
\rightline{\eightrm OSU-M-93-3}\vskip-3pt
\rightline{\eightrm UTTG-25-93}\vskip-3pt
\rightline{\eightrm hep-th/9412117}\vskip-3pt
\rightline{\eightrm December 13, 1994}
\vskip.7truein
\centerline{\seventeenrm Mirror Symmetry for \cy\ Hypersurfaces}
\vskip.1in
\centerline{\seventeenrm  in Weighted \smash{\bigcp}\ and}
\vskip.1in
\centerline{\seventeenrm \hphantom{\ \raise6pt\hbox{*}}Extensions of
            Landau--Ginzburg Theory\ \smash{\raise6pt\hbox{*}}}
\vfootnote{\eightrm *}{\eightrm Supported in part
       by the Robert A. Welch Foundation, the Swiss
       National Science Foundation, a WorldLab Fellowship,
       N.S.F. grants DMS-9311386 and PHY-9009850 and a grant
       from the Friends of The Institute for Advanced Study.}
\vskip.4truein
\centerline{\csc Philip~Candelas$^{1,2,3}$,
                 ~Xenia~de la Ossa$^{2,4}$~~and Sheldon Katz$^5$}
\vskip.3truein\bigskip
\centerline{
\vtop{\hsize = 2.0truein
\centerline{$^1$\it Theory Group}
\centerline{\it Department of Physics}
\centerline{\it University of Texas}
\centerline{\it Austin, TX 78712, USA}}
\qquad
\vtop{\hsize = 2.0truein
\centerline{$^2$\it Institute for Advanced Study}
\centerline{\it School of Natural Sciences}
\centerline{\it Olden Lane}
\centerline{\it Princeton, NJ 08540, USA}}
\qquad
\vtop{\hsize = 2.0truein
\centerline {$^3$\it Theory Division}
\centerline {\it  CERN}
\centerline {\it CH-1211 Geneva 23}
\centerline {\it Switzerland}}}
\vskip0.3truein
\centerline{
\vtop{\hsize = 2.0truein
\centerline{$^4$\it Institut de Physique}
\centerline{\it Universit\'e\:\ de\:\ Neuch\^atel}
\centerline{\it CH-2000 Neuch\^atel}
\centerline{\it Switzerland}}
\qquad\qquad
\vtop{\hsize = 2.0truein
\centerline{$^5$\it Department of Mathematics}
\centerline{\it Oklahoma State University}
\centerline{\it Stillwater, OK 74078}
\centerline{\it USA}}}
%
%%%%%%%%%%%%%%%%%%%%%%%%%%%%%%%%%%%%%%%%%%%%%%%%%
\vskip.5truein
\nobreak\vbox{\centerline{\bf ABSTRACT}}
\vskip.25truein
\vbox{\baselineskip 11pt\noindent Recently two groups have listed
all sets of weights $\vk = (k_1,\ldots,k_5)$ such that the weighted
projective space \smash{$\cp4^{\vk}$} admits a transverse \cy\ hypersurface.
It was noticed that the corresponding \cy\ manifolds do not form a mirror
symmetric set since some 850 of the 7555 manifolds have Hodge numbers
$(b_{11},b_{21})$ whose mirrors do not occur in the list.  By means of
Batyrev's construction we have checked that each of the 7555 manifolds
does indeed have a mirror.  The `missing mirrors' are constructed as
hypersurfaces in toric varieties.  We show that many of these manifolds
may be interpreted as non-transverse hypersurfaces in weighted
$\cp4$'s, \ie, hypersurfaces for which $dp$ vanishes at a point other than
the origin. This falls outside the usual range of Landau--Ginzburg theory.
Nevertheless Batyrev's procedure provides a way of making sense of these
theories.}
\newpage
\contents
\vskip15pt
\item{1.~}Introduction
\bigskip
\item{2.~}Toric Considerations
\itemitem{\it 2.1~}{\it Newton polyhedra and Batyrev's construction}
\itemitem{\it 2.2~}{\it Application to weighted projective spaces}
\bigskip
\item{3.~}A Generalized Transposition Rule
\itemitem{\it 3.1~}{\it The Berglund--H\"ubsch rule}
\itemitem{\it 3.2~}{\it The Berglund-H\"ubsch cases}
\itemitem{\it 3.3~}{\it A non-transverse example}
\itemitem{\it 3.4~}{\it A cautionary note}
\bigskip
\item{4.~}Manifolds with No \LG\ Phase
\itemitem{\it 4.1~}{\it A manifold whose mirror does not appear in the list}
\itemitem{\it 4.2~}{\it Phases of the model}
\bigskip
\item{5.~}Some Observations on Fractional Transformations
\itemitem{\it 5.1~}{\it A simple identification}
\itemitem{\it 5.2~}{\it Isomorphism of $\M_1$ and $\M_2$}
\itemitem{\it 5.3~}{\it Chiral rings}
\bigskip
\item{A.~}Plot of the Hodge Numbers
\newpage

%%%%%%%%%%%%%%%%%%%%%%%%%%%%%%%%%%%%%%%%%%%%%%%%%%%%%%%%%%%%%%%%%%%%%%%%%%%%
\headline={\ifproofmode\hfil\eightrm draft:\ \today\
\hourandminute\else\hfil\fi}
\pageno=1
\footline={\rm\hfil\folio\hfil}
\section{intro}{Introduction}
The considerations of this article arose in relation to the construction by
Klemm and Schimmrigk, and Kreuzer and Skarke~
\REFS\rKlSc{A.~Klemm and R.~Schimmrigk: \npb{411} (1994) 559, hep-th/9204060.}
\REFSCON\rKrSk{M.~Kreuzer and H.~Skarke: \npb{388} (1993) 113,
hep-th/9205004.}\refsend\
of a complete list, comprising
7555 cases, of all sets of weights $\vk=(k_1,k_2,k_3,k_4,k_5)$ such that the
weighted projective space \smash{$\IP_4^{\vk}$} admits a transverse
polynomial $p$ of degree \smash{$d=\sum_{j=1}^5 k_j$}.
That is the equations $dp=0$,
taken to hold in $\IC^5$, are satisfied only when all five of the coordinates
$x_j$ vanish. In this case it is known that the singularities of the weighted
space can be resolved and that the resulting hypersurface, specified by the
equation $p=0$, is a \cym
\REFS\rcls{P.~Candelas, M.~Lynker and R.~Schimmrigk: \npb{341}\ (1990) 383.}
\REFSCON\rgry{B.~R.~Greene, S.~S.~Roan and S.~T-.~Yau: \cmp{142} (1991) 142.}
\refsend.
One reason for being interested in this list is
that it manifests a compelling mirror symmetry. If one lists the Hodge numbers
$(b_{11},b_{21})$ corresponding to these manifolds then in some 90\%\ of cases
where some value of $(b_{11},b_{21})$ occurs in the list the reflected numbers
$(b_{21},b_{11})$ also occur. The list however does not manifest a complete
symmetry leading to the question ``Where are the missing mirrors?''
\REF\rKrii{M.~Kreuzer: \plb{314} (1993) 31, hep-th/9303015.}
\cite{\rKrSk,\rKrii}.
The context
in which this should be discussed is toric geometry since it is the methods of
toric geometry that permit the singularities of the ambient weighted
projective spaces
%, which properly speaking do not exist except as toric
%varieties,
to be resolved. Within toric geometry there is a powerful method,
due to Batyrev, for constructing the mirrors of a certain class of manifolds
\Ref\rBat{V.~V.~Batyrev: Duke Math.\ Journ.\ {\bf 69} (1993) 349.}.
We will
outline this construction in the following but for the present it suffices to
remark that there is a natural way to associate a four-dimensional polyhedron
to a \cy\ hypersurface in certain toric varieties. In many cases the polyhedra
associated
to \cy\ hypersurfaces have a property termed reflexivity by Batyrev.
Batyrev shows that a \cym\ can be constructed from each reflexive
polyhedron, $\D$, and observes that if $\D$ is reflexive then
the dual polyhedron, $\nabla$, is also reflexive. Hence a \cym\ $\W$ may be
constructed from $\nabla$. The new manifold has its Hodge numbers
$(b_{11},b_{21})$ reflected relative to those of $\M$. It is generally
assumed that the $\W$ so constructed is the mirror of $\M$ although this has
not been checked at the level of superconformal theories. It has not been
shown that every \cym\ gives rise to a reflexive polyhedron. However if a
given \cym\ is associated to a reflexive polyhedron then the mirror may be
constructed from the dual polyhedron.
We have checked by means of a computer program that all 7555 manifolds of the
list are indeed associated to reflexive polyhedra and so have mirrors in
virtue of Batyrev's construction
\REF\rTh{S.~Theisen, private communication.}
\Footnote{We note
that A.~Klemm has independently checked the reflexivity of the polyhedra
corresponding to the members of the list whose reflected Hodge numbers do not
appear\cite\rTh.}.

Mirror symmetry was discovered `empirically' by the generation of many models
$\IP_4^{\vk}[d]$ admitting transverse polynomials\cite{\rcls} and,
contemporaneously, by the work of Greene and Plesser
\Ref\rgp{B.~R.~Greene and M.~R.~Plesser: \npb{338} (1990) 15.}\
who explicitly constructed a class of mirror pairs and showed that the
mirror pairs corresponded to the same superconformal theory. Although
manifesting a striking mirror symmetry the list produced in \cite{\rcls}
was not perfectly symmetric and this was thought to be due, in part, to
the fact that the list of weights admitting transverse hypersurfaces was
incomplete. It was intriguing therefore when the complete list
of Refs.\cite{\rKlSc,\rKrSk} manifested an asymmetry that was greater
rather than less than the earlier list. The fact that we report here
is that each manifold of the list nevertheless has a mirror.  The
cases that were missing correspond to manifolds that cannot be realized as
transverse hypersurfaces in a weighted \cp4\ but are to be understood as
hypersurfaces in a toric variety.
In at least some cases it is possible to think of the
`missing mirrors'  as hypersurfaces in a weighted \cp4\ for which
the defining polynomial is not transverse, that is, the equations $dp=0$ are
satisfied at some point(s) apart from $x_j=0$. The condition of transversality
that was used to construct the list was employed because it was known to
guarantee that the singularities of the hypersurface $p=0$ could be resolved.
This criterion is overly strong since it can happen that a zero of $dp$ lies
on a coordinate plane where the embedding $\IP_4^\vk$ is singular. In some
cases the singularity of the embedding space can be repaired in such a way as
to produce a smooth \cym.

In the context of Witten's linear sigma models
\Ref\rWit{E.~Witten: \npb{403} (1993) 159, hep-th/9301042.},
the `missing mirrors' do
not have a Landau--Ginzburg phase (because of the non--transversality of
of $p$) but instead have interesting new phases which may be
considered as extensions of Landau--Ginzburg theories. An interesting feature
is that the description of these models requires the introduction of extra
coordinates and extra gauge symmetries associated to the blowing up of the
singularities of the ambient space.  This is in fact a general feature
of toric geometry (following D.~Cox
\Ref\rCox{D.~Cox: The Homogeneous Coordinate Ring of a Toric Variety, to
appear and alg-geom/9210008}) which can be naturally implemented
in Witten's linear sigma model.  In many
cases it is possible to eliminate these extra fields and present the model
as a
hypersurface in a $\IP_4^\vk$.  The `missing mirrors' are cases for which
this is not possible. Strictly speaking, in the great majority of cases these
extra fields should be retained in order to obtain a full description of the
phases of the model.  This is true even for models which can be represented as
hypersurfaces in a~$\IP_4^\vk$.

To underline the point that toric geometry and
Batyrev's construction are the correct way to understand mirror symmetry
we show for the manifolds of the list how the
Berglund--H\"ubsch transposition rule for finding the mirror of a given
manifold is a special case of Batyrev's method. It is perhaps worth
remarking also that this procedure provides a useful way of computing
the Hodge numbers of a \cy\ hypersurface in a weighted \cp4, numbers that were
previously calculated via \LG\ theory.

The layout of the paper is the following: in Section~2 we recall Batyrev's
procedure and describe its application to the list of weights. In Section~3
we show that the transposition rule of Berglund and H\"ubsch follows as a
special case of Batyrev's construction. In Section~4 we study a manifold
whose mirror does not appear in the list. If the mirror is interpreted as a
hypersurface in a weighted \cp4\ then the weights associated with the mirror
are such that the hypersurface cannot be transverse (this is the reason
that the mirror was not listed) thus, in this case, the mirror does not
have a \LG\ phase. The methods of toric geometry however afford a good
description of this manifold. We describe in
detail the chiral ring of this manifold and the exotic phases of the
corresponding theory. Section~5 is concerned with an
illustration of the application of toric methods to manifolds that are related
by birational transformations. The existence of such
transformations between manifolds is a pervasive phenomenon and the reason we
include this here is that these transformations tend to relate manifolds whose
Newton Polyhedra are similar and we wish to illustrate the fact that toric
geometry provides a natural framework in which this can be discussed.

We present, in an appendix, a plot of the Hodge numbers of the manifolds of the
list and of their mirrors which is now (by construction) symmetric.
\newpage
\section{toric}{Toric Considerations}
\vskip-20pt
\subsection{Newton polyhedra and Batyrev's construction}
Consider a weighted projective space $\IP_r^{(k_1,\ldots,k_{r+1})}$, and
let $d=k_1+\ldots +k_{r+1}$.  To understand hypersurfaces of degree $d$
as \cys, we apply the ideas of Batyrev\cite{\rBat}\ and
Aspinwall, Greene and Morrison
\Ref\ragm{P. S. Aspinwall, B. R. Greene, and D. R. Morrison:
Int.\ Math.\ Res.\ Notices (1993) 319, alg-geom/9309007.},
which we shall
briefly review below. The basic idea is to construct the
Newton polyhedron associated to degree $d$ monomials, and note that this
is often a reflexive polyhedron.

Let ${\bf m}=(m_1,\ldots,m_{r+1})$ be a degree vector and let
$(x_1,\ldots,x_{r+1})$ be the  homogeneous coordinates of the weighted
projective space. We denote by $x^\vm$ the monomial $x_1^{m_1}x_2^{m_2}\ldots
x_{r+1}^{m_{r+1}}$ and, as previously, we denote the weight vector by $\vk$.
The general polynomial of degree $d$ is then a linear combination
 $$
p~=~\sum_\vm c_\vm x^\vm $$
of monomials $x^\vm$ for which $\vm.\vk=d$. We shall sometimes speak of
a monomial $\vm$ as an abbreviation for the monomial $x^\vm$. The convex
hull of all $\vm$'s of degree $d$ forms the {\it Newton polyhedron\/}, $\D$, of
$p$.

If we naively formed the Newton polyhedron as the convex
hull of the set of exponents of all degree $d$ monomials, we would
typically get the point $\vone=(1,\ldots,1)$ (corresponding to the monomial
$x_1\cdots x_{r+1}$) as an interior point.  We therefore translate this vector
to the origin by subtracting $\vone$.  So given the degree $d$
monomial $x^\vm$
(which satisfies $\vk.\vm=d$),
we associate the vector $(a_1,\dots,a_{r+1})=(m_1-1,\ldots,m_{r+1}-1).$
Since we have $\vk.{\bf a}=0$, we define the lattice
 $$
\L~=~\{\,{\bf a}\in\IZ^{r+1}\mid~\vk.{\bf a}=0\,\}.$$
There is correspondingly the dual lattice
 $$
\rV~=~\IZ^{r+1}/(\IZ\cdot\vk).$$
We put $\L_{\IR}=\L\otimes\IR$ and $\rV_{\IR}=\rV\otimes\IR$; these are the
vector spaces in which the lattices are embedded.

The Newton polyhedron is therefore identified with
 $$
\D=\hbox{the convex hull of }\{\, {\bf a}\in \L\mid a_i\ge -1\ \forall i\,\}~.
$$
Note that $\L$ is a lattice of
rank $r$. If one of the weights $k_i$ ($k_1$ say) has the value unity
then we can use the equation $\vk.{\bf a}=0$ to solve for $a_1$ and take
$(a_2,\ldots,a_{r+1})$ as coordinates for~$\L$ (in this case, our computer
program will make this choice of coordinates up to a sign change).
If none of the weights is
unity then we may of course still find coordinates for $\L$ though the
procedure is more involved.

\noindent A polyhedron, $\D$, is {\it reflexive\/} if the following three
conditions obtain:
\vskip20truept
\item{i.~~}The vertices of $\D$ are integral, \ie correspond to vectors
$\vm$ whose components are integers.

\item{ii.~~}There is precisely one integral point interior to $\D$.

\item{iii.~~}The `distance' of any facet (a codimension~1 face) of
$\D$ from the interior point is 1.
\vskip20truept
\noindent By `distance' in (iii) is meant the following:
We may choose the unique interior point as the origin of coordinates. Let
$(y_2,y_3,\ldots,y_{r+1})$ be coordinates for $\L_\IR$ (if $k_1=1$ these can
be taken to be the quantities $(a_2,\ldots,a_{r+1})$~). The equation of a
facet of $\D$ has the form
 $$
l_2y_2 + l_3 y_3 +\ldots +l_{r+1}y_{r+1}~=~\d~.$$
Since the vectors $\vm$ lie on an integral lattice the quantities
$(l_2,\ldots,l_{r+1},\d)$ are rational and hence, by multiplying through by a
suitable integer if necessary, can be taken to be integers with no common
factor. Also, $\d$ may be taken positive. With this understanding the
`distance' of this face from the origin is $\d$.

In the remainder of this paper, we will always assume that $\D$ has been
translated if necessary so that the origin becomes its unique integral
interior point.

As a simple example consider the quintic threefold $\cp4[5]$. Here all the
weights are unity and $\D$ is the set of integral points $(m_2,m_3,m_4,m_5)$
such that
 $$
0\leq m_i~,~~~i=2,\ldots,5~~~~\hbox{and}~~~~\sum_{i=2}^5m_i \leq 5 $$
which is a simplex. An interior point is such that these inequalities are
satisfied with strict inequality. When this is the case we have
 $$
1\leq m_i~,~~~i=2,\ldots,5~~~~\hbox{and}~~~~\sum_{i=2}^5m_i \leq 4 $$
which has the unique solution $m_2=m_3=m_4=m_5=1$. If we take the interior
point as the origin and write $a_i=m_i - 1$ then we see that the five facets
of the simplex are given by the five equations
 $$\eqalign{
{}-a_2~&=~1\cr
{}-a_3~&=~1\cr
{}-a_4~&=~1\cr
{}-a_5~&=~1\cr
a_2 + a_3 + a_4 + a_5~&=~1~.\cr}$$
The unique interior point corresponds to the monomial $x_1x_2x_3x_4x_5$ and
it is the case in general that the unique interior point corresponds to the
product of the homogeneous coordinates when $\D$ is reflexive.

One of the key points of Batyrev's construction is that to a convex polyhedron
$\D$ which has the origin as an interior point we may associate a dual, or
polar polyhedron $\nabla$:
 $$
\nabla = \Big\{~{\bf y}~\Big|~{\bf x}.{\bf y}\geq -1, ~\forall~{\bf x}\in \D~
\Big\}~.$$
If $\D$ is reflexive then so is $\nabla$ and Batyrev has shown that we may
associate a family of \cys\ to $\nabla$.  These \cys\ are hypersurfaces
in a toric variety $X_\nabla$ whose fan consists of the set of cones over
the faces of $\nabla$.  The hypersurfaces are associated to sections of the
anticanonical bundle of $X_\nabla$.  While
$X_\nabla$ need not be smooth, it is
Gorenstein, which means that the canonical bundle (which is a priori only
defined on the smooth locus of $X_\nabla$)
extends to a bundle on all of $X_\nabla$.  Thus sections of the anticanonical
bundle will still give \cys.

The Hodge numbers $(b_{11},b_{21})$ of
a hypersurface $\M$ of this family may be calculated directly in terms of data
derived from the Newton
polyhedron. Let $\hbox{pts}(\D)$ denote the number of integral points of
$\D$ and let $\D_r$ denote the set of $r$-dimensional faces of $\D$. Write also
$\hbox{int}(\th)$ for the number of integral points interior to a face, $\th$,
of $\D$ and define similar quantities with $\D$ and $\nabla$ interchanged.
Duality provides a unique correspondence between an $r$-dimensional face,
$\th$, of $\D$ and a $(3-r)$-dimensional face $\tilde\th$ of $\nabla$. With
this notation the formulae
\REFS\rbatmir{V. Batyrev, J.\ Alg.\ Geom. {\bf 3} (1994) 493--535,
alg-geom/9310003.}
\REFSCON\rmondiv{P. S. Aspinwall, B. R. Greene and D. R. Morrison, Int.\
Math.\ Res.\ Notices (1993) 319--337, alg-geom/9309007.}
\refsend\
for the Hodge numbers are
 $$\eqalign{
b_{21}(\D) &= \hbox{pts}(\D) - \sum_{\th\in\D_3} \hbox{int}(\th) +
\sum_{\th\in\D_2} \hbox{int}(\th)\,\hbox{int}(\tilde\th) - 5~,\cropen{5pt}
b_{11}(\D) &= \hbox{pts}(\nabla) -
\sum_{\tilde\th\in\nabla_3} \hbox{int}(\tilde\th) +
\sum_{\tilde\th\in\nabla_2} \hbox{int}(\tilde\th)\,\hbox{int}(\th) - 5~.\cr}$$
{}Here, the expressions $b_{i1}(\D)$ denote the appropriate Hodge number of the
\cy\ hypersurfaces of $X_{\nabla}$.  The notation emphasizes the role of
$\D$ as the Newton polyhedron of the \cym\ (in the toric context, $\D$
arises as the Newton polyhedron associated to sections of the anticanonical
bundle on $X_{\nabla}$).
{}From these expressions it is clear that $b_{11}$ and $b_{21}$ are exchanged
under the operation $\D\leftrightarrow\nabla$.
\subsection{Application to weighted projective spaces}
With a computer program, we can check that Newton polyhedron is reflexive
in all 7555~cases corresponding to transverse hypersurfaces in a weighted
\cp4. For each weight vector $\vk$ of the list the program makes a list of
all possible monomials, constructs the corresponding Newton polyhedron and
checks that it is reflexive. We insist here that this check was
highly nontrivial since,  as mentioned in the introduction, there was no
theorem that the polyhedra associated to these examples had to be reflexive,
except for the few cases for which the weights admit a polynomial of Fermat
type.

Let $w_1,\ldots w_5$ be the elements of the dual lattice $\rV$ (with $r=4$)
induced by the standard coordinate vectors of $\IZ^5$. Recall that the fan for
$\cp4^{\vk}$ is the simplicial fan with edges spanned by $w_1,\ldots,w_5$.
Since $w_i\in \nabla\cap \rV$
\Ref\rBKi{P.~Berglund and S.~Katz: \npb{420} (1994) 289, hep-th/9311014.},
the edges of the cones of the fan of
$\cp4^{\vk}$ are a subset of the edges of the fan of $X_\nabla$; hence
$X_\nabla$ is
birational to $\cp4^{\vk}$; it follows the the \cy\ hypersurfaces in
$X_\nabla$ are birational to the original \cy\ hypersurfaces in
$\cp4^{\vk}$. So Batyrev's construction is indeed an appropriate
one to use.  It would not have made geometric sense to work directly with
$\cp4^{\vk}$, since the hypersurfaces would have had unacceptable
singularities.

We note that this construction generalizes examples that have appeared
previously in the literature
\REFS\rcdfkm{P.~Candelas, X.~de~la~Ossa, A.~Font, S.~Katz and D.R.~Morrison:
Mirror Symmetry for Two Parameter Models -- I, \npb{416} (1994) 481,
hepth/9308083.}
\REFSCON\rhkty{S.~Hosono, A.~Klemm, S.~Theisen, and S.-T.~Yau:
Mirror Symmetry, Mirror Map, and Applications to Calabi-Yau Hypersurfaces,
HUTMP-93-0801, hep-th/9308122.}
\REFSCON\rcfkm{P.~Candelas, A.~Font, S.~Katz and D.R.~Morrison: Mirror
Symmetry for Two Parameter Models -- II, UTTG-25-93, IASSNS-HEP-94/12,
OSU-M-94-1, hep-th/9403187.}
\refsend.
We illustrate the procedure with two examples the first corresponding to
weights that admit a transverse polynomial and the second to weights that
do not.
\subsubsection{$\vk=(1,1,1,2,2)$}
Consider first an example taken from the list: the weighted projective space
$\IP_4^{(1,1,1,2,2)}[7]$. This is not of Fermat type and was not, prior to this
analysis, known to correspond to a reflexive polyhedron. The program lists
120 monomials and finds among them the 9 vertices
 $$\eqalign{
v_1\,:&~~  (  -6,\- 1,\- 1,\- 1)\cr
v_2\,:&~~  (\- 1,\- 0,  -2,\- 1)\cr
v_3\,:&~~  (\- 1,\- 0,\- 1,  -2)\cr
v_4\,:&~~  (\- 1,\- 1,  -2,\- 1)\cr
v_5\,:&~~  (\- 1,\- 1,\- 1,\- 1)\cr
v_6\,:&~~  (\- 1,\- 1,\- 1,  -2)\cr
v_7\,:&~~  (\- 0,\- 1,\- 1,  -2)\cr
v_8\,:&~~  (\- 1,  -6,\- 1,\- 1)\cr
v_9\,:&~~  (\- 0,\- 1,  -2,\- 1)~.\cr}$$
These, as discussed previously, are expressed in terms of the coordinates
$(a_2,a_3,a_4,a_5)$ for~$\L$. These vertices define a polyhedron, $\D$, with
the six facets
 $$\eqalign{
f_1\,:&~~\cr
f_2\,:&~~\cr
f_3\,:&~~\cr
f_4\,:&~~\cr
f_5\,:&~~\cr
f_6\,:&~~\cr}
\eqalign{
x_1 &= 1~,\cr
x_2 &= 1~,\cr
x_3 &= 1~,\cr
x_4 &= 1~,\cr
- x_3 - x_4  &= 1~,\cr
- x_1 - x_2 - 2\, x_3 - 2\, x_4 &= 1~,\cr}\hskip100pt
\eqalign{
&\{2, 3, 4, 5, 6, 8\}~,\cr
&\{1, 4, 5, 6, 7, 9\}~,\cr
&\{1, 3, 5, 6, 7, 8\}~,\cr
&\{1, 2, 4, 5, 8, 9\}~,\cr
&\{2, 3, 4, 6, 7, 9\}~,\cr
&\{1, 2, 3, 7, 8, 9\}~,\cr} $$
where the lists on the right correspond to the vertices that are incident on
each facet. The origin is the only integral interior point so we see that the
polyhedron is reflexive. The dual polyhedron has vertices corresponding to
the facets of $\D$
 $$\eqalign{
\tilde f_1\,:&~~  (\- 1,\- 0,\- 0,\- 0)~,\cr
\tilde f_2\,:&~~  (\- 0,\- 1,\- 0,\- 0)~,\cr
\tilde f_3\,:&~~  (\- 0,\- 0,\- 1,\- 0)~,\cr
\tilde f_4\,:&~~  (\- 0,\- 0,\- 0,\- 1)~,\cr
\tilde f_5\,:&~~  (\- 0,\- 0,  -1,  -1)~,\cr
\tilde f_6\,:&~~  (  -1,  -1,  -2,  -2)~.\cr} $$
The coordinates of each vertex being given by the coefficients in the
equation of the corresponding facet of $\D$. The equations of the facets of the
dual likewise correspond to the vertices of $\D$
 $$\eqalign{
\tilde v_1\,:&~~\cr
\tilde v_2\,:&~~\cr
\tilde v_3\,:&~~\cr
\tilde v_4\,:&~~\cr
\tilde v_5\,:&~~\cr
\tilde v_6\,:&~~\cr
\tilde v_7\,:&~~\cr
\tilde v_8\,:&~~\cr
\tilde v_9\,:&~~\cr}
\eqalign{
  6\, y_1 - y_2 - y_3 - y_4&= 1~,\cr
      - y_1 + 2\, y_3 - y_4&= 1~,\cr
      - y_1 - y_3 + 2\, y_4&= 1~,\cr
- y_1 - y_2 + 2\, y_3 - y_4&= 1~,\cr
    - y_1 - y_2 - y_3 - y_4&= 1~,\cr
- y_1 - y_2 - y_3 + 2\, y_4&= 1~,\cr
      - y_2 - y_3 + 2\, y_4&= 1~,\cr
- y_1 + 6\, y_2 - y_3 - y_4&= 1~,\cr
      - y_2 + 2\, y_3 - y_4&= 1~,\cr}\hskip100pt
\eqalign{
&\{2, 3, 4, 6\}~,\cr
&\{1, 4, 5, 6\}~,\cr
&\{1, 3, 5, 6\}~,\cr
&\{1, 2, 4, 5\}~,\cr
&\{1, 2, 3, 4\}~,\cr
&\{1, 2, 3, 5\}~,\cr
&\{2, 3, 5, 6\}~,\cr
&\{1, 3, 4, 6\}~,\cr
&\{2, 4, 5, 6\}~.\cr} $$

Note that the vertices $\tilde f_1, \tilde f_2, \tilde f_3, \tilde f_4,
\tilde f_6$
of $\nabla$ determine the simplicial fan of $\IP_4^{(1,1,1,2,2)}$.  The
remaining vertex $\tilde f_5$ lies in the interior of the cone spanned by
$\tilde f_1, \tilde f_2, \tilde f_6$, so the fan for $X_\nabla$ is obtained
by subdividing this cone and all cones which contain it.  This geometrically
corresponds to blowing up the curve $x_1=x_2=x_3=0$, which is the singular
locus of of $\IP_4^{(1,1,1,2,2)}$.  This illustrates the general point that
$X_\nabla$ is birational to $\IP_4^\vk$ and is less singular.
The reflexivity of $\D$ is in fact the starting point for the calculation of
the instanton numbers for this model
\Ref\rBKK{P.~Berglund, S.~Katz, and A.~Klemm, Mirror Symmetry for Generic
Hypersurfaces in Weighted Projective Spaces, in preparation.}.
Many of the toric calculations in that work were done using our program
as well as a similar program written later by A.~Klemm.
\subsubsection{$\vk=(1,1,1,1,5)$}
An example of a space that does not admit any transverse polynomial
is  $\IP_4^{(1,1,1,1,5)}[9]$. Since the homogeneous coordinate $X_5$ has
weight 5 a polynomial of degree 9 must have the form
 $$
p = F_9 + X_5 G_4 $$
with $F_9$ a polynomial of degree 9 and $G_4$ a polynomial of degree 4 in the
variables $(X_1,X_2,X_3,X_4)$. It is clear that all the derivatives of such a
$p$ vanish at the point $(0,0,0,0,1)$.  For this space the program lists 255
monomials and finds among them the vertices
 $$\eqalign{
v_1\,:&~~  (  -8,\- 1,\- 1,\- 1)~,\cr
v_2\,:&~~  (  -3,\- 1,\- 1,\- 0)~,\cr
v_3\,:&~~  (\- 1,  -3,\- 1,\- 0)~,\cr
v_4\,:&~~  (\- 1,\- 1,  -8,\- 1)~,\cr
v_5\,:&~~  (\- 1,\- 1,\- 1,\- 0)~,\cr
v_6\,:&~~  (\- 1,  -8,\- 1,\- 1)~,\cr
v_7\,:&~~  (\- 1,\- 1,  -3,\- 0)~,\cr
v_8\,:&~~  (\- 1,\- 1,\- 1,\- 1)~.\cr} $$
However if we examine the facets of the polyhedron we find
 $$\eqalign{
f_1\,:&~~\cr
f_2\,:&~~\cr
f_3\,:&~~\cr
f_4\,:&~~\cr
f_5\,:&~~\cr
f_6\,:&~~\cr}
\eqalign{
x_1&= 1~,\cr
x_2&= 1~,\cr
x_3&= 1~,\cr
x_4&= 1~,\cr
x_4&= 0~,\cr
- x_1 - x_2 - x_3 - 5\, x_4&= 1~,\cr}\hskip100pt
\eqalign{
&\{3, 4, 5, 6, 7, 8\}~,\cr
&\{1, 2, 4, 5, 7, 8\}~,\cr
&\{1, 2, 3, 5, 6, 8\}~,\cr
&\{1, 4, 6, 8\}~,\cr
&\{2, 3, 5, 7\}~,\cr
&\{1, 2, 3, 4, 6, 7\}~.\cr} $$
The polyhedron is not reflexive owing to the fact that there is no interior
point (an interior point would have to have $0<x_4<1$, which is impossible).
The origin now lies in the facet~$f_5$.
\newpage
\section{BHrule}{A Generalized Transposition Rule}
\vskip-20pt
\subsection{Generalization of the Berglund--H\"ubsch rule}
In this section we generalize the transposition rule of Berglund and
H\"ubsch
\Ref\rbh{P.~Berglund and T.~H\"ubsch: \npb{393} (1993) 377, hep-th/9201014.}.
For a review and examples see~
\Ref\rBKii{P.~Berglund and S.~Katz: Mirror Symmetry Constructions: A review,
IASSNS-HEP-94/38, OSU-M-94-2,
to appear in ``Essays on Mirror Manifolds, II'', hep-th/9406008.}.

Suppose that, as previously, one starts with a weighted projective space
$\IP_r^\vk$ whose Newton polyhedron
$\D$ is reflexive.  Suppose that one is also given
$r+1$~monomials $\vm_1,\ldots,\vm_{r+1}$ of degree $d$.
Let ${\bf a}_i=\vm_i-\vone$, so that $\va_i\in \L$.  Suppose in addition that
the $\va_i$ span $\L_\IR$.  Note that we do not require that the
general polynomial formed from these $r+1$~monomials be transverse.

Form the matrix $M=(\vm_1^T,\ldots,\vm_{r+1}^T)$ of exponents of the terms of
the polynomial $p~=~\sum_\vm c_\vm x^\vm $, this
is an $(r+1)\times (r+1)$ matrix (we
think of $\vm$ and $\vk$ as row vectors). Then $\vk M=d\vone$.  Equivalently,
consider the matrix $A$ obtained from $M$ by subtracting~1 from each entry, to
correspond to the translated polyhedron.  Then $\vk A=\vzero$, the zero~vector.

Our assumptions imply that $A$ has rank~$r$, since $\L$ has rank $r$
and the $\va_i$ span $\L_\IR$.  This implies that
there are uniquely determined (up to an overall sign)
relatively prime integers $\khat_i$ such that
 $$
\sum_{i=1}^{r+1}\khat_i\va_i=\vzero~.\eqlabel{newwts}$$
In other words, we have $\vkhat A^T={\bf 0}$,
where $\vkhat$ is the
vector $(\khat_1,\ldots,\khat_{r+1})$.  This can of course be rephrased as
 $$
\sum_{i=1}^{r+1}\khat_i\vm_i=\dhat\vone~,\eqlabel{newdeg}$$
where $\dhat=\sum_i\khat_i$.
We make the final assumption that the
$\khat_i$ all have the same sign, and in particular may be chosen to be all
positive.
%This is equivalent to assuming that the $\va_i$ do not all lie on
%the same side of any hyperplane passing through the origin.

With these assumptions,
our assertion is that the mirror manifold is obtained from
the original equation by the transposition rule.  That is, one transposes
$M$ to get $r+1$ new monomials in $\IP_r^\vkhat$, forms
their sum to get the transposed polynomial $\phat$, takes an appropriate
orbifold, and resolves singularities to get the mirror manifold.

More precisely, we are asserting that the conformal field theory
derived from the superpotential corresponding to $p$ is identified
via mirror symmetry with an orbifold of the theory derived from $\phat$.
While we do not have a field-theoretic proof of this assertion (see however
\Ref\rKreuzer{M.~Kreuzer: Phys.\ Lett.\ {\bf B328} (1994) 312,
hepth/9402114.}),
our confidence is based on two observations:
we can identify the symmetries of these theories, and the respective
theories are associated with a pair of polar polyhedra.

Recall that the fan for the toric~$r+1$-fold determined by the polar polyhedron
$\nabla$ is just the normal fan of $\D$, which is the
collection of cones over the proper faces of $\D$.  To find the mirror family,
this fan must further be subdivided, using all of the lattice points of
$\D$ to span new edges.  Note that this fan is a
refinement of the fan $F$ obtained from coning the proper
faces of the simplex spanned by the $r+1$~chosen lattice points.

Now the fan $\Sigma$ for $\IP_r^\vkhat$ naturally lives
inside
 $$
\rV'=\IR^{r+1}/\IR\cdot\vkhat~.\eqlabel{V'}$$
This $\Sigma$ is determined by coning the proper faces of the simplex spanned
by the vertices $\vw_i'$, where $\vw_i'$ is the element of $\rV'$ represented
by  the standard basis vector $e_i=(0,\ldots,0,1,0,\ldots,0)$ of
$\IC^{r+1}$.  There is clearly
a map from $\Sigma$ to $F$ induced by the linear map
sending $\vw_i'$ to $\va_i$.  By simple considerations of toric geometry
this corresponds to a finite quotient mapping
\Ref\rFul{W.~Fulton: Introduction to Toric Varieties, Princeton University
Press, 1993.}.
The process of refinement of $F$ to get the subdivided
normal fan corresponds to a birational transformation.  In summary, the
mirror family sits inside a partially desingularized orbifold of
$\IP_r^\vkhat$.

We now recall from
\cite{\rBKi}\ that to the points $e_i$ of $\rV$ correspond
monomials in the toric variety determined by $\nabla$, and one obtains
a polynomial from adding up these terms.  We can now observe that when
referred back to $\IP_r^\vkhat$ as described above, this
coincides with the transposed polynomial $\phat$.
In other words, we must take the toric hypersurface
given by Batyrev's procedure, then pull the equations back to
$\IP_r^\vkhat$, and check that the transposed monomials occur
among the monomials so obtained.  This can be done directly using the toric
description, since for a weighted projective space, the exponent of a
monomial belonging to a particular variable can be calculated by taking
the inner product of the lattice point corresponding to the monomial with the
standard basis vector corresponding to the variable (the one for which the
position of the ``1'' is determined by the subscript of the
variable).  This immediately gives the
desired result.  (More precisely, we obtain the columns of $A^T$ by
this procedure, then add~$\vone$ to get $M^T$.)  Examples appear
in \cite{\rBKii}.

The final thing to do is to verify that the group of geometric symmetries has
the claimed order.  Of course, the toric method gives the group explicitly,
so we have given more information than noticed by Berglund and H\"ubsch
(but see~
\Ref\rBH{P.~Berglund and M.~Henningson: Landau-Ginzburg Orbifolds, Mirror
Symmetry and the Elliptic Genus, IASSNS-HEP-93/92, hep-th/9401029.}).
To do this, we must show that mirror symmetry exchanges the groups of
geometric and quantum symmetries.  This follows from several observations.
\vskip20pt
\item{1.} The order of the
group of symmetries of the theory is just the
determinant of $M$.  Equivalently, this is also the index of the sublattice
$K$ of $\IZ^{r+1}$ spanned by the $\vm_i$.

\item{2.} Let $e$ be the index (in $\L$) of the sublattice $L$
spanned by the $\va_i$.  Then \hbox{${\rm det}(M)=d\dhat e$}.

\item{3.} The group of quantum symmetries of the manifold corresponding to
$p$ is $\IZ_d$.  The group of geometric symmetries of the manifold
corresponding to $\phat$ has order \hbox{${\rm det}(M)/\dhat=de$}.

\item{4.} The order of the orbifold given by the toric procedure is just $e$.
\vskip20pt
\noindent To establish these facts: for~1, we observe that the group of
symmetries is just $\IZ^{r+1}/K$ (the roots of unity needed to define the
symmetries arise from describing the homomorphisms from $\IZ^{r+1}/K$ to
$\IC^*$ in coordinates); 3 then follows immediately from~1.
Observation~2 is established by exhibiting coset representatives for $K$
as follows.

Choose vectors $\vga_i$
for $1\le i\le d$ such that $\vk.\vga_i=i$.  Put $\vgb_j=j\vone$ for
$1\le j\le\dhat$.  Pick coset representatives $\vgc_k$ for $1\le k\le e$ of
$L$ in $\L$.
Then the set of all vectors
$\vga_i+\vgb_j+\vgc_k$ has the desired cardinality, and is seen to be a set
of coset representatives of $K$ as follows.
To see that these vectors span all of $\IZ^{r+1}/K$, we pick an arbitrary
vector $\vv\in\IZ^{r+1}$, and write $\vk\cdot\vv=qd+i$ with $1\le i\le d$ and
$q$ integral.  Then $\vv-\vga_i-q\vm_1\in\L$.  So for some $k$ we have that
$$\vv-\vga_i-q\vm_1-\vgc_k=\sum_lr_l\va_l=\sum_lr_l(\vm_l-\vone)
\eqlabel{reduct}$$
for some integers $r_l$.  Equation~\eqref{newdeg} says that $\dhat\vone\in K$;
so can multiply out the right hand side of~\eqref{reduct}, and see that modulo
$K$ it must be
equal to $\vgb_j$ for some $j$.  Thus $\vv$ is congruent to
$\vga_i+\vgb_j+\vgc_k$ modulo $K$.
On the other hand, suppose that
$$\vga_i+\vgb_j+\vgc_k\equiv\vga_{i'}+\vgb_{j'}+\vgc_{k'}
\qquad{\rm modulo}\ K.
\eqlabel{cong}
$$
Since $\vk\cdot K=d\IZ$, premultiplying~\eqref{cong} by $\vk$ is
well-defined modulo $d$.  This gives \hbox{$i\equiv i'$}
\hbox{${\rm mod}\ d$}, which  implies that $i=i'$. This in turn implies that
$$\vgb_j+\vgc_k\equiv\vgb_{j'}+\vgc_{k'}\qquad{\rm modulo}\ K.
\eqlabel{simpcong}$$
Furthermore, $K$ is a sublattice of $L+\IZ\cdot\vone$; so the congruence
in \eqref{simpcong} holds modulo $L+\IZ\cdot\vone$.  This implies that
$\vgc_k\equiv \vgc_{k'}\ {\rm modulo}\ L+\IZ\cdot\vone$.  But $\vgc_k$
and $\vgc_{k'}$ lie in $\L$, and $(L+\IZ\cdot\vone)\cap\L = L$;  hence
$\vgc_k\equiv \vgc_{k'}\ {\rm modulo}\ L$.  Thus $k=k'$ as desired,
establishing~2.

Observation~4 follows from
toric generalities \cite{\rFul}.
\subsection{The Berglund-H\"ubsch cases}
In this subsection, we show that the results of Berglund and H\"ubsch follow
immediately from the previous considerations.

Recall that the polynomials under consideration are sums of expressions of the
following type:
\vskip5pt
\vskip-20pt
 $$\eqalign{
x^{\b}\iffigmode~=\else\fi&
\cropen{1pt}
x_1^{\b_1}x_2+x_2^{\b_2}x_3+\ldots
+x_{n-1}^{\b_{n-1}}x_n+x_n^{\b_n}\iffigmode~=\else\fi&
\cropen{4pt}
x_1^{\b_1}x_2+x_2^{\b_2}x_3+\ldots +x_{n-1}^{\b_{n-1}}x_n+x_n^{\b_n}x_1
\iffigmode~=\else\fi&\cr}
{}~~
\iffigmode%
\def\polys{\hbox{\def\epsfsize##1##2{0.4##1}\hskip0pt\vbox{%
\vskip0pt\epsfxsize=3.5truein\epsfbox{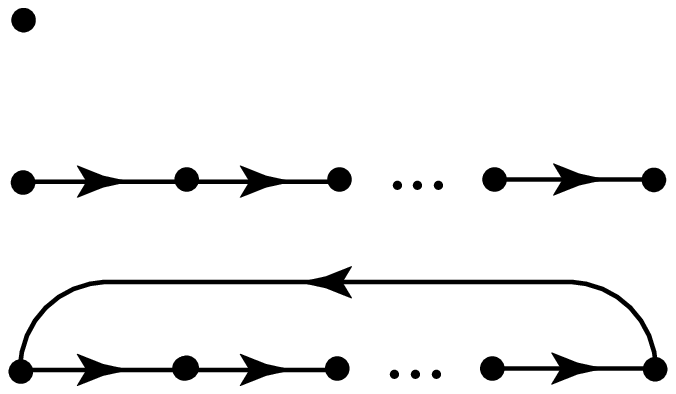}}}}\lower27pt\hbox{\polys}
\else\fi
\eqlabel{terms}$$
\vskip5pt
\noindent The polynomial $p$ is transverse for these cases.
Berglund and H\"ubsch assume further that $r=4$.
Since we now know by the computer program that $\D$ is reflexive in this case,
we can use Batyrev's construction of the mirror.

In these cases, the matrix $M$ has a simple block diagonal form, and
each of the blocks is easily seen to be nonsingular.  This implies that the
monomials $m_1,\ldots,m_5$ are linearly independent, and so span all of
$\IR^5$. Hence their translates $a_1,\ldots,a_5$ span all of $\L_\IR$.

It only remains to check that
the weights $\khat_i$ may all be taken to be positive.
This is simplified by the following claim:
if $\vv$ satisfies
$\vv M^T = c \vone$ for any constant $c$, then $\vv A^T=\vzero$.
As a consequence, this says that we must merely find
positive weights $\vkhat$ which
result in identical degrees for the five transposed monomials, and these
weights are in fact the desired weights.

To see the claim, we note that $M^T$ is also invertible, hence the equation
$\vv M^T = c \vone$ has a unique solution for $\vv$.
But since $A^T$ has rank~4 by the discussion at the beginning of this section,
there is a unique solution of $\vkhat A^T = \vzero$
for $\vkhat$ (up to multiple).  For such a $\vkhat$, we have
$(c/\dhat)\vkhat M^T = c\vone$; the uniqueness noted above shows that
$\vv=(c/\dhat)\vkhat$, which implies that $\vv A^T=\vzero$.

Now $M^T$ also has a block diagonal form.  Suppose that we can find
positive weights for the variables in each block such that
each of the transposed monomials in a block acquires the same weight.
Then we can rescale the weights in each block relative to the other blocks
to ensure that the weights of the monomials from all blocks agree with
each other.  In other words, we reduce the problem to consideration of each
of the three types given in~\eqref{terms}.

We finally check that we can find such weights for each of the three
types of blocks.  The first case is trivial, and the second is similarly
straightforward.  The third case results from calculation.  We illustrate
the calculation for the most difficult case, the ``loop'' case with $n=5$.

Here we have to solve the system of equations
 $$
\khat_1\b_1+\khat_2=\khat_2\b_2+\khat_3=\khat_3\b_3+\khat_4=
\khat_4\b_4+\khat_5=\khat_5\b_5+\khat_1~. \eqlabel{wteqn}$$
The solution is given by
 $$
\khat_i=\l^{-1}\Big[1+\b_{i-1}(\b_{i-2}-1)+\b_{i-1}\b_{i-2}\b_{i-3}(\b_{i-4}-1)
\Big]~,
\eqlabel{wtsoln}$$
where we think of the subscripts in $\b_i$ as indexed  by $\IZ_5$ and $\l$ is
chosen so as to render the $\khat_i$ mutually prime. Now each $\b_i>1$, since
otherwise the degree of each monomial in \eqref{terms} could not be equal to
$\sum_{i=1}^5k_i$, so it is clear the form of~\eqref{wtsoln} that all of the
$\khat_i$ are positive.

Finally, we would like to remark that the examples in the list that had no
known mirror did not fall within the Berglund--H\"ubsch cases.  At the time,
the transversality of the polynomial was required to have
a well defined Landau--Ginzburg theory
and transversal polynomials for these examples always have more than five
monomials.  These models then seemed to necessitate a non-square
matrix $M$ and therefore the transposition rule could not be applied.  Of
course, we now understand that it is not necessary to insist on the
transversality of the polynomial and that we only need to choose monomials
that span  the lattice  $\L_\IR$.
\subsection{A non-transverse example}
Consider the example $\IP_4[5]$, with polynomial
 $$
p=x_1^3x_2x_3+x_2^5+x_3^5+x_4^5+x_5^5~.$$
This polynomial is not transverse at $(1,0,0,0,0)$.
The transposed polynomial is
 $$
\phat=y_1^3+y_1y_2^5+y_1y_3^5+y_4^5+y_5^5$$
in $\IP_4^{(5,2,2,3,3)}[15]$.
The matrix $M$ has determinant~1875, while $d=5$ and $\dhat=15$.  Thus the
group of geometric symmetries of the transposed polynomial has order
$1875/15=125$.  To reduce this to the group of order~5 which is the group
of quantum symmetries of the original manifold, we have to take an orbifold
by a group of order~25, and this group can easily be written down explicitly
if desired.  This is seen to coincide with the toric description.
\subsection{A cautionary note}
We have given above a proof of the Berglund--H\"ubsch rule which shows that the
rule is applicable outside the domain in which it was originally stated.
Berglund and H\"ubsch required that it be possible to write a transverse
polynomial with five monomials. We have seen that it is sufficient to choose
five monomials that form a basis for $\L$. There is however a catch: which is
that it is important to keep in mind that a set of weights $\hat{\vk}$ may
arise which does not permit the existence of any transverse polynomial of
degree $\hat d = \sum\hat k_i$.
More generally, the transposed polynomial of a non-transverse polynomial
need not be transverse, thus defining a singular hypersurface in the class
\smash{$\IP^{\hat{\vk}}[\dhat]$} which should be resolved to become a smooth
(or at least less singular) hypersurface in a toric variety.
The point we want to make here is that it might happen that this
singular hypersurface corresponds to a point in the moduli space which lies in
the common boundary of the moduli spaces for two (or more) inequivalent
resolutions of the singularity. Said differently a set of weights does not
necessarily specify a unique family of \cys\ if the weights do not admit a
transverse polynomial. Moreover a given hypersurface in a toric variety need
not correspond to a hypersurface in a weighted projective space for any set of
weights.
The difficulty arises only if we insist on thinking in terms of hypersurfaces
in weighted~\cp4. As hypersurfaces in toric varieties specified by polyhedra
the varieties are well defined.

Perhaps this can be clarified by the following example. Consider the manifolds
 $$\eqalign{
\M_1&=\IP^{(7,41,247,590,885)}[1770]~,\qquad (b_{11},b_{21})=(294,36)\cr
\M_2&=\IP^{(4,41,147,343,494)}[1029]~,\qquad (b_{11},b_{21})=(293,38)\cr}$$
which have reflexive Newton polyhedra and
different Hodge numbers. (These examples were also missing a mirror
previously.) An indiscriminate application of the
Berglund--H\"ubsch rule to the polynomials
 $$\eqalign{
p_1~&=~x_1^{247} x_2 + x_2^{43} x_1 + x_3^7 x_2          + x_4^3 +x_5^2\cr
p_2~&=~x_1^{247} x_2 + x_2^{25} x_1 + x_3^7\hphantom{x_2}+ x_4^3 +x_5^2 x_2\cr}
$$
would seem to show that both of these manifolds
correspond to a mirror with weights $\hat{\vk}=(1,5,36,84,126)$.
Neither $p_1$ nor $p_2$ are transverse but can be made so by adding a
suitable monomial (one is enough in these examples).   Now
 $$\ca{W}=\IP^{(1,5,36,84,126)}[252]\quad \hbox{has} \quad
(b_{11},b_{21})=(36,294)~,$$
the Hodge numbers being calculated from the monomials and polyhedron
corresponding to weights $\hat{\vk}$. Note that the weights
$(1,5,36,84,126)$ are such that they do not admit any transverse polynomial
of degree $\hat d = 252$.  By studying the polyhedra  for
$\M_1$ and $\ca{W}$, it is possible to show that the mirror class
$\ca{W}_1$ for $\M_1$ actually coincides with $\ca{W}$.
The singular space defined by the transposed polynomial of $p_2$ in the class
$\ca{W}=\IP^{(1,5,36,84,126)}[252]$ defines a point in the moduli
space of $\ca{W}_2$ that resides in the common boundary of the
moduli spaces of each of $\ca{W}_2$ and $\ca{W}_1$ and
there is a resolution of this singularity
which produces the class $\ca{W}_2$.  To our knowledge, the class
$\ca{W}_2$ cannot be
described as a hypersurface in a weighted projective space but is a more
general hypersurface in a toric variety.
\newpage
\section{nonpoly}{Manifolds with No \LG\ Phase}
\vskip-20pt
\subsection{A manifold whose mirror does not appear in the list}
Consider the manifold $\M=\cp4^{(21,37,108,295,424)}[885]$ which has
$b_{11}=295$ and $b_{21}=7$. This is a manifold that appears in the lists of
Klemm and Schimmrigk, and Kreuzer and Skarke. No mirror of this manifold
appears
in the list. It is shown in
\Ref{\rperiods}{P.~Berglund, P.~Candelas, X.~de la Ossa, A.~Font, T.~H\"ubsch,
D.~Jan\v{c}ic and F.~Quevedo: \npb{419} (1994) 352, hep-th/9308005.}~
that the periods of a given \cym\ \M\ are most easily written as hypergeometric
functions in terms of the weights $\hat{\vk}$ of the mirror of \M. Since we are
able to write the periods directly we may read off the weights of the mirror.
Alternatively we may apply the Berglund--H\"ubsch procedure. Both procedures
give
the same result and suggest that the mirror,
$\ca{W}$, is, in some sense, the manifold $\ca{W}=\cp4^{(1,1,5,14,21)}[42]$.
The
problem is that the coordinate $x_3$ in $\ca{W}$ has weight 5 so we cannot
write
down a transverse polynomial. (This is why $\ca{W}$ was not listed.) The
best we can do is to write down a polynomial such as
 $$
p~=~x_1^{42}+x_2^{42}+x_3^8 x_1 x_2+x_4^3+x_5^2~.$$
This polynomial fails to be transverse at the point $(0,0,1,0,0)$. Since this
point does not lie in the algebraic torus we might hope to be able to proceed
via the Newton polyhedron. We find that the Newton polyhedron is reflexive and
that $\ca{W}$ has its Hodge numbers exchanged relative to \M.

In the choices made by the computer program, the vertices of $\D$ have
coordinates
$$\matrix{
  (\- 1,\-  0,\-  0,\-  0),&\hskip20pt (  -1,\-  3,\-  4,\-  5),\hskip20pt
& (\- 0,   -2,\-  2,   -1),\cropen{5pt}
  (\- 0,   -1,   -1,\-  0),&\hskip20pt (\- 0,\-  1,\-  0,\-  0),\hskip20pt
& (\- 0,\-  2,\-  2,\-  3).\cr }$$

We claim the this coincides with the polar of the Newton polyhedron of
$\cp4^{(1,1,5,14,21)}$ after a coordinate change.  The polar polyhedron has
vertices
$$
\def\skip{\hphantom{0}}
\matrix{
  (  -1,  -5, -14, -21), &\hskip20pt (\- 1,\- 0,\- 0,\- 0),\hskip20pt
& (\- 0,\- 1,\- 0,\-\skip 0),   \cropen{5pt}
  (\- 0,\- 0,\-\skip 1,\-\skip 0), &\hskip20pt (\- 0,\- 0,\- 0,\- 1),\hskip20pt
& (\- 0 -3, -8, -12). }$$
Since the set of vertices consisting of all except the
first and last vertices are linearly independent for each of these
two polyhedra, there is a unique integral linear transformation taking
one set to the other set.  We need only notice that the first
and last vertices also correspond under this transformation, thereby
identifying the polyhedra as claimed.

Note that our polyhedron contains numerous other lattice points; in other
words, $\IP_4^{(1,1,5,14,21)}$ needs more blowups than the one determined
by the insertion of the generator \hbox{$(\- 0, -3, -8, -12)$} into the fan
in order to  get the
full 7~parameter theory.  We will simplify our discussion of this example
by not performing these further blowups, thereby constraining ourselves
to a 2~parameter subfamily of the boundary of the moduli space.

We describe chiral rings by the procedure of Batyrev and Cox
\Ref\rBatCox{V.~Batyrev and D.~Cox: On the Hodge Structure of Projective
Hypersurfaces in Toric Varieties, alg-geom/9306011.}.
We must first
identify the homogeneous coordinate rings of our toric varieties.  We associate
coordinates $X_1,\ldots,X_6$ to the 6~edges of the fan (in the order written),
and
denote by $S$ the polynomial ring that they generate.  We need to weight them.
While they are weighted by divisor classes according to
Ref.\cite\rCox, we do not
need this geometry for our purposes and content ourselves to describe the
weights as certain special multidegrees.  To do this, we note the relations
 $$\matrix{1\,v_1 &+& 1\,v_2 &+& 5\,v_3 &+& 14\,v_4 &+& 21\,v_5 &+& 0\,v_6 &=&
0 \cr
          0\,v_1 &+& 0\,v_2 &+& 3\,v_3 &+& 8\,v_4 &+& 12\,v_5 &+& 1\,v_6 &=& 0
}$$
among the vertices of the polyhedron, numbered in the order listed above.
These relations tell us that the weights are as follows.
 $$\matrix{ X_1 & (1,0) \cr
            X_2 & (1,0) \cr
            X_3 & (5,3) \cr
            X_4 & (14,8) \cr
            X_5 & (21,12) \cr
            X_6 & (0,1) }\eqlabel{qnumb}$$
The anticanonical class as always has weight equal to the sum of the weights
of all of the edges, in this case $(42,24)$.
The equation of $\ca{W}$ becomes
 $$
f=X_1^{42}X_6^{24} + X_2^{42}X_6^{24} + X_3^8X_1X_2 + X_4^3 + X_5^2~.$$
Note that we needed extra factors of $X_6$ to make the equation homogeneous.
The chiral ring consists of the parts of
the quotient ring $S/J_f$ of weights
 $$
(0,0)~,~~~~~(42,24)~,~~~~~(84,48)~,~~~~~(126,72)~.$$
The monomial of top degree may be taken to be
$X_1^{41}X_2^{41}X_3^6X_4X_6^{46}$.
\subsection{Phases of the model}
The phases for  the model\Footnote{Much of the analysis of the phases of this
model emerged in discussions with R.~Plesser.}
$\IP_4^{(1,1,5,14,21)}[42]$ can be obtained by requiring the vanishing of
scalar potential $U$ of the corresponding linear sigma model\cite\rWit.
For the present model this is given by
 $$
  U = - {1\over 2} \sum_a {1\over e_a^2} D_a^2 + |f|^2 +
              |X_0|^2 \sum_{i=1}^6 \Big| \pd{f}{X_i}\Big|^2 ~, $$
where $X_0$ is the fiber coordinate on the canonical bundle
and the $D_a$ are the $D$-components of vector superfields $V_a$. The gauge
symmetry in our case is $U(1)\times U(1)$ with charges for the chiral fields
$X_0, X_1, \ldots, X_6$ given in \eq\eqref{qnumb}.  Using their
equations of motion
in the linear sigma model action, the $D_a$'s are given by
 $$\eqalign{
D_1 &= |X_1|^2 + |X_2|^2 + 5 |X_3|^2 + 14 |X_4|^2 + 21 |X_5|^2
                                    - 42 |X_0|^2 - r_1 \cropen{5pt}
D_2 &= 3 |X_3|^2 + 8 |X_4|^2 + 12 |X_5|^2 +  |X_6|^2
                                    - 24 |X_0|^2 - r_2~.\cr}$$

The scalar potential vanishes only when the $D$-terms vanish and
when $X_0 = 0$ and either $f = 0$ or $df = 0$.
Thus the minima of $U$ correspond to three possible branches
 $$\eqalign{
a)&\hskip40pt  X_0 = 0~\hbox{and}~f = 0\cr
b)&\hskip40pt  X_1 = X_2 = X_4 = X_5 = 0~\hbox{and}~X_0~\hbox{is not zero}\cr
c)&\hskip40pt  X_3 = X_6 = X_4 = X_5 = 0~\hbox{and}~X_0~\hbox{is not zero.}\cr}
$$
{}{}From these, we find that the phases are
 $$\eqalign{
  I&\hskip40pt     0 < {4 r_1\over 7} < r_2 < {3 r_1\over 5}\cr
 II&\hskip40pt     0 < r_2 < {4 r_1\over 7}\cr
III&\hskip40pt   r_2 < 0~\hbox{and}~r_2 < {4 r_1\over 7}\cr
 IV&\hskip40pt   r_1 < 0~\hbox{and}~r_2 > {4 r_1\over 7}\cr
  V&\hskip40pt     0 < {3 r_1\over 5} < r_2~.\cr}$$

It is easy to see that branch $a$ covers phases $I$, $II$ and $V$,
branch $b$ covers phase $IV$ and $V$ and branch $c$ covers phase $III$.
This gives the phase diagram shown in Figure~1.
The phases have the following interpretation:
\midinsert
\iffigmode
\def\phases{\hbox{\hskip20pt\vbox{\vskip30pt\epsfxsize=3.5truein%
\epsfbox{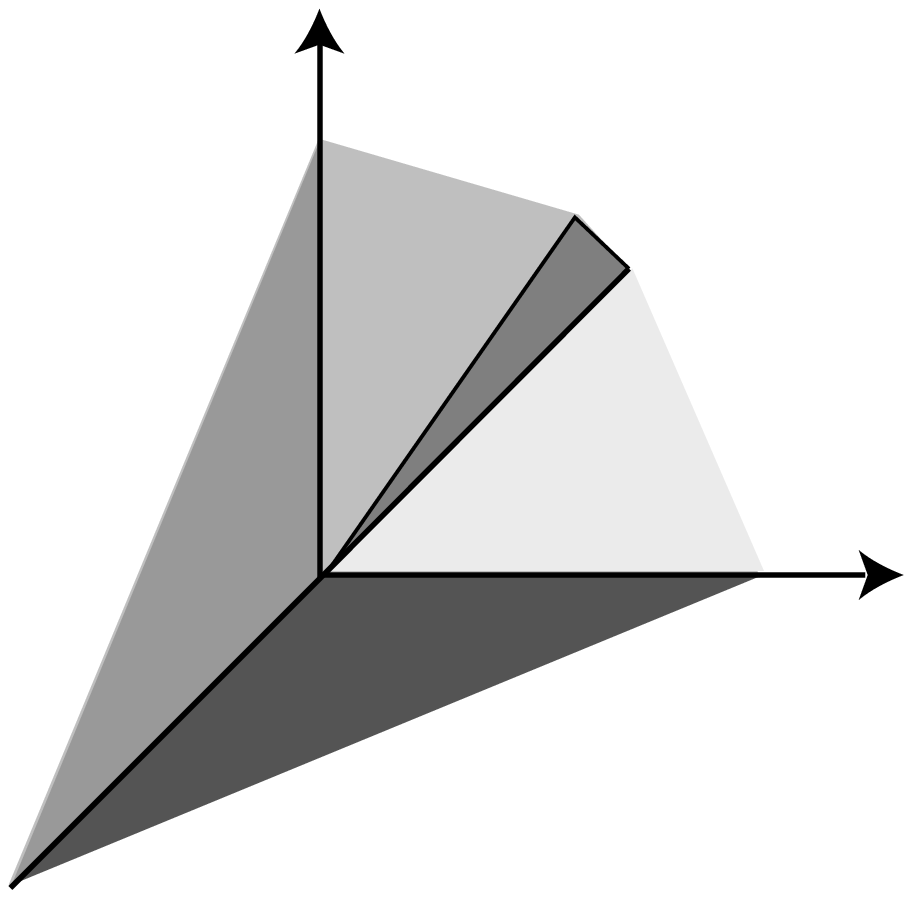}}}}
\figbox{\phases\vskip15pt}{\figlabel{phases}}{The phases of the theory}
\place{4.1}{3.7}{\twelvemath I}
\place{4.5}{3.0}{\twelvemath II}
\place{3.3}{1.5}{\twelvemath III}
\place{2.0}{2.6}{\twelvemath IV}
\place{3.4}{4.0}{\twelvemath V}
\place{5.2}{2.2}{$\hbox{\twelvemath r}_{\hbox{\twelvemath 1}}$}
\place{2.7}{4.5}{$\hbox{\twelvemath r}_{\hbox{\twelvemath 2}}$}
\else
\figbox{{\vrule height 4truein width 0pt}\vskip15pt}{\figlabel{phases}}{The
phases of the theory}
\fi
\endinsert
\subsubsection{Phases $I$ and $II$}
$X_0 = 0$ and $f = 0$, and
 $$\eqalign{
&\hbox{for phase $I$}\hskip40pt~ 0 < {4 r_1\over 7} < r_2 < {3 r_1\over 5}
                                        \quad {\rm or}\cr
&\hbox{for phase $II$}\hskip40pt 0 < r_2 < {4 r_1\over 7}~.\cr}$$
The sets of coordinates that do not vanish simultaneously in
phase $I$ are $(X_1,X_2,X_4,X_5)$ and $(X_3,X_6)$.  For phase $II$
we just interchange (1,2) with (3,6).  Both of these phases
correspond to \cy\ hypersurfaces
which are related to each other by a (non-simple) flop. Note that
the point $X_1=X_2=X_4=X_5=0$ at which the polynomial fails to be
transverse is forbidden in both phases, \ie\ the singularity of $p$
does not belong to the toric variety in which the \cy\ hypersurface
is embedded.  In fact, it is precisely in this sense that the \cy\
hypersurface
$\IP_4^{(1,1,5,14,21)}[42]$ makes sense.

The boundary between phases $I$ and $II$ ($r_1 > 0$ and $r_2 =  {4 r_1\over
7}$)
corresponds to a singular \cym\ with a conifold
singularity at $X_1 = X_2 = X_3 = X_6 = 0$,
which gives phase $I$ or phase $II$ depending on the choice of blowup.
\subsubsection{Phase $III$}
$$
X_3 = X_6 = X_4 = X_5 = 0~,\quad X_0\neq 0\quad\quad {\rm and}
\quad\quad r_2 < 0~,\quad  r_2 < {4 r_1\over 7}~.$$
 This phase corresponds to
a hybrid of a $\IP_3^{(1,3,8,12)}[24]\,(=\hbox{K3})$ \LG\ orbifold
fibered over the \cp1\ defined by the coordinates $(X_1,X_2)$.
The effective potential for the \LG\ orbifold is
 $$
W_{\hbox{eff}} = \sqrt{|r_2|\over 24} (c_1 X_6^{24} + c_2 X_3^8 + X_4^3 +
X_5^2)$$
and, obviously, the quantum symmetry is $\IZ_{24}$.

\noindent In the boundary between phases $II$ and $III$,
the $\IZ_{24}$ quantum symmetry
is promoted to a U(1) gauge symmetry so the hybrid corresponds to a
{\it gauged}  \LG\ model fibered over the \cp1\ and with effective
potential
 $$
W_{\hbox{eff}} = X_0 (c_1 X_6^{24} + c_2 X_3^8 + X_4^3 + X_5^2)$$
where the fields $X_0, X_6, X_3, X_4, X_5$ have $U(1)$
 charges $(-24,1,3,8,12)$.  We can also describe the boundary by the mapping
to $\IP^1$ given by $(X_1,X_2)$; we approach the boundary by letting the
size of the fibers approach zero.
\subsubsection{Phase $IV$}
 $$
    X_1 = X_2 = X_4 = X_5 = 0~,\quad X_0\neq 0 \quad\quad {\rm and}
\quad\quad r_1 < 0~, r_2 > {4 r_1\over 7}~.$$
%We do not understand this phase,  the case
This phase is completely
new and does not admit as nice an interpretation as the phases already
described.
%It looks like a bad fibration of a theory with coordinates
%$X_1, X_2, X_4, X_5$ over a
%\cp1\ defined by $(X_3,X_6)$. The theory defined by
%$X_1, X_2, X_4, X_5$ has an effective potential with no definite degree,
%in other word it is {\it not} quasihomogeneous.
%For what it is worth, the
%effective potential looks like
% $$
%W_{\hbox{eff}} =\langle X_0\rangle (c_1(X_1^{42} + X_2^{42}) + c_2 X_1 X_2 +
% %X_4^3 +
%X_5^2)~. $$

\noindent The boundary between this phase and phase $III$ is the closest we can
get to a \LG\ orbifold.  Since for this boundary $r_1 < 0$ and
$r_2 = {4 r_1\over 7}$ and all the fields vanish except $X_0$, we get
a singular \LG\ orbifold with $\IZ_{42} \times \IZ_{24}$ quantum symmetry.
\subsubsection{Phase $V$}
 $$ 0 < {3 r_1\over 5} < r_2  $$
This phase is very strange too.
First notice that branches $a$ and $c$ overlap over this phase.
The interpretation here is that we get a
singular \cy\ $\IP_4^{(1,1,5,14,21)}$ (from branch $a$)
with the point singularity at $X_1 = X_2 = X_4 = X_5 = 0$ replaced by
the strange model of phase $IV$ (from branch $c$).

\noindent The boundary between phases $IV$ and $V$ corresponds to
the \cy\ in phase $V$ shrinking to a point.

\noindent The boundary between phases $V$ and $I$ is a singular
\cy\ with singularity at $X_1 = X_2 = X_4 = X_5 = 0$.
Aspinwall, Greene and Morrison\cite{\ragm} have called an ``exoflop''
the process of crossing this type of boundary between a smooth \cy\ phase
like phase $I$ and a phase like $V$ which consists of the \cy\ with the
point singularity replaced by a hybrid model.
Along the locus $X_1 = X_2 = X_4 = X_5 = 0$, we note that $X_3$ cannot
also vanish in this phase.  There remains a $\IP^1$ determined by the
variables $X_0, X_6$.  This $\IP^1$ gets flopped {\it outside\/} the
original \cy, hence the term exoflop.
\newpage
\section{frac}{Some Observations on Fractional Transformations}
\vskip-20pt
\subsection{A simple identification}
We wish to discuss, in the context of another example, some issues
associated with fractional transformations. Consider the following pair of
\cys\ which we have taken from
Ref.\Ref{\rLynkSch}{M.~Lynker and R.~Schimmrigk: Phys.\ Lett.\ {\bf 249B}
(1990) 237.}
 $$\eqalign{
\M_1&\in \cp4^{(1,1,1,1,3)}[7]~\hphantom{4}:\cropen{5pt}
\M_2&\in \cp4^{(1,2,2,2,7)}[14]~:\cr}\hskip50pt
\eqalign{
p_1&=x_1^{7\hphantom{4}} + x_2^7 + x_3^7 + x_4^7 + x_1x_5^2~.\cropen{5pt}
p_2&=y_1^{14} + y_2^7 + y_3^7\, + y_4^7 + y_5^2~.\cr}$$
Both $\M_1$ and $\M_2$ have $b_{11}=2$ and $b_{21}=122$ and it is tempting to
identify the two manifolds via the transformation
 $$\eqalign{
x_1~&=~y_1^2\cr
x_i~&=~y_i~~,~~i=2,3,4, \cr
x_5~&=~{y_5\over y_1}\cr}\eqlabel{transf}$$
It is easy to check that while all 122 complex structure
deformations of $\M_1$ can be realised as polynomial deformations of $p_1$
the same is not true of $\M_2$. Only 107 of the parameters of $\M_2$ can be
realised as polynomial deformations of $p_2$. This is not surprising in virtue
of the identification \eqref{transf} since the 15 missing deformations are of
the form
 $$
qx_5~=~q{y_5\over y_1}\eqlabel{twisted}$$
with $q$ a quartic in the variables $(x_2,x_3,x_4)$. Note that in this case the
$\IZ_2$ ambiguity of the transformation is part of the projective equivalence.

Below we shall give a brief toric description of this birational relation.  The
point is that when the weighted hypersurfaces are desingularized as required by
the general procedure, that the manifolds $\M_1$ and $\M_2$ are indeed
isomorphic. (This is the case despite the fact that $\M_1$ and $\M_2$ have
distinct Newton polyhedra.)  The relation of this example to that of the
previous section is that, apart from the fact that we can treat the present
case by the same methods as the previous one, is that one of the questions
that we are asking is how to represent the non-polynomial deformations of
$p_2$.
We are motivated in part by the following question. Suppose that we had started
from $\M_2$ and that in virtue of the Landau-Ginzburg formalism, or the results
of Berglund and H\"ubsch
\ref{P.~Berglund and T.~H\"ubsch: \npb{411} (1994) 223, hep-th/9303158.},
we had learnt that the missing deformations were of
the form \eqref{twisted}. How would we see that the ``correct'' way to
represent
the deformations is by making the change of variables \eqref{transf}?
The example of the previous section was an extreme example for which the
polynomial was not transverse for any choice of the parameters so that none of
the elements of the chiral ring could be represented as deformations of a
transverse polynomial.
\subsection{Isomorphism of $\M_1$ and $\M_2$}
The vertices
of the Newton polyhedron $\D_1$ for $\cp4^{(1,1,1,1,3)}[7]$ are
 $$\matrix{
(  -1,-1,-1,-1),&~~(-1,-1,\- 0,\- 1),&~~(  -1,-1,\- 6,  -1),&~~(-1,\- 6,-1,-1),
\cropen{3pt}
(\- 6,-1,-1,-1),&~~(-1,-1,  -1,\- 1),&~~(\- 0,-1,  -1,\- 1),&~~(-1,\- 0,-1,\-
1).  \cr} $$
which yield the vertices of the polar polyhedron $\nabla_1$ as
 $$\matrix{
(-1,-1,-1,-3),&~~(\- 1,\- 0,\- 0,\- 0),&~~(\- 0,\- 1,\- 0,\- 0),  \cropen{3pt}
(\- 0,\- 0,\- 1,\- 0),&~~(\- 0,\- 0,\- 0,\- 1),&~~(\- 0,\- 0,\- 0,-1).\cr}$$
We note that $\nabla_1$ has no other lattice points besides the origin.
It is clear from looking at the first 5~vertices that the normal fan of
$\nabla_1$ describes a toric variety $X_1$ birational to $\cp4^{(1,1,1,1,3)}$.
In fact, $X_1$ is just a blowup of $\cp4^{(1,1,1,1,3)}$ at the point
$(0,0,0,0,1)$; the exceptional set is a $\cp3$.  Numbering the edges from
1~to~6, we note first the fan of $\cp4^{(1,1,1,1,3)}$ has maximal cones
spanned by the set of edges numbered
 $$
\{2,3,4,5\}~,~~~\{1,3,4,5\}~,~~~\{1,2,4,5\}~,~~~\{1,2,3,5\}~,~~~\{1,2,3,4\}~.$$
Since the edge $(\- 0,\- 0,\- 0,-1)$ lies in the interior of the last cone,
to get the fan for $X_1$ from the fan for $\cp4^{(1,1,1,1,3)}$,
the last cone is replaced by the cones spanned by edges numbered
 $$
\{2,3,4,6\}~,~~~\{1,3,4,6\}~,~~~\{1,2,4,6\}~,~~~\{1,2,3,6\}~.$$

Turning next to $\cp4^{(1,2,2,2,7)}$, the vertices of the Newton polyhedron
$\D_2$ are
 $$\displaylines{
(-1,-1,-1,-1), ~~~ (-1,-1,-1,\- 1), ~~~ (-1,-1,\- 6,-1),\cropen{3pt}
         ~~ (-1,\- 6,-1,-1), ~~~ (\- 6,-1,-1,-1).\cr }$$
so those of $\nabla_2$ are
$$\displaylines{
(-2,-2,-2,-7), ~~~ (\- 1,\- 0,\- 0,\- 0), ~~~ (\- 0,\- 1,\- 0,\- 0),
\cropen{3pt}
{}~(\- 0,\- 0,\- 1,\- 0), ~~~ (\- 0,\- 0,\- 0,\- 1).\cr}$$
($\D_2$ and $\nabla_2$ are both simplicial; this corresponds to the
existence
of a Fermat hypersurface).  In this case, $\nabla_2$ has 3~more lattice points
in addition to vertices and the origin:
$\{(\- 0,\- 0,\- 0,-1),\ (-1,-1,-1,-4),\ (-1,-1,-1,-3)\}$.
The first two of these lie in the interior of the facet of $\nabla_2$ dual to
the vertex $(-1,-1,-1,\- 1)$; the last lies on the edge spanned by the last
two vertices of $\nabla_2$.  To get Calabi-Yau hypersurfaces, we take a
subdivision of the normal fan of $\nabla_2$.
A subdivision yielding a toric variety $X_2$ can in fact be obtained by
further subdividing the fan for $X_1$.
We first insert the edge spanned by $(-1,-1,-1,-4)$.  Since this is
just $(-1,-1,-1,-3)+(\- 0,\- 0,\- 0,-1)$, we see that we are blowing up the
$\cp2$ where the proper transform of $x_1=0$ meets the exceptional divisor.
On cones, we replace each cone containing $(-1,-1,-1,-3)$ and
$(\- 0,\- 0,\- 0,-1)$ by two cones, the first one containing instead
$(-1,-1,-1,-3)$ and $(-1,-1,-1,-4)$, while the second cone replaces them by
$(-1,-1,-1,-4)$ and $(\- 0,\- 0,\- 0,-1)$. A similar procedure applies for the
insertion of $(-2,-2,-2,-7)$ (since this is
$(-1,-1,-1,-3)+(-1,-1,-1,-4)$)---here
we are blowing up the new proper transform of $x_1=0$ with the newest
exceptional divisor.

In a similar fashion, one sees that the same fan can be obtained from
the fan for $\cp4^{(1,2,2,2,7)}$ by first blowing up the the locus
$x_1=x_5=0$ (this inserts the edge\break $(-1,-1,-1,-3)$ between
$(-2,-2,-2,-7)$
and $(\- 0,\- 0,\- 0,-1)$), then resolving the point\break $(0,0,0,0,1)$
by placing the edge $(\- 0,\- 0,\- 0,-1)$ inside the cone spanned by
 $$
(-2,-2,-2,-7)~,~~~(\- 1,\- 0,\- 0,\- 0)~,~~~(\- 0,\- 1,\- 0,\- 0)~,~~~
(\- 0,\- 0,\- 1,\- 0)~,$$
then blowing up
the intersection of the proper transform of $x_1=0$ with the last exceptional
divisor (corresponding to the insertion the edge $(-1,-1,-1,-4)$ between
$(-2,-2,-2,-7)$ and $(0,\- 0,\- 0,-1)$).
Either way, the result is a fan with 14~maximal cones.

The upshot of all this is that
there naturally results an everywhere defined map \hbox{$X_2\to X_1$}.
The equations given  above describe these in terms of the coordinates
of the weighted projective space.

While the $\L$ lattices used in the two examples are a priori different,
our choice of coordinates gives a natural way to identify them.  A similar
assertion holds for the $\rV$ lattices.  With these identifications,
we note the inclusions $\D_2\subset \D_1$ and $\nabla_1\subset \nabla_2$.

So the points of $\D_2\cap\L$ correspond not only to monomials for $\M_2$,
but also a subset of the monomials for $\M_1$.  A closer investigation of the
geometry reveals
that these monomials are precisely the ones with the property that when
a polynomial is formed from them yielding a hypersurface
$\M_1\subset X_1$, the pullback of $\M_1$
to $X_2$ contains both of the expectional
divisors of the map $X_2\to X_1$.
This implies that if the proper transform $\M_2\subset X_2$
of $\M_1$ is smooth, then $\M_2$ is a Calabi-Yau
hypersurface.  In fact more is true: for the generic hypersurface
$\M_1$ formed from these monomials, the manifolds $\M_1$ and $\M_2$ are
actually isomorphic\Footnote{This has been strikingly underscored by a
calculation in \cite{\rBKK}, where the instanton numbers of the two
models are computed, and are seen to coincide.}.  In this
way, we can be certain that we can identify chiral rings.
\subsection{Chiral rings}
We can again describe the chiral rings by the procedure of \cite{\rBatCox}.
Actually, this
will give only the part of the chiral ring corresponding to the polynomial
deformations.  The reason why \cite{\rBatCox} does not apply to give the
entire chiral ring is that the hypersurface in the blown-up
toric variety is not ample (this was pointed out to us
by Batyrev).  Nevertheless, we
continue to refer to this subring as the chiral ring.

We start with $X_1$.  We note the relations
 $$\matrix{1\,v_1 &+& 1\,v_2 &+& 1\,v_3 &+& 1\,v_4 &+& 3\,v_5 &+& 0\,v_6 &=&
0\cr
           0\,v_1 &+& 0\,v_2 &+& 0\,v_3 &+& 0\,v_4 &+& 1\,v_5 &+& 1\,v_6 &=& 0}
$$
These relations tell us that $X_1,\ldots,X_4$ has weight $(1,0)$, that
$X_5$ has weight $(3,1)$, and that $X_6$ has weight $(0,1)$.

The anticanonical class has weight $(7,2)$.  We write the equation of $\M_1$
in homogeneous coordinates and get
 $$f=X_1^7X_6^2 + X_2^7X_6^2 + X_3^7X_6^2 + X_4^7X_6^2 + X_1X_5^2.$$
The Jacobian ideal $J_f$ is the ideal of partial derivatives of $f$
with respect to the $X_i$.
The results of \cite{\rBatCox} imply that the chiral ring consists of the parts
of
the quotient ring $S/J_f$ of weights $(0,0),\ (7,2),\ (14,4),\ (21,6)$.

The situation for $\M_2$ is easier, since $\nabla_2$ is simplicial.
The chiral ring is contained in $S'/J_{p_2}$, where $S'$ is the polynomial ring
in $y_1,\ldots,y_5$ and $J_{p_2}$ is the Jacobian ideal of $p_2$.  The chiral
ring is given by the parts of $S'/J_{p_2}$ of degrees 0, 14, 28, and~42.

The reason why this simpler description of the chiral ring (let us for the
moment call this the ``naive'' chiral ring) suffices rests
on two points.
First of all, if we had put in the extra 3~vertices (as the one extra vertex
was added for $\M_1$), we would have obtained 3~new variables, and modified
$p_2$ to get a polynomial $g$ involving the new variables.  The
polynomials on the toric variety ``restrict'' to polynomials on the
weighted projective space by
setting the new variables to~1 (in particular, $g$ restricts to $p_2$).
The chain rule shows that the restriction of $J_g$ is contained in $J_{p_2}$.
In other words, restriction gives a mapping from the chiral ring to the
naive chiral ring.
Secondly, Since $p_2$ is transverse, the chiral
ring as we have written it down automatically satisfies Poincar\'e duality.
The parts of these rings corresponding to $H^{2,1}$ are isomorphic,
as they each have $\D_2$ as
a basis by construction.  Poincar\'e duality then shows that the restriction
map is an isomorphism between the chiral ring and the naive chiral ring,
justifying our identification.

By the geometric reasoning, the natural maps should induce an inclusion
of chiral rings $S'/J_{p_2}\hookrightarrow S/J_f$ (more precisely, after
restricting to the parts of the relevant (multi)degrees).  This can be checked
directly.  So if we want to incorporate the non-polynomial deformations
of $\cp4^{(1,2,2,2,7)}$ into $S'/J_{p_2}$, we know the answer explicitly---
it is just $S/J_f$.

It may help the reader to observe that when the inclusion $\D_2\subset \D_1$
is interpreted via monomials on the respective toric varieties, the monomial
$y_1^*y_2^ay_3^by_4^cy_5^d$ is identified with
$X_1^*X_2^aX_3^bX_4^cX_5^dX_6^*$, where some exponents (denoted with a $*$)
are intentionally supressed to emphasize the coincidence of the remaining
exponents; the supressed exponents can be recovered by considering
(multi)degrees.

We note also that the 15 `missing' deformations of $p_2$ arise in this approach
because of the blowup of the $\cp2$ with equations $x_1=x_5=0$
given by insertion of $(-1,-1,-1,-3)$.  The generic weight~14 polynomial
intersects $\cp2$ in a degree~7 curve which has genus~15.  The blowup
can be seen by general considerations to add~15 to the dimension of
$H^{2,1}$, thereby inducing 15~new  deformations.  In toric language, this
can be seen in \cite{\rBat}.
\vskip50pt
\acknowledgements
It is a pleasure to acknowledge fruitful discussions with V.~Batyrev,
E.~Derrick, M.~Kreuzer, R.~Schimmrigk, and C.~Vafa.
We would like to thank R.~Plesser for numerous discussions on the
linear sigma model and also for sharing with us work in progress.
We also owe special thanks to M.~Kreuzer for discussions and for help in
choosing examples which were missing a mirror.
\vskip2.3truein
\chapno=-1
\section{plot}{Plot of the Hodge Numbers}
On the following page we plot the Hodge numbers of each manifold of the list
together with the Hodge numbers of the mirrors. The Euler number,
$\chi=2(b_{1,1}-b_{2,1})$,
is plotted horizontally and $b_{1,1}+b_{2,1}$ is plotted vertically. The plot,
which is similar to the plots of \cite{\rcls} and \cite{\rKlSc}, is now
symmetric by construction.
\newpage
 %
%\listreferences\bye
 %
\vsize=9.3truein
\null\vskip-.4truein

\def\plot#1#2{\vskip\parskip

                  \vbox{\hrule width\hsize

                        \hbox{\kern-0.2pt\vrule height#1

                              \vbox{\hfill}\kern-0.6pt

                              \vrule}\hrule width\hsize}

    \setbox0=\hbox{#2} \dimen0=\wd0 \divide\dimen0 by 2

    \setbox0=\hbox{\kern-\dimen0 #2}

    \dimen3=#1}

\def\hmark{\kern-0.2pt\lower10pt\hbox{\vrule height 5pt}}

\def\leftscalemark{\vbox{\hrule width5pt}}

\def\rightscalemark{\kern-5pt\vbox{\hrule width5pt}}

\def\Place#1#2#3{

    \count10=#1 \advance\count10 by 960

    \dimen1=\hsize \divide\dimen1 by 1920 \multiply\dimen1 by \count10

    \dimen2=\dimen3 \divide\dimen2 by 550 \multiply\dimen2 by #2

    \vbox to 0pt{\kern-\parskip\kern-20truept\kern-\dimen2

    \hbox{\kern\dimen1#3}\vss}\nointerlineskip}

\def\datum#1#2{\Place{#1}{#2}{\copy0}}

\plot{9.0truein}{{\seventeenrm .}}
\nobreak

\Place{-960}{50}{\leftscalemark~~50}

\Place{-960}{100}{\leftscalemark~~100}

\Place{-960}{150}{\leftscalemark~~150}

\Place{-960}{200}{\leftscalemark~~200}

\Place{-960}{250}{\leftscalemark~~250}

\Place{-960}{300}{\leftscalemark~~300}

\Place{-960}{350}{\leftscalemark~~350}

\Place{-960}{400}{\leftscalemark~~400}

\Place{-960}{450}{\leftscalemark~~450}

\Place{-960}{500}{\leftscalemark~~500}

\Place{960}{50}{\rightscalemark\vphantom{0}}

\Place{960}{100}{\rightscalemark\vphantom{0}}

\Place{960}{150}{\rightscalemark\vphantom{0}}

\Place{960}{200}{\rightscalemark\vphantom{0}}

\Place{960}{250}{\rightscalemark\vphantom{0}}

\Place{960}{300}{\rightscalemark\vphantom{0}}

\Place{960}{350}{\rightscalemark\vphantom{0}}

\Place{960}{400}{\rightscalemark\vphantom{0}}

\Place{960}{450}{\rightscalemark\vphantom{0}}

\Place{960}{500}{\rightscalemark\vphantom{0}}

\Place{-960}{0}{\hmark\lower18pt\hbox{-960}}

\Place{-720}{0}{\hmark\lower18pt\hbox{-720}}

\Place{-480}{0}{\hmark\lower18pt\hbox{-480}}

\Place{-240}{0}{\hmark\lower18pt\hbox{-240}}

\Place{0}{0}{\hmark\lower18pt\hbox{0}}

\Place{240}{0}{\hmark\lower18pt\hbox{240}}

\Place{480}{0}{\hmark\lower18pt\hbox{480}}

\Place{720}{0}{\hmark\lower18pt\hbox{720}}

\Place{960}{0}{\hmark\lower18pt\hbox{960}}

\Place{-720}{550}{\hmark}

\Place{-480}{550}{\hmark}

\Place{-240}{550}{\hmark}

\Place{0}{550}{\hmark}

\Place{240}{550}{\hmark}

\Place{480}{550}{\hmark}

\Place{720}{550}{\hmark}

\Place{960}{550}{\hmark}

\nobreak

\datum{-960}{ 502}
\datum{-900}{ 474}
\datum{-840}{ 446}
\datum{-804}{ 430}
\datum{-744}{ 402}
\datum{-732}{ 386}
\datum{-720}{ 394}
\datum{-672}{ 374}
\datum{-660}{ 366}
\datum{-648}{ 358}
\datum{-636}{ 342}
\datum{-624}{ 330}
\datum{-624}{ 358}
\datum{-612}{ 330}
\datum{-612}{ 346}
\datum{-588}{ 346}
\datum{-576}{ 302}
\datum{-576}{ 314}
\datum{-564}{ 322}
\datum{-564}{ 330}
\datum{-564}{ 340}
\datum{-552}{ 306}
\datum{-540}{ 274}
\datum{-540}{ 298}
\datum{-540}{ 334}
\datum{-528}{ 278}
\datum{-528}{ 286}
\datum{-528}{ 318}
\datum{-528}{ 334}
\datum{-516}{ 302}
\datum{-516}{ 330}
\datum{-512}{ 286}
\datum{-510}{ 331}
\datum{-504}{ 276}
\datum{-504}{ 312}
\datum{-492}{ 256}
\datum{-480}{ 246}
\datum{-480}{ 262}
\datum{-480}{ 278}
\datum{-480}{ 286}
\datum{-480}{ 306}
\datum{-480}{ 334}
\datum{-476}{ 270}
\datum{-468}{ 286}
\datum{-468}{ 306}
\datum{-456}{ 234}
\datum{-456}{ 248}
\datum{-456}{ 256}
\datum{-456}{ 262}
\datum{-456}{ 264}
\datum{-456}{ 272}
\datum{-456}{ 302}
\datum{-450}{ 303}
\datum{-450}{ 331}
\datum{-444}{ 234}
\datum{-444}{ 330}
\datum{-432}{ 238}
\datum{-432}{ 242}
\datum{-432}{ 266}
\datum{-432}{ 274}
\datum{-432}{ 334}
\datum{-426}{ 265}
\datum{-420}{ 218}
\datum{-420}{ 230}
\datum{-420}{ 248}
\datum{-420}{ 250}
\datum{-420}{ 306}
\datum{-420}{ 334}
\datum{-416}{ 262}
\datum{-408}{ 212}
\datum{-408}{ 224}
\datum{-408}{ 232}
\datum{-408}{ 240}
\datum{-408}{ 268}
\datum{-396}{ 214}
\datum{-396}{ 222}
\datum{-396}{ 262}
\datum{-396}{ 340}
\datum{-390}{ 303}
\datum{-384}{ 204}
\datum{-384}{ 218}
\datum{-384}{ 222}
\datum{-384}{ 232}
\datum{-384}{ 234}
\datum{-384}{ 242}
\datum{-384}{ 250}
\datum{-384}{ 262}
\datum{-380}{ 198}
\datum{-376}{ 214}
\datum{-372}{ 194}
\datum{-372}{ 202}
\datum{-372}{ 226}
\datum{-372}{ 258}
\datum{-372}{ 262}
\datum{-372}{ 306}
\datum{-372}{ 346}
\datum{-368}{ 204}
\datum{-364}{ 194}
\datum{-364}{ 204}
\datum{-360}{ 190}
\datum{-360}{ 192}
\datum{-360}{ 194}
\datum{-360}{ 206}
\datum{-360}{ 212}
\datum{-360}{ 228}
\datum{-360}{ 258}
\datum{-360}{ 306}
\datum{-356}{ 202}
\datum{-356}{ 220}
\datum{-356}{ 234}
\datum{-354}{ 259}
\datum{-352}{ 190}
\datum{-348}{ 186}
\datum{-348}{ 198}
\datum{-348}{ 226}
\datum{-348}{ 238}
\datum{-348}{ 258}
\datum{-344}{ 224}
\datum{-342}{ 233}
\datum{-340}{ 198}
\datum{-336}{ 178}
\datum{-336}{ 188}
\datum{-336}{ 194}
\datum{-336}{ 198}
\datum{-336}{ 202}
\datum{-336}{ 206}
\datum{-336}{ 222}
\datum{-336}{ 230}
\datum{-336}{ 312}
\datum{-336}{ 358}
\datum{-330}{ 181}
\datum{-330}{ 221}
\datum{-330}{ 261}
\datum{-324}{ 168}
\datum{-324}{ 182}
\datum{-324}{ 184}
\datum{-324}{ 222}
\datum{-324}{ 230}
\datum{-324}{ 232}
\datum{-324}{ 262}
\datum{-322}{ 193}
\datum{-320}{ 170}
\datum{-320}{ 174}
\datum{-320}{ 190}
\datum{-320}{ 198}
\datum{-320}{ 206}
\datum{-320}{ 222}
\datum{-318}{ 197}
\datum{-316}{ 262}
\datum{-312}{ 166}
\datum{-312}{ 172}
\datum{-312}{ 174}
\datum{-312}{ 178}
\datum{-312}{ 180}
\datum{-312}{ 190}
\datum{-312}{ 192}
\datum{-312}{ 196}
\datum{-312}{ 224}
\datum{-312}{ 318}
\datum{-306}{ 169}
\datum{-306}{ 177}
\datum{-306}{ 217}
\datum{-304}{ 176}
\datum{-300}{ 168}
\datum{-300}{ 178}
\datum{-300}{ 180}
\datum{-300}{ 194}
\datum{-300}{ 198}
\datum{-300}{ 218}
\datum{-300}{ 222}
\datum{-300}{ 226}
\datum{-300}{ 230}
\datum{-296}{ 150}
\datum{-296}{ 162}
\datum{-296}{ 166}
\datum{-296}{ 174}
\datum{-294}{ 159}
\datum{-294}{ 187}
\datum{-294}{ 259}
\datum{-292}{ 158}
\datum{-288}{ 146}
\datum{-288}{ 152}
\datum{-288}{ 158}
\datum{-288}{ 162}
\datum{-288}{ 166}
\datum{-288}{ 170}
\datum{-288}{ 182}
\datum{-288}{ 186}
\datum{-288}{ 188}
\datum{-288}{ 190}
\datum{-288}{ 206}
\datum{-288}{ 214}
\datum{-288}{ 222}
\datum{-288}{ 226}
\datum{-288}{ 258}
\datum{-288}{ 270}
\datum{-288}{ 374}
\datum{-286}{ 163}
\datum{-286}{ 177}
\datum{-284}{ 166}
\datum{-280}{ 148}
\datum{-280}{ 150}
\datum{-280}{ 166}
\datum{-276}{ 150}
\datum{-276}{ 154}
\datum{-276}{ 162}
\datum{-276}{ 172}
\datum{-276}{ 174}
\datum{-276}{ 186}
\datum{-276}{ 192}
\datum{-276}{ 212}
\datum{-276}{ 214}
\datum{-276}{ 222}
\datum{-276}{ 234}
\datum{-276}{ 262}
\datum{-276}{ 330}
\datum{-272}{ 148}
\datum{-272}{ 150}
\datum{-272}{ 166}
\datum{-272}{ 178}
\datum{-270}{ 223}
\datum{-266}{ 169}
\datum{-264}{ 144}
\datum{-264}{ 148}
\datum{-264}{ 150}
\datum{-264}{ 154}
\datum{-264}{ 158}
\datum{-264}{ 160}
\datum{-264}{ 162}
\datum{-264}{ 164}
\datum{-264}{ 172}
\datum{-264}{ 178}
\datum{-264}{ 184}
\datum{-264}{ 188}
\datum{-264}{ 214}
\datum{-264}{ 228}
\datum{-264}{ 262}
\datum{-260}{ 134}
\datum{-260}{ 150}
\datum{-258}{ 151}
\datum{-258}{ 161}
\datum{-258}{ 163}
\datum{-256}{ 134}
\datum{-256}{ 156}
\datum{-256}{ 158}
\datum{-256}{ 166}
\datum{-256}{ 174}
\datum{-256}{ 234}
\datum{-252}{ 130}
\datum{-252}{ 138}
\datum{-252}{ 142}
\datum{-252}{ 158}
\datum{-252}{ 162}
\datum{-252}{ 166}
\datum{-252}{ 174}
\datum{-252}{ 178}
\datum{-252}{ 202}
\datum{-252}{ 206}
\datum{-252}{ 278}
\datum{-248}{ 134}
\datum{-248}{ 174}
\datum{-246}{ 167}
\datum{-246}{ 185}
\datum{-246}{ 237}
\datum{-244}{ 162}
\datum{-242}{ 139}
\datum{-242}{ 189}
\datum{-240}{ 124}
\datum{-240}{ 126}
\datum{-240}{ 130}
\datum{-240}{ 134}
\datum{-240}{ 138}
\datum{-240}{ 140}
\datum{-240}{ 142}
\datum{-240}{ 144}
\datum{-240}{ 146}
\datum{-240}{ 150}
\datum{-240}{ 158}
\datum{-240}{ 162}
\datum{-240}{ 166}
\datum{-240}{ 174}
\datum{-240}{ 178}
\datum{-240}{ 188}
\datum{-240}{ 206}
\datum{-240}{ 218}
\datum{-240}{ 226}
\datum{-240}{ 232}
\datum{-240}{ 268}
\datum{-240}{ 394}
\datum{-236}{ 162}
\datum{-236}{ 172}
\datum{-236}{ 182}
\datum{-234}{ 131}
\datum{-234}{ 133}
\datum{-234}{ 141}
\datum{-234}{ 217}
\datum{-232}{ 126}
\datum{-232}{ 134}
\datum{-232}{ 158}
\datum{-230}{ 129}
\datum{-228}{ 126}
\datum{-228}{ 130}
\datum{-228}{ 132}
\datum{-228}{ 138}
\datum{-228}{ 140}
\datum{-228}{ 142}
\datum{-228}{ 146}
\datum{-228}{ 170}
\datum{-228}{ 176}
\datum{-228}{ 178}
\datum{-228}{ 196}
\datum{-228}{ 202}
\datum{-228}{ 346}
\datum{-224}{ 122}
\datum{-224}{ 124}
\datum{-224}{ 126}
\datum{-224}{ 134}
\datum{-224}{ 138}
\datum{-224}{ 142}
\datum{-224}{ 146}
\datum{-224}{ 152}
\datum{-224}{ 170}
\datum{-224}{ 178}
\datum{-222}{ 133}
\datum{-222}{ 135}
\datum{-220}{ 126}
\datum{-220}{ 158}
\datum{-220}{ 218}
\datum{-216}{ 116}
\datum{-216}{ 120}
\datum{-216}{ 124}
\datum{-216}{ 126}
\datum{-216}{ 128}
\datum{-216}{ 132}
\datum{-216}{ 134}
\datum{-216}{ 136}
\datum{-216}{ 138}
\datum{-216}{ 140}
\datum{-216}{ 142}
\datum{-216}{ 144}
\datum{-216}{ 150}
\datum{-216}{ 152}
\datum{-216}{ 154}
\datum{-216}{ 158}
\datum{-216}{ 164}
\datum{-216}{ 168}
\datum{-216}{ 174}
\datum{-216}{ 180}
\datum{-216}{ 182}
\datum{-216}{ 188}
\datum{-216}{ 192}
\datum{-216}{ 212}
\datum{-216}{ 240}
\datum{-216}{ 244}
\datum{-216}{ 274}
\datum{-216}{ 292}
\datum{-212}{ 150}
\datum{-212}{ 158}
\datum{-210}{ 113}
\datum{-210}{ 117}
\datum{-210}{ 131}
\datum{-210}{ 133}
\datum{-210}{ 143}
\datum{-210}{ 145}
\datum{-210}{ 151}
\datum{-210}{ 171}
\datum{-210}{ 221}
\datum{-210}{ 241}
\datum{-208}{ 108}
\datum{-208}{ 118}
\datum{-208}{ 126}
\datum{-208}{ 138}
\datum{-208}{ 178}
\datum{-204}{ 104}
\datum{-204}{ 108}
\datum{-204}{ 118}
\datum{-204}{ 120}
\datum{-204}{ 124}
\datum{-204}{ 126}
\datum{-204}{ 130}
\datum{-204}{ 134}
\datum{-204}{ 142}
\datum{-204}{ 164}
\datum{-204}{ 168}
\datum{-204}{ 178}
\datum{-204}{ 190}
\datum{-204}{ 222}
\datum{-202}{ 133}
\datum{-200}{ 102}
\datum{-200}{ 106}
\datum{-200}{ 108}
\datum{-200}{ 118}
\datum{-200}{ 128}
\datum{-200}{ 130}
\datum{-200}{ 134}
\datum{-200}{ 136}
\datum{-200}{ 148}
\datum{-200}{ 150}
\datum{-200}{ 156}
\datum{-200}{ 168}
\datum{-198}{ 113}
\datum{-198}{ 125}
\datum{-198}{ 131}
\datum{-198}{ 139}
\datum{-196}{ 106}
\datum{-196}{ 150}
\datum{-196}{ 172}
\datum{-194}{ 113}
\datum{-194}{ 119}
\datum{-192}{ 102}
\datum{-192}{ 106}
\datum{-192}{ 108}
\datum{-192}{ 110}
\datum{-192}{ 112}
\datum{-192}{ 114}
\datum{-192}{ 116}
\datum{-192}{ 118}
\datum{-192}{ 122}
\datum{-192}{ 124}
\datum{-192}{ 126}
\datum{-192}{ 128}
\datum{-192}{ 130}
\datum{-192}{ 134}
\datum{-192}{ 138}
\datum{-192}{ 142}
\datum{-192}{ 150}
\datum{-192}{ 154}
\datum{-192}{ 162}
\datum{-192}{ 166}
\datum{-192}{ 170}
\datum{-192}{ 178}
\datum{-192}{ 184}
\datum{-192}{ 190}
\datum{-192}{ 198}
\datum{-192}{ 206}
\datum{-192}{ 218}
\datum{-192}{ 226}
\datum{-192}{ 250}
\datum{-190}{ 129}
\datum{-186}{ 97}
\datum{-186}{ 115}
\datum{-186}{ 117}
\datum{-186}{ 123}
\datum{-186}{ 127}
\datum{-186}{ 151}
\datum{-186}{ 167}
\datum{-184}{ 102}
\datum{-184}{ 106}
\datum{-184}{ 110}
\datum{-184}{ 112}
\datum{-184}{ 116}
\datum{-184}{ 120}
\datum{-184}{ 122}
\datum{-184}{ 134}
\datum{-184}{ 136}
\datum{-184}{ 148}
\datum{-184}{ 168}
\datum{-184}{ 174}
\datum{-184}{ 180}
\datum{-180}{ 98}
\datum{-180}{ 100}
\datum{-180}{ 104}
\datum{-180}{ 106}
\datum{-180}{ 108}
\datum{-180}{ 110}
\datum{-180}{ 112}
\datum{-180}{ 114}
\datum{-180}{ 118}
\datum{-180}{ 120}
\datum{-180}{ 124}
\datum{-180}{ 126}
\datum{-180}{ 134}
\datum{-180}{ 136}
\datum{-180}{ 138}
\datum{-180}{ 142}
\datum{-180}{ 144}
\datum{-180}{ 148}
\datum{-180}{ 154}
\datum{-180}{ 158}
\datum{-180}{ 168}
\datum{-180}{ 174}
\datum{-180}{ 184}
\datum{-180}{ 194}
\datum{-180}{ 226}
\datum{-180}{ 228}
\datum{-180}{ 286}
\datum{-180}{ 366}
\datum{-176}{ 98}
\datum{-176}{ 100}
\datum{-176}{ 102}
\datum{-176}{ 106}
\datum{-176}{ 108}
\datum{-176}{ 114}
\datum{-176}{ 126}
\datum{-176}{ 132}
\datum{-176}{ 150}
\datum{-176}{ 174}
\datum{-174}{ 97}
\datum{-174}{ 143}
\datum{-174}{ 157}
\datum{-174}{ 167}
\datum{-172}{ 106}
\datum{-172}{ 114}
\datum{-172}{ 122}
\datum{-172}{ 144}
\datum{-170}{ 93}
\datum{-170}{ 117}
\datum{-168}{ 88}
\datum{-168}{ 90}
\datum{-168}{ 94}
\datum{-168}{ 96}
\datum{-168}{ 98}
\datum{-168}{ 100}
\datum{-168}{ 102}
\datum{-168}{ 106}
\datum{-168}{ 108}
\datum{-168}{ 110}
\datum{-168}{ 112}
\datum{-168}{ 114}
\datum{-168}{ 116}
\datum{-168}{ 118}
\datum{-168}{ 120}
\datum{-168}{ 122}
\datum{-168}{ 124}
\datum{-168}{ 126}
\datum{-168}{ 128}
\datum{-168}{ 134}
\datum{-168}{ 138}
\datum{-168}{ 140}
\datum{-168}{ 142}
\datum{-168}{ 144}
\datum{-168}{ 148}
\datum{-168}{ 152}
\datum{-168}{ 164}
\datum{-168}{ 168}
\datum{-168}{ 174}
\datum{-168}{ 178}
\datum{-168}{ 184}
\datum{-168}{ 256}
\datum{-166}{ 127}
\datum{-164}{ 94}
\datum{-162}{ 97}
\datum{-162}{ 115}
\datum{-162}{ 121}
\datum{-162}{ 131}
\datum{-162}{ 133}
\datum{-162}{ 137}
\datum{-162}{ 185}
\datum{-160}{ 90}
\datum{-160}{ 94}
\datum{-160}{ 96}
\datum{-160}{ 98}
\datum{-160}{ 102}
\datum{-160}{ 108}
\datum{-160}{ 110}
\datum{-160}{ 124}
\datum{-160}{ 126}
\datum{-160}{ 150}
\datum{-160}{ 156}
\datum{-160}{ 170}
\datum{-160}{ 178}
\datum{-160}{ 198}
\datum{-160}{ 314}
\datum{-156}{ 88}
\datum{-156}{ 92}
\datum{-156}{ 94}
\datum{-156}{ 102}
\datum{-156}{ 106}
\datum{-156}{ 108}
\datum{-156}{ 110}
\datum{-156}{ 112}
\datum{-156}{ 114}
\datum{-156}{ 118}
\datum{-156}{ 120}
\datum{-156}{ 122}
\datum{-156}{ 124}
\datum{-156}{ 126}
\datum{-156}{ 132}
\datum{-156}{ 148}
\datum{-156}{ 150}
\datum{-156}{ 154}
\datum{-156}{ 178}
\datum{-156}{ 210}
\datum{-156}{ 232}
\datum{-156}{ 234}
\datum{-156}{ 430}
\datum{-154}{ 89}
\datum{-154}{ 97}
\datum{-152}{ 96}
\datum{-152}{ 102}
\datum{-152}{ 108}
\datum{-152}{ 110}
\datum{-152}{ 130}
\datum{-152}{ 134}
\datum{-152}{ 150}
\datum{-150}{ 93}
\datum{-150}{ 97}
\datum{-150}{ 101}
\datum{-150}{ 103}
\datum{-150}{ 109}
\datum{-150}{ 123}
\datum{-150}{ 133}
\datum{-150}{ 151}
\datum{-150}{ 153}
\datum{-150}{ 181}
\datum{-148}{ 92}
\datum{-148}{ 96}
\datum{-148}{ 106}
\datum{-148}{ 158}
\datum{-146}{ 193}
\datum{-144}{ 76}
\datum{-144}{ 78}
\datum{-144}{ 82}
\datum{-144}{ 84}
\datum{-144}{ 86}
\datum{-144}{ 88}
\datum{-144}{ 90}
\datum{-144}{ 92}
\datum{-144}{ 94}
\datum{-144}{ 96}
\datum{-144}{ 98}
\datum{-144}{ 100}
\datum{-144}{ 102}
\datum{-144}{ 104}
\datum{-144}{ 106}
\datum{-144}{ 110}
\datum{-144}{ 112}
\datum{-144}{ 114}
\datum{-144}{ 118}
\datum{-144}{ 122}
\datum{-144}{ 124}
\datum{-144}{ 126}
\datum{-144}{ 130}
\datum{-144}{ 134}
\datum{-144}{ 138}
\datum{-144}{ 140}
\datum{-144}{ 148}
\datum{-144}{ 150}
\datum{-144}{ 152}
\datum{-144}{ 154}
\datum{-144}{ 158}
\datum{-144}{ 164}
\datum{-144}{ 172}
\datum{-144}{ 174}
\datum{-144}{ 182}
\datum{-144}{ 188}
\datum{-144}{ 190}
\datum{-144}{ 206}
\datum{-144}{ 214}
\datum{-142}{ 127}
\datum{-142}{ 129}
\datum{-140}{ 84}
\datum{-140}{ 88}
\datum{-140}{ 94}
\datum{-140}{ 96}
\datum{-140}{ 98}
\datum{-140}{ 110}
\datum{-140}{ 114}
\datum{-140}{ 130}
\datum{-140}{ 138}
\datum{-140}{ 148}
\datum{-140}{ 158}
\datum{-138}{ 79}
\datum{-138}{ 85}
\datum{-138}{ 97}
\datum{-138}{ 103}
\datum{-138}{ 105}
\datum{-138}{ 125}
\datum{-138}{ 151}
\datum{-136}{ 78}
\datum{-136}{ 82}
\datum{-136}{ 94}
\datum{-136}{ 102}
\datum{-136}{ 108}
\datum{-136}{ 112}
\datum{-136}{ 118}
\datum{-136}{ 126}
\datum{-136}{ 136}
\datum{-136}{ 138}
\datum{-136}{ 182}
\datum{-134}{ 87}
\datum{-132}{ 72}
\datum{-132}{ 74}
\datum{-132}{ 76}
\datum{-132}{ 78}
\datum{-132}{ 80}
\datum{-132}{ 84}
\datum{-132}{ 86}
\datum{-132}{ 90}
\datum{-132}{ 92}
\datum{-132}{ 94}
\datum{-132}{ 96}
\datum{-132}{ 100}
\datum{-132}{ 102}
\datum{-132}{ 104}
\datum{-132}{ 114}
\datum{-132}{ 126}
\datum{-132}{ 128}
\datum{-132}{ 148}
\datum{-132}{ 150}
\datum{-132}{ 172}
\datum{-132}{ 178}
\datum{-132}{ 202}
\datum{-132}{ 238}
\datum{-132}{ 302}
\datum{-130}{ 93}
\datum{-130}{ 109}
\datum{-128}{ 74}
\datum{-128}{ 78}
\datum{-128}{ 82}
\datum{-128}{ 86}
\datum{-128}{ 90}
\datum{-128}{ 92}
\datum{-128}{ 94}
\datum{-128}{ 98}
\datum{-128}{ 102}
\datum{-128}{ 108}
\datum{-128}{ 110}
\datum{-128}{ 118}
\datum{-128}{ 124}
\datum{-128}{ 126}
\datum{-128}{ 142}
\datum{-126}{ 69}
\datum{-126}{ 77}
\datum{-126}{ 79}
\datum{-126}{ 85}
\datum{-126}{ 87}
\datum{-126}{ 89}
\datum{-126}{ 91}
\datum{-126}{ 99}
\datum{-126}{ 101}
\datum{-126}{ 109}
\datum{-126}{ 115}
\datum{-126}{ 117}
\datum{-126}{ 133}
\datum{-126}{ 143}
\datum{-126}{ 179}
\datum{-124}{ 84}
\datum{-124}{ 94}
\datum{-124}{ 106}
\datum{-124}{ 136}
\datum{-124}{ 150}
\datum{-124}{ 162}
\datum{-120}{ 68}
\datum{-120}{ 70}
\datum{-120}{ 72}
\datum{-120}{ 74}
\datum{-120}{ 76}
\datum{-120}{ 78}
\datum{-120}{ 80}
\datum{-120}{ 82}
\datum{-120}{ 86}
\datum{-120}{ 88}
\datum{-120}{ 90}
\datum{-120}{ 92}
\datum{-120}{ 94}
\datum{-120}{ 96}
\datum{-120}{ 98}
\datum{-120}{ 100}
\datum{-120}{ 102}
\datum{-120}{ 104}
\datum{-120}{ 106}
\datum{-120}{ 108}
\datum{-120}{ 110}
\datum{-120}{ 112}
\datum{-120}{ 114}
\datum{-120}{ 116}
\datum{-120}{ 118}
\datum{-120}{ 122}
\datum{-120}{ 124}
\datum{-120}{ 128}
\datum{-120}{ 130}
\datum{-120}{ 132}
\datum{-120}{ 134}
\datum{-120}{ 136}
\datum{-120}{ 138}
\datum{-120}{ 140}
\datum{-120}{ 142}
\datum{-120}{ 146}
\datum{-120}{ 148}
\datum{-120}{ 150}
\datum{-120}{ 152}
\datum{-120}{ 154}
\datum{-120}{ 156}
\datum{-120}{ 158}
\datum{-120}{ 168}
\datum{-120}{ 178}
\datum{-120}{ 190}
\datum{-120}{ 196}
\datum{-120}{ 212}
\datum{-120}{ 248}
\datum{-120}{ 276}
\datum{-120}{ 278}
\datum{-116}{ 72}
\datum{-116}{ 108}
\datum{-116}{ 124}
\datum{-114}{ 67}
\datum{-114}{ 77}
\datum{-114}{ 79}
\datum{-114}{ 85}
\datum{-114}{ 91}
\datum{-114}{ 97}
\datum{-114}{ 103}
\datum{-114}{ 113}
\datum{-114}{ 115}
\datum{-114}{ 127}
\datum{-114}{ 145}
\datum{-114}{ 147}
\datum{-114}{ 169}
\datum{-114}{ 187}
\datum{-114}{ 197}
\datum{-112}{ 70}
\datum{-112}{ 74}
\datum{-112}{ 76}
\datum{-112}{ 78}
\datum{-112}{ 80}
\datum{-112}{ 86}
\datum{-112}{ 88}
\datum{-112}{ 90}
\datum{-112}{ 92}
\datum{-112}{ 94}
\datum{-112}{ 96}
\datum{-112}{ 102}
\datum{-112}{ 108}
\datum{-112}{ 110}
\datum{-112}{ 118}
\datum{-112}{ 124}
\datum{-112}{ 126}
\datum{-112}{ 138}
\datum{-112}{ 150}
\datum{-112}{ 160}
\datum{-110}{ 73}
\datum{-110}{ 83}
\datum{-110}{ 119}
\datum{-110}{ 141}
\datum{-110}{ 193}
\datum{-108}{ 66}
\datum{-108}{ 68}
\datum{-108}{ 70}
\datum{-108}{ 74}
\datum{-108}{ 76}
\datum{-108}{ 78}
\datum{-108}{ 80}
\datum{-108}{ 82}
\datum{-108}{ 84}
\datum{-108}{ 86}
\datum{-108}{ 88}
\datum{-108}{ 90}
\datum{-108}{ 94}
\datum{-108}{ 96}
\datum{-108}{ 100}
\datum{-108}{ 106}
\datum{-108}{ 108}
\datum{-108}{ 112}
\datum{-108}{ 118}
\datum{-108}{ 120}
\datum{-108}{ 122}
\datum{-108}{ 130}
\datum{-108}{ 150}
\datum{-108}{ 166}
\datum{-108}{ 178}
\datum{-108}{ 186}
\datum{-108}{ 190}
\datum{-108}{ 212}
\datum{-106}{ 69}
\datum{-106}{ 117}
\datum{-106}{ 129}
\datum{-106}{ 133}
\datum{-104}{ 62}
\datum{-104}{ 68}
\datum{-104}{ 74}
\datum{-104}{ 76}
\datum{-104}{ 78}
\datum{-104}{ 80}
\datum{-104}{ 82}
\datum{-104}{ 84}
\datum{-104}{ 86}
\datum{-104}{ 90}
\datum{-104}{ 92}
\datum{-104}{ 100}
\datum{-104}{ 102}
\datum{-104}{ 106}
\datum{-104}{ 108}
\datum{-104}{ 126}
\datum{-104}{ 132}
\datum{-104}{ 140}
\datum{-104}{ 154}
\datum{-102}{ 61}
\datum{-102}{ 67}
\datum{-102}{ 73}
\datum{-102}{ 81}
\datum{-102}{ 83}
\datum{-102}{ 93}
\datum{-102}{ 97}
\datum{-102}{ 99}
\datum{-102}{ 113}
\datum{-100}{ 62}
\datum{-100}{ 66}
\datum{-100}{ 70}
\datum{-100}{ 76}
\datum{-100}{ 80}
\datum{-100}{ 88}
\datum{-100}{ 96}
\datum{-100}{ 98}
\datum{-100}{ 118}
\datum{-100}{ 120}
\datum{-100}{ 168}
\datum{-98}{ 87}
\datum{-96}{ 54}
\datum{-96}{ 58}
\datum{-96}{ 60}
\datum{-96}{ 62}
\datum{-96}{ 64}
\datum{-96}{ 66}
\datum{-96}{ 70}
\datum{-96}{ 74}
\datum{-96}{ 76}
\datum{-96}{ 78}
\datum{-96}{ 80}
\datum{-96}{ 82}
\datum{-96}{ 84}
\datum{-96}{ 86}
\datum{-96}{ 88}
\datum{-96}{ 90}
\datum{-96}{ 92}
\datum{-96}{ 94}
\datum{-96}{ 96}
\datum{-96}{ 98}
\datum{-96}{ 102}
\datum{-96}{ 106}
\datum{-96}{ 108}
\datum{-96}{ 110}
\datum{-96}{ 112}
\datum{-96}{ 114}
\datum{-96}{ 118}
\datum{-96}{ 120}
\datum{-96}{ 122}
\datum{-96}{ 126}
\datum{-96}{ 134}
\datum{-96}{ 138}
\datum{-96}{ 142}
\datum{-96}{ 146}
\datum{-96}{ 154}
\datum{-96}{ 158}
\datum{-96}{ 166}
\datum{-96}{ 186}
\datum{-96}{ 190}
\datum{-96}{ 204}
\datum{-96}{ 212}
\datum{-96}{ 250}
\datum{-96}{ 286}
\datum{-96}{ 342}
\datum{-96}{ 402}
\datum{-94}{ 87}
\datum{-92}{ 56}
\datum{-92}{ 64}
\datum{-92}{ 86}
\datum{-92}{ 94}
\datum{-92}{ 98}
\datum{-92}{ 150}
\datum{-90}{ 53}
\datum{-90}{ 61}
\datum{-90}{ 63}
\datum{-90}{ 65}
\datum{-90}{ 67}
\datum{-90}{ 71}
\datum{-90}{ 73}
\datum{-90}{ 79}
\datum{-90}{ 83}
\datum{-90}{ 93}
\datum{-90}{ 95}
\datum{-90}{ 97}
\datum{-90}{ 99}
\datum{-90}{ 105}
\datum{-90}{ 109}
\datum{-90}{ 131}
\datum{-90}{ 133}
\datum{-90}{ 141}
\datum{-90}{ 171}
\datum{-88}{ 54}
\datum{-88}{ 62}
\datum{-88}{ 66}
\datum{-88}{ 68}
\datum{-88}{ 70}
\datum{-88}{ 78}
\datum{-88}{ 82}
\datum{-88}{ 84}
\datum{-88}{ 102}
\datum{-88}{ 104}
\datum{-88}{ 108}
\datum{-88}{ 128}
\datum{-88}{ 132}
\datum{-88}{ 158}
\datum{-86}{ 57}
\datum{-86}{ 77}
\datum{-86}{ 87}
\datum{-86}{ 127}
\datum{-86}{ 137}
\datum{-84}{ 52}
\datum{-84}{ 54}
\datum{-84}{ 58}
\datum{-84}{ 60}
\datum{-84}{ 64}
\datum{-84}{ 66}
\datum{-84}{ 68}
\datum{-84}{ 70}
\datum{-84}{ 72}
\datum{-84}{ 74}
\datum{-84}{ 78}
\datum{-84}{ 80}
\datum{-84}{ 82}
\datum{-84}{ 84}
\datum{-84}{ 88}
\datum{-84}{ 90}
\datum{-84}{ 94}
\datum{-84}{ 96}
\datum{-84}{ 98}
\datum{-84}{ 100}
\datum{-84}{ 102}
\datum{-84}{ 106}
\datum{-84}{ 114}
\datum{-84}{ 124}
\datum{-84}{ 130}
\datum{-84}{ 132}
\datum{-84}{ 134}
\datum{-84}{ 138}
\datum{-84}{ 154}
\datum{-84}{ 162}
\datum{-84}{ 164}
\datum{-84}{ 166}
\datum{-84}{ 172}
\datum{-84}{ 174}
\datum{-84}{ 178}
\datum{-84}{ 190}
\datum{-84}{ 194}
\datum{-84}{ 262}
\datum{-84}{ 322}
\datum{-80}{ 54}
\datum{-80}{ 58}
\datum{-80}{ 60}
\datum{-80}{ 62}
\datum{-80}{ 66}
\datum{-80}{ 68}
\datum{-80}{ 70}
\datum{-80}{ 72}
\datum{-80}{ 74}
\datum{-80}{ 76}
\datum{-80}{ 78}
\datum{-80}{ 94}
\datum{-80}{ 96}
\datum{-80}{ 98}
\datum{-80}{ 100}
\datum{-80}{ 102}
\datum{-80}{ 106}
\datum{-80}{ 118}
\datum{-80}{ 122}
\datum{-80}{ 138}
\datum{-80}{ 146}
\datum{-78}{ 61}
\datum{-78}{ 79}
\datum{-78}{ 83}
\datum{-78}{ 89}
\datum{-78}{ 91}
\datum{-78}{ 93}
\datum{-78}{ 95}
\datum{-78}{ 109}
\datum{-78}{ 111}
\datum{-78}{ 153}
\datum{-78}{ 163}
\datum{-78}{ 175}
\datum{-76}{ 68}
\datum{-76}{ 74}
\datum{-76}{ 94}
\datum{-76}{ 100}
\datum{-76}{ 112}
\datum{-76}{ 124}
\datum{-76}{ 142}
\datum{-74}{ 121}
\datum{-74}{ 177}
\datum{-72}{ 46}
\datum{-72}{ 48}
\datum{-72}{ 50}
\datum{-72}{ 52}
\datum{-72}{ 54}
\datum{-72}{ 56}
\datum{-72}{ 58}
\datum{-72}{ 60}
\datum{-72}{ 62}
\datum{-72}{ 64}
\datum{-72}{ 66}
\datum{-72}{ 68}
\datum{-72}{ 70}
\datum{-72}{ 74}
\datum{-72}{ 76}
\datum{-72}{ 78}
\datum{-72}{ 80}
\datum{-72}{ 82}
\datum{-72}{ 84}
\datum{-72}{ 86}
\datum{-72}{ 88}
\datum{-72}{ 90}
\datum{-72}{ 92}
\datum{-72}{ 94}
\datum{-72}{ 96}
\datum{-72}{ 98}
\datum{-72}{ 100}
\datum{-72}{ 102}
\datum{-72}{ 104}
\datum{-72}{ 112}
\datum{-72}{ 116}
\datum{-72}{ 118}
\datum{-72}{ 120}
\datum{-72}{ 124}
\datum{-72}{ 126}
\datum{-72}{ 128}
\datum{-72}{ 132}
\datum{-72}{ 136}
\datum{-72}{ 140}
\datum{-72}{ 142}
\datum{-72}{ 144}
\datum{-72}{ 148}
\datum{-72}{ 150}
\datum{-72}{ 158}
\datum{-72}{ 160}
\datum{-72}{ 164}
\datum{-72}{ 174}
\datum{-72}{ 182}
\datum{-72}{ 198}
\datum{-72}{ 212}
\datum{-70}{ 53}
\datum{-70}{ 73}
\datum{-70}{ 75}
\datum{-70}{ 89}
\datum{-70}{ 95}
\datum{-70}{ 101}
\datum{-70}{ 135}
\datum{-68}{ 58}
\datum{-68}{ 64}
\datum{-68}{ 78}
\datum{-68}{ 86}
\datum{-68}{ 110}
\datum{-68}{ 126}
\datum{-68}{ 150}
\datum{-66}{ 61}
\datum{-66}{ 63}
\datum{-66}{ 65}
\datum{-66}{ 69}
\datum{-66}{ 73}
\datum{-66}{ 75}
\datum{-66}{ 79}
\datum{-66}{ 87}
\datum{-66}{ 89}
\datum{-66}{ 97}
\datum{-66}{ 129}
\datum{-66}{ 137}
\datum{-66}{ 159}
\datum{-64}{ 48}
\datum{-64}{ 52}
\datum{-64}{ 54}
\datum{-64}{ 60}
\datum{-64}{ 62}
\datum{-64}{ 66}
\datum{-64}{ 70}
\datum{-64}{ 72}
\datum{-64}{ 76}
\datum{-64}{ 78}
\datum{-64}{ 80}
\datum{-64}{ 82}
\datum{-64}{ 86}
\datum{-64}{ 90}
\datum{-64}{ 102}
\datum{-64}{ 106}
\datum{-64}{ 110}
\datum{-64}{ 118}
\datum{-64}{ 124}
\datum{-64}{ 134}
\datum{-64}{ 154}
\datum{-64}{ 166}
\datum{-62}{ 89}
\datum{-62}{ 105}
\datum{-62}{ 147}
\datum{-60}{ 46}
\datum{-60}{ 48}
\datum{-60}{ 52}
\datum{-60}{ 54}
\datum{-60}{ 56}
\datum{-60}{ 58}
\datum{-60}{ 60}
\datum{-60}{ 62}
\datum{-60}{ 64}
\datum{-60}{ 66}
\datum{-60}{ 68}
\datum{-60}{ 70}
\datum{-60}{ 74}
\datum{-60}{ 78}
\datum{-60}{ 82}
\datum{-60}{ 84}
\datum{-60}{ 88}
\datum{-60}{ 90}
\datum{-60}{ 94}
\datum{-60}{ 96}
\datum{-60}{ 98}
\datum{-60}{ 106}
\datum{-60}{ 110}
\datum{-60}{ 112}
\datum{-60}{ 120}
\datum{-60}{ 122}
\datum{-60}{ 124}
\datum{-60}{ 126}
\datum{-60}{ 132}
\datum{-60}{ 144}
\datum{-60}{ 166}
\datum{-60}{ 178}
\datum{-60}{ 196}
\datum{-60}{ 218}
\datum{-60}{ 222}
\datum{-60}{ 234}
\datum{-60}{ 358}
\datum{-60}{ 474}
\datum{-58}{ 43}
\datum{-58}{ 49}
\datum{-58}{ 129}
\datum{-56}{ 50}
\datum{-56}{ 52}
\datum{-56}{ 54}
\datum{-56}{ 58}
\datum{-56}{ 62}
\datum{-56}{ 66}
\datum{-56}{ 68}
\datum{-56}{ 70}
\datum{-56}{ 72}
\datum{-56}{ 74}
\datum{-56}{ 82}
\datum{-56}{ 84}
\datum{-56}{ 86}
\datum{-56}{ 90}
\datum{-56}{ 96}
\datum{-56}{ 108}
\datum{-56}{ 114}
\datum{-56}{ 116}
\datum{-56}{ 124}
\datum{-56}{ 126}
\datum{-56}{ 134}
\datum{-56}{ 150}
\datum{-56}{ 166}
\datum{-56}{ 180}
\datum{-56}{ 214}
\datum{-54}{ 41}
\datum{-54}{ 43}
\datum{-54}{ 47}
\datum{-54}{ 49}
\datum{-54}{ 57}
\datum{-54}{ 59}
\datum{-54}{ 61}
\datum{-54}{ 63}
\datum{-54}{ 71}
\datum{-54}{ 73}
\datum{-54}{ 77}
\datum{-54}{ 79}
\datum{-54}{ 83}
\datum{-54}{ 85}
\datum{-54}{ 87}
\datum{-54}{ 103}
\datum{-54}{ 117}
\datum{-54}{ 133}
\datum{-54}{ 147}
\datum{-54}{ 157}
\datum{-52}{ 64}
\datum{-52}{ 108}
\datum{-52}{ 116}
\datum{-50}{ 45}
\datum{-50}{ 53}
\datum{-50}{ 63}
\datum{-50}{ 73}
\datum{-50}{ 93}
\datum{-48}{ 40}
\datum{-48}{ 42}
\datum{-48}{ 44}
\datum{-48}{ 46}
\datum{-48}{ 48}
\datum{-48}{ 50}
\datum{-48}{ 52}
\datum{-48}{ 54}
\datum{-48}{ 56}
\datum{-48}{ 58}
\datum{-48}{ 60}
\datum{-48}{ 62}
\datum{-48}{ 64}
\datum{-48}{ 66}
\datum{-48}{ 68}
\datum{-48}{ 70}
\datum{-48}{ 72}
\datum{-48}{ 74}
\datum{-48}{ 76}
\datum{-48}{ 78}
\datum{-48}{ 80}
\datum{-48}{ 82}
\datum{-48}{ 84}
\datum{-48}{ 86}
\datum{-48}{ 88}
\datum{-48}{ 90}
\datum{-48}{ 92}
\datum{-48}{ 94}
\datum{-48}{ 96}
\datum{-48}{ 98}
\datum{-48}{ 100}
\datum{-48}{ 102}
\datum{-48}{ 106}
\datum{-48}{ 108}
\datum{-48}{ 110}
\datum{-48}{ 112}
\datum{-48}{ 118}
\datum{-48}{ 122}
\datum{-48}{ 124}
\datum{-48}{ 134}
\datum{-48}{ 138}
\datum{-48}{ 142}
\datum{-48}{ 158}
\datum{-48}{ 166}
\datum{-48}{ 168}
\datum{-48}{ 178}
\datum{-48}{ 202}
\datum{-48}{ 230}
\datum{-48}{ 266}
\datum{-46}{ 153}
\datum{-44}{ 44}
\datum{-44}{ 52}
\datum{-44}{ 56}
\datum{-44}{ 58}
\datum{-44}{ 64}
\datum{-44}{ 66}
\datum{-44}{ 78}
\datum{-44}{ 80}
\datum{-44}{ 92}
\datum{-44}{ 94}
\datum{-44}{ 108}
\datum{-44}{ 134}
\datum{-42}{ 43}
\datum{-42}{ 49}
\datum{-42}{ 53}
\datum{-42}{ 55}
\datum{-42}{ 61}
\datum{-42}{ 65}
\datum{-42}{ 67}
\datum{-42}{ 73}
\datum{-42}{ 79}
\datum{-42}{ 81}
\datum{-42}{ 83}
\datum{-42}{ 85}
\datum{-42}{ 89}
\datum{-42}{ 91}
\datum{-42}{ 93}
\datum{-42}{ 97}
\datum{-42}{ 115}
\datum{-42}{ 137}
\datum{-40}{ 36}
\datum{-40}{ 46}
\datum{-40}{ 48}
\datum{-40}{ 50}
\datum{-40}{ 52}
\datum{-40}{ 54}
\datum{-40}{ 56}
\datum{-40}{ 58}
\datum{-40}{ 62}
\datum{-40}{ 68}
\datum{-40}{ 70}
\datum{-40}{ 74}
\datum{-40}{ 78}
\datum{-40}{ 80}
\datum{-40}{ 86}
\datum{-40}{ 90}
\datum{-40}{ 94}
\datum{-40}{ 110}
\datum{-40}{ 116}
\datum{-40}{ 118}
\datum{-40}{ 120}
\datum{-40}{ 130}
\datum{-40}{ 138}
\datum{-40}{ 148}
\datum{-40}{ 160}
\datum{-38}{ 45}
\datum{-36}{ 42}
\datum{-36}{ 46}
\datum{-36}{ 50}
\datum{-36}{ 52}
\datum{-36}{ 54}
\datum{-36}{ 56}
\datum{-36}{ 58}
\datum{-36}{ 60}
\datum{-36}{ 62}
\datum{-36}{ 66}
\datum{-36}{ 68}
\datum{-36}{ 70}
\datum{-36}{ 72}
\datum{-36}{ 74}
\datum{-36}{ 76}
\datum{-36}{ 78}
\datum{-36}{ 82}
\datum{-36}{ 84}
\datum{-36}{ 88}
\datum{-36}{ 94}
\datum{-36}{ 100}
\datum{-36}{ 102}
\datum{-36}{ 104}
\datum{-36}{ 106}
\datum{-36}{ 108}
\datum{-36}{ 110}
\datum{-36}{ 112}
\datum{-36}{ 114}
\datum{-36}{ 118}
\datum{-36}{ 120}
\datum{-36}{ 126}
\datum{-36}{ 130}
\datum{-36}{ 134}
\datum{-36}{ 142}
\datum{-36}{ 144}
\datum{-36}{ 156}
\datum{-36}{ 162}
\datum{-36}{ 172}
\datum{-36}{ 184}
\datum{-36}{ 202}
\datum{-36}{ 212}
\datum{-36}{ 214}
\datum{-36}{ 222}
\datum{-36}{ 314}
\datum{-34}{ 45}
\datum{-34}{ 69}
\datum{-34}{ 83}
\datum{-32}{ 38}
\datum{-32}{ 42}
\datum{-32}{ 46}
\datum{-32}{ 48}
\datum{-32}{ 50}
\datum{-32}{ 52}
\datum{-32}{ 54}
\datum{-32}{ 56}
\datum{-32}{ 60}
\datum{-32}{ 62}
\datum{-32}{ 66}
\datum{-32}{ 70}
\datum{-32}{ 74}
\datum{-32}{ 76}
\datum{-32}{ 78}
\datum{-32}{ 82}
\datum{-32}{ 84}
\datum{-32}{ 86}
\datum{-32}{ 94}
\datum{-32}{ 104}
\datum{-32}{ 110}
\datum{-32}{ 114}
\datum{-32}{ 118}
\datum{-32}{ 190}
\datum{-30}{ 33}
\datum{-30}{ 43}
\datum{-30}{ 49}
\datum{-30}{ 51}
\datum{-30}{ 53}
\datum{-30}{ 61}
\datum{-30}{ 63}
\datum{-30}{ 69}
\datum{-30}{ 71}
\datum{-30}{ 73}
\datum{-30}{ 77}
\datum{-30}{ 79}
\datum{-30}{ 83}
\datum{-30}{ 91}
\datum{-30}{ 93}
\datum{-30}{ 103}
\datum{-30}{ 133}
\datum{-30}{ 163}
\datum{-28}{ 36}
\datum{-28}{ 48}
\datum{-28}{ 54}
\datum{-28}{ 58}
\datum{-28}{ 62}
\datum{-28}{ 72}
\datum{-28}{ 98}
\datum{-28}{ 146}
\datum{-28}{ 186}
\datum{-26}{ 63}
\datum{-26}{ 89}
\datum{-24}{ 36}
\datum{-24}{ 38}
\datum{-24}{ 40}
\datum{-24}{ 42}
\datum{-24}{ 44}
\datum{-24}{ 46}
\datum{-24}{ 48}
\datum{-24}{ 50}
\datum{-24}{ 52}
\datum{-24}{ 54}
\datum{-24}{ 56}
\datum{-24}{ 58}
\datum{-24}{ 60}
\datum{-24}{ 62}
\datum{-24}{ 64}
\datum{-24}{ 66}
\datum{-24}{ 68}
\datum{-24}{ 70}
\datum{-24}{ 72}
\datum{-24}{ 74}
\datum{-24}{ 76}
\datum{-24}{ 78}
\datum{-24}{ 80}
\datum{-24}{ 82}
\datum{-24}{ 84}
\datum{-24}{ 86}
\datum{-24}{ 88}
\datum{-24}{ 90}
\datum{-24}{ 92}
\datum{-24}{ 94}
\datum{-24}{ 96}
\datum{-24}{ 98}
\datum{-24}{ 100}
\datum{-24}{ 102}
\datum{-24}{ 112}
\datum{-24}{ 114}
\datum{-24}{ 116}
\datum{-24}{ 120}
\datum{-24}{ 128}
\datum{-24}{ 130}
\datum{-24}{ 132}
\datum{-24}{ 134}
\datum{-24}{ 142}
\datum{-24}{ 144}
\datum{-24}{ 148}
\datum{-24}{ 160}
\datum{-24}{ 162}
\datum{-24}{ 164}
\datum{-24}{ 166}
\datum{-24}{ 184}
\datum{-24}{ 232}
\datum{-22}{ 41}
\datum{-20}{ 40}
\datum{-20}{ 44}
\datum{-20}{ 46}
\datum{-20}{ 48}
\datum{-20}{ 56}
\datum{-20}{ 58}
\datum{-20}{ 64}
\datum{-20}{ 68}
\datum{-20}{ 74}
\datum{-20}{ 84}
\datum{-20}{ 90}
\datum{-20}{ 108}
\datum{-20}{ 114}
\datum{-20}{ 116}
\datum{-20}{ 198}
\datum{-20}{ 234}
\datum{-18}{ 37}
\datum{-18}{ 39}
\datum{-18}{ 45}
\datum{-18}{ 47}
\datum{-18}{ 49}
\datum{-18}{ 55}
\datum{-18}{ 61}
\datum{-18}{ 63}
\datum{-18}{ 65}
\datum{-18}{ 67}
\datum{-18}{ 69}
\datum{-18}{ 79}
\datum{-18}{ 81}
\datum{-18}{ 83}
\datum{-18}{ 85}
\datum{-18}{ 89}
\datum{-18}{ 91}
\datum{-18}{ 97}
\datum{-18}{ 103}
\datum{-18}{ 105}
\datum{-18}{ 109}
\datum{-18}{ 115}
\datum{-18}{ 149}
\datum{-16}{ 42}
\datum{-16}{ 46}
\datum{-16}{ 48}
\datum{-16}{ 50}
\datum{-16}{ 52}
\datum{-16}{ 54}
\datum{-16}{ 56}
\datum{-16}{ 58}
\datum{-16}{ 60}
\datum{-16}{ 62}
\datum{-16}{ 64}
\datum{-16}{ 66}
\datum{-16}{ 70}
\datum{-16}{ 72}
\datum{-16}{ 74}
\datum{-16}{ 76}
\datum{-16}{ 78}
\datum{-16}{ 82}
\datum{-16}{ 84}
\datum{-16}{ 86}
\datum{-16}{ 92}
\datum{-16}{ 94}
\datum{-16}{ 122}
\datum{-16}{ 150}
\datum{-14}{ 45}
\datum{-14}{ 47}
\datum{-14}{ 61}
\datum{-14}{ 137}
\datum{-12}{ 36}
\datum{-12}{ 38}
\datum{-12}{ 42}
\datum{-12}{ 44}
\datum{-12}{ 48}
\datum{-12}{ 50}
\datum{-12}{ 52}
\datum{-12}{ 54}
\datum{-12}{ 56}
\datum{-12}{ 58}
\datum{-12}{ 64}
\datum{-12}{ 66}
\datum{-12}{ 70}
\datum{-12}{ 72}
\datum{-12}{ 76}
\datum{-12}{ 78}
\datum{-12}{ 80}
\datum{-12}{ 82}
\datum{-12}{ 86}
\datum{-12}{ 88}
\datum{-12}{ 90}
\datum{-12}{ 94}
\datum{-12}{ 96}
\datum{-12}{ 98}
\datum{-12}{ 100}
\datum{-12}{ 104}
\datum{-12}{ 108}
\datum{-12}{ 116}
\datum{-12}{ 122}
\datum{-12}{ 128}
\datum{-12}{ 130}
\datum{-12}{ 146}
\datum{-12}{ 156}
\datum{-10}{ 43}
\datum{-10}{ 53}
\datum{-10}{ 63}
\datum{-10}{ 73}
\datum{-10}{ 83}
\datum{-10}{ 141}
\datum{-8}{ 40}
\datum{-8}{ 42}
\datum{-8}{ 46}
\datum{-8}{ 50}
\datum{-8}{ 52}
\datum{-8}{ 54}
\datum{-8}{ 56}
\datum{-8}{ 58}
\datum{-8}{ 60}
\datum{-8}{ 62}
\datum{-8}{ 66}
\datum{-8}{ 68}
\datum{-8}{ 70}
\datum{-8}{ 78}
\datum{-8}{ 82}
\datum{-8}{ 90}
\datum{-8}{ 96}
\datum{-8}{ 98}
\datum{-8}{ 102}
\datum{-8}{ 116}
\datum{-6}{ 29}
\datum{-6}{ 33}
\datum{-6}{ 37}
\datum{-6}{ 43}
\datum{-6}{ 45}
\datum{-6}{ 49}
\datum{-6}{ 55}
\datum{-6}{ 57}
\datum{-6}{ 61}
\datum{-6}{ 65}
\datum{-6}{ 67}
\datum{-6}{ 71}
\datum{-6}{ 73}
\datum{-6}{ 77}
\datum{-6}{ 83}
\datum{-6}{ 87}
\datum{-6}{ 97}
\datum{-6}{ 99}
\datum{-6}{ 117}
\datum{-6}{ 137}
\datum{-6}{ 151}
\datum{-4}{ 38}
\datum{-4}{ 44}
\datum{-4}{ 50}
\datum{-4}{ 54}
\datum{-4}{ 64}
\datum{-4}{ 66}
\datum{-4}{ 68}
\datum{-4}{ 82}
\datum{-4}{ 92}
\datum{-4}{ 96}
\datum{-4}{ 102}
\datum{-4}{ 110}
\datum{-4}{ 156}
\datum{0}{ 36}
\datum{0}{ 38}
\datum{0}{ 40}
\datum{0}{ 42}
\datum{0}{ 44}
\datum{0}{ 46}
\datum{0}{ 48}
\datum{0}{ 50}
\datum{0}{ 52}
\datum{0}{ 54}
\datum{0}{ 56}
\datum{0}{ 58}
\datum{0}{ 62}
\datum{0}{ 64}
\datum{0}{ 66}
\datum{0}{ 68}
\datum{0}{ 70}
\datum{0}{ 74}
\datum{0}{ 76}
\datum{0}{ 78}
\datum{0}{ 82}
\datum{0}{ 86}
\datum{0}{ 88}
\datum{0}{ 94}
\datum{0}{ 98}
\datum{0}{ 104}
\datum{0}{ 106}
\datum{0}{ 110}
\datum{0}{ 112}
\datum{0}{ 114}
\datum{0}{ 118}
\datum{0}{ 122}
\datum{0}{ 124}
\datum{0}{ 126}
\datum{0}{ 130}
\datum{0}{ 134}
\datum{0}{ 138}
\datum{0}{ 142}
\datum{0}{ 150}
\datum{0}{ 154}
\datum{0}{ 156}
\datum{0}{ 158}
\datum{0}{ 162}
\datum{0}{ 166}
\datum{0}{ 170}
\datum{0}{ 174}
\datum{0}{ 178}
\datum{0}{ 182}
\datum{0}{ 190}
\datum{0}{ 194}
\datum{0}{ 206}
\datum{0}{ 214}
\datum{0}{ 222}
\datum{0}{ 238}
\datum{0}{ 242}
\datum{0}{ 246}
\datum{0}{ 262}
\datum{0}{ 286}
\datum{0}{ 298}
\datum{0}{ 302}
\datum{0}{ 358}
\datum{0}{ 446}
\datum{0}{ 502}
\datum{4}{ 38}
\datum{4}{ 44}
\datum{4}{ 50}
\datum{4}{ 54}
\datum{4}{ 64}
\datum{4}{ 66}
\datum{4}{ 68}
\datum{4}{ 82}
\datum{4}{ 92}
\datum{4}{ 96}
\datum{4}{ 102}
\datum{4}{ 110}
\datum{4}{ 156}
\datum{6}{ 29}
\datum{6}{ 33}
\datum{6}{ 37}
\datum{6}{ 43}
\datum{6}{ 45}
\datum{6}{ 49}
\datum{6}{ 55}
\datum{6}{ 57}
\datum{6}{ 61}
\datum{6}{ 65}
\datum{6}{ 67}
\datum{6}{ 71}
\datum{6}{ 73}
\datum{6}{ 77}
\datum{6}{ 83}
\datum{6}{ 87}
\datum{6}{ 97}
\datum{6}{ 99}
\datum{6}{ 117}
\datum{6}{ 137}
\datum{6}{ 151}
\datum{8}{ 40}
\datum{8}{ 42}
\datum{8}{ 46}
\datum{8}{ 50}
\datum{8}{ 52}
\datum{8}{ 54}
\datum{8}{ 56}
\datum{8}{ 58}
\datum{8}{ 60}
\datum{8}{ 62}
\datum{8}{ 66}
\datum{8}{ 68}
\datum{8}{ 70}
\datum{8}{ 78}
\datum{8}{ 82}
\datum{8}{ 90}
\datum{8}{ 96}
\datum{8}{ 98}
\datum{8}{ 102}
\datum{8}{ 116}
\datum{10}{ 43}
\datum{10}{ 53}
\datum{10}{ 63}
\datum{10}{ 73}
\datum{10}{ 83}
\datum{10}{ 141}
\datum{12}{ 36}
\datum{12}{ 38}
\datum{12}{ 42}
\datum{12}{ 44}
\datum{12}{ 48}
\datum{12}{ 50}
\datum{12}{ 52}
\datum{12}{ 54}
\datum{12}{ 56}
\datum{12}{ 58}
\datum{12}{ 64}
\datum{12}{ 66}
\datum{12}{ 70}
\datum{12}{ 72}
\datum{12}{ 76}
\datum{12}{ 78}
\datum{12}{ 80}
\datum{12}{ 82}
\datum{12}{ 86}
\datum{12}{ 88}
\datum{12}{ 90}
\datum{12}{ 94}
\datum{12}{ 96}
\datum{12}{ 98}
\datum{12}{ 100}
\datum{12}{ 104}
\datum{12}{ 108}
\datum{12}{ 116}
\datum{12}{ 122}
\datum{12}{ 128}
\datum{12}{ 130}
\datum{12}{ 146}
\datum{12}{ 156}
\datum{14}{ 45}
\datum{14}{ 47}
\datum{14}{ 61}
\datum{14}{ 137}
\datum{16}{ 42}
\datum{16}{ 46}
\datum{16}{ 48}
\datum{16}{ 50}
\datum{16}{ 52}
\datum{16}{ 54}
\datum{16}{ 56}
\datum{16}{ 58}
\datum{16}{ 60}
\datum{16}{ 62}
\datum{16}{ 64}
\datum{16}{ 66}
\datum{16}{ 70}
\datum{16}{ 72}
\datum{16}{ 74}
\datum{16}{ 76}
\datum{16}{ 78}
\datum{16}{ 82}
\datum{16}{ 84}
\datum{16}{ 86}
\datum{16}{ 92}
\datum{16}{ 94}
\datum{16}{ 122}
\datum{16}{ 150}
\datum{18}{ 37}
\datum{18}{ 39}
\datum{18}{ 45}
\datum{18}{ 47}
\datum{18}{ 49}
\datum{18}{ 55}
\datum{18}{ 61}
\datum{18}{ 63}
\datum{18}{ 65}
\datum{18}{ 67}
\datum{18}{ 69}
\datum{18}{ 79}
\datum{18}{ 81}
\datum{18}{ 83}
\datum{18}{ 85}
\datum{18}{ 89}
\datum{18}{ 91}
\datum{18}{ 97}
\datum{18}{ 103}
\datum{18}{ 105}
\datum{18}{ 109}
\datum{18}{ 115}
\datum{18}{ 149}
\datum{20}{ 40}
\datum{20}{ 44}
\datum{20}{ 46}
\datum{20}{ 48}
\datum{20}{ 56}
\datum{20}{ 58}
\datum{20}{ 64}
\datum{20}{ 68}
\datum{20}{ 74}
\datum{20}{ 84}
\datum{20}{ 90}
\datum{20}{ 108}
\datum{20}{ 114}
\datum{20}{ 116}
\datum{20}{ 198}
\datum{20}{ 234}
\datum{22}{ 41}
\datum{24}{ 36}
\datum{24}{ 38}
\datum{24}{ 40}
\datum{24}{ 42}
\datum{24}{ 44}
\datum{24}{ 46}
\datum{24}{ 48}
\datum{24}{ 50}
\datum{24}{ 52}
\datum{24}{ 54}
\datum{24}{ 56}
\datum{24}{ 58}
\datum{24}{ 60}
\datum{24}{ 62}
\datum{24}{ 64}
\datum{24}{ 66}
\datum{24}{ 68}
\datum{24}{ 70}
\datum{24}{ 72}
\datum{24}{ 74}
\datum{24}{ 76}
\datum{24}{ 78}
\datum{24}{ 80}
\datum{24}{ 82}
\datum{24}{ 84}
\datum{24}{ 86}
\datum{24}{ 88}
\datum{24}{ 90}
\datum{24}{ 92}
\datum{24}{ 94}
\datum{24}{ 96}
\datum{24}{ 98}
\datum{24}{ 100}
\datum{24}{ 102}
\datum{24}{ 112}
\datum{24}{ 114}
\datum{24}{ 116}
\datum{24}{ 120}
\datum{24}{ 128}
\datum{24}{ 130}
\datum{24}{ 132}
\datum{24}{ 134}
\datum{24}{ 142}
\datum{24}{ 144}
\datum{24}{ 148}
\datum{24}{ 160}
\datum{24}{ 162}
\datum{24}{ 164}
\datum{24}{ 166}
\datum{24}{ 184}
\datum{24}{ 232}
\datum{26}{ 63}
\datum{26}{ 89}
\datum{28}{ 36}
\datum{28}{ 48}
\datum{28}{ 54}
\datum{28}{ 58}
\datum{28}{ 62}
\datum{28}{ 72}
\datum{28}{ 98}
\datum{28}{ 146}
\datum{28}{ 186}
\datum{30}{ 33}
\datum{30}{ 43}
\datum{30}{ 49}
\datum{30}{ 51}
\datum{30}{ 53}
\datum{30}{ 61}
\datum{30}{ 63}
\datum{30}{ 69}
\datum{30}{ 71}
\datum{30}{ 73}
\datum{30}{ 77}
\datum{30}{ 79}
\datum{30}{ 83}
\datum{30}{ 91}
\datum{30}{ 93}
\datum{30}{ 103}
\datum{30}{ 133}
\datum{30}{ 163}
\datum{32}{ 38}
\datum{32}{ 42}
\datum{32}{ 46}
\datum{32}{ 48}
\datum{32}{ 50}
\datum{32}{ 52}
\datum{32}{ 54}
\datum{32}{ 56}
\datum{32}{ 60}
\datum{32}{ 62}
\datum{32}{ 66}
\datum{32}{ 70}
\datum{32}{ 74}
\datum{32}{ 76}
\datum{32}{ 78}
\datum{32}{ 82}
\datum{32}{ 84}
\datum{32}{ 86}
\datum{32}{ 94}
\datum{32}{ 104}
\datum{32}{ 110}
\datum{32}{ 114}
\datum{32}{ 118}
\datum{32}{ 190}
\datum{34}{ 45}
\datum{34}{ 69}
\datum{34}{ 83}
\datum{36}{ 42}
\datum{36}{ 46}
\datum{36}{ 50}
\datum{36}{ 52}
\datum{36}{ 54}
\datum{36}{ 56}
\datum{36}{ 58}
\datum{36}{ 60}
\datum{36}{ 62}
\datum{36}{ 66}
\datum{36}{ 68}
\datum{36}{ 70}
\datum{36}{ 72}
\datum{36}{ 74}
\datum{36}{ 76}
\datum{36}{ 78}
\datum{36}{ 82}
\datum{36}{ 84}
\datum{36}{ 88}
\datum{36}{ 94}
\datum{36}{ 100}
\datum{36}{ 102}
\datum{36}{ 104}
\datum{36}{ 106}
\datum{36}{ 108}
\datum{36}{ 110}
\datum{36}{ 112}
\datum{36}{ 114}
\datum{36}{ 118}
\datum{36}{ 120}
\datum{36}{ 126}
\datum{36}{ 130}
\datum{36}{ 134}
\datum{36}{ 142}
\datum{36}{ 144}
\datum{36}{ 156}
\datum{36}{ 162}
\datum{36}{ 172}
\datum{36}{ 184}
\datum{36}{ 202}
\datum{36}{ 212}
\datum{36}{ 214}
\datum{36}{ 222}
\datum{36}{ 314}
\datum{38}{ 45}
\datum{40}{ 36}
\datum{40}{ 46}
\datum{40}{ 48}
\datum{40}{ 50}
\datum{40}{ 52}
\datum{40}{ 54}
\datum{40}{ 56}
\datum{40}{ 58}
\datum{40}{ 62}
\datum{40}{ 68}
\datum{40}{ 70}
\datum{40}{ 74}
\datum{40}{ 78}
\datum{40}{ 80}
\datum{40}{ 86}
\datum{40}{ 90}
\datum{40}{ 94}
\datum{40}{ 110}
\datum{40}{ 116}
\datum{40}{ 118}
\datum{40}{ 120}
\datum{40}{ 130}
\datum{40}{ 138}
\datum{40}{ 148}
\datum{40}{ 160}
\datum{42}{ 43}
\datum{42}{ 49}
\datum{42}{ 53}
\datum{42}{ 55}
\datum{42}{ 61}
\datum{42}{ 65}
\datum{42}{ 67}
\datum{42}{ 73}
\datum{42}{ 79}
\datum{42}{ 81}
\datum{42}{ 83}
\datum{42}{ 85}
\datum{42}{ 89}
\datum{42}{ 91}
\datum{42}{ 93}
\datum{42}{ 97}
\datum{42}{ 115}
\datum{42}{ 137}
\datum{44}{ 44}
\datum{44}{ 52}
\datum{44}{ 56}
\datum{44}{ 58}
\datum{44}{ 64}
\datum{44}{ 66}
\datum{44}{ 78}
\datum{44}{ 80}
\datum{44}{ 92}
\datum{44}{ 94}
\datum{44}{ 108}
\datum{44}{ 134}
\datum{46}{ 153}
\datum{48}{ 40}
\datum{48}{ 42}
\datum{48}{ 44}
\datum{48}{ 46}
\datum{48}{ 48}
\datum{48}{ 50}
\datum{48}{ 52}
\datum{48}{ 54}
\datum{48}{ 56}
\datum{48}{ 58}
\datum{48}{ 60}
\datum{48}{ 62}
\datum{48}{ 64}
\datum{48}{ 66}
\datum{48}{ 68}
\datum{48}{ 70}
\datum{48}{ 72}
\datum{48}{ 74}
\datum{48}{ 76}
\datum{48}{ 78}
\datum{48}{ 80}
\datum{48}{ 82}
\datum{48}{ 84}
\datum{48}{ 86}
\datum{48}{ 88}
\datum{48}{ 90}
\datum{48}{ 92}
\datum{48}{ 94}
\datum{48}{ 96}
\datum{48}{ 98}
\datum{48}{ 100}
\datum{48}{ 102}
\datum{48}{ 106}
\datum{48}{ 108}
\datum{48}{ 110}
\datum{48}{ 112}
\datum{48}{ 118}
\datum{48}{ 122}
\datum{48}{ 124}
\datum{48}{ 134}
\datum{48}{ 138}
\datum{48}{ 142}
\datum{48}{ 158}
\datum{48}{ 166}
\datum{48}{ 168}
\datum{48}{ 178}
\datum{48}{ 202}
\datum{48}{ 230}
\datum{48}{ 266}
\datum{50}{ 45}
\datum{50}{ 53}
\datum{50}{ 63}
\datum{50}{ 73}
\datum{50}{ 93}
\datum{52}{ 64}
\datum{52}{ 108}
\datum{52}{ 116}
\datum{54}{ 41}
\datum{54}{ 43}
\datum{54}{ 47}
\datum{54}{ 49}
\datum{54}{ 57}
\datum{54}{ 59}
\datum{54}{ 61}
\datum{54}{ 63}
\datum{54}{ 71}
\datum{54}{ 73}
\datum{54}{ 77}
\datum{54}{ 79}
\datum{54}{ 83}
\datum{54}{ 85}
\datum{54}{ 87}
\datum{54}{ 103}
\datum{54}{ 117}
\datum{54}{ 133}
\datum{54}{ 147}
\datum{54}{ 157}
\datum{56}{ 50}
\datum{56}{ 52}
\datum{56}{ 54}
\datum{56}{ 58}
\datum{56}{ 62}
\datum{56}{ 66}
\datum{56}{ 68}
\datum{56}{ 70}
\datum{56}{ 72}
\datum{56}{ 74}
\datum{56}{ 82}
\datum{56}{ 84}
\datum{56}{ 86}
\datum{56}{ 90}
\datum{56}{ 96}
\datum{56}{ 108}
\datum{56}{ 114}
\datum{56}{ 116}
\datum{56}{ 124}
\datum{56}{ 126}
\datum{56}{ 134}
\datum{56}{ 150}
\datum{56}{ 166}
\datum{56}{ 180}
\datum{56}{ 214}
\datum{58}{ 43}
\datum{58}{ 49}
\datum{58}{ 129}
\datum{60}{ 46}
\datum{60}{ 48}
\datum{60}{ 52}
\datum{60}{ 54}
\datum{60}{ 56}
\datum{60}{ 58}
\datum{60}{ 60}
\datum{60}{ 62}
\datum{60}{ 64}
\datum{60}{ 66}
\datum{60}{ 68}
\datum{60}{ 70}
\datum{60}{ 74}
\datum{60}{ 78}
\datum{60}{ 82}
\datum{60}{ 84}
\datum{60}{ 88}
\datum{60}{ 90}
\datum{60}{ 94}
\datum{60}{ 96}
\datum{60}{ 98}
\datum{60}{ 106}
\datum{60}{ 110}
\datum{60}{ 112}
\datum{60}{ 120}
\datum{60}{ 122}
\datum{60}{ 124}
\datum{60}{ 126}
\datum{60}{ 132}
\datum{60}{ 144}
\datum{60}{ 166}
\datum{60}{ 178}
\datum{60}{ 196}
\datum{60}{ 218}
\datum{60}{ 222}
\datum{60}{ 234}
\datum{60}{ 358}
\datum{60}{ 474}
\datum{62}{ 89}
\datum{62}{ 105}
\datum{62}{ 147}
\datum{64}{ 48}
\datum{64}{ 52}
\datum{64}{ 54}
\datum{64}{ 60}
\datum{64}{ 62}
\datum{64}{ 66}
\datum{64}{ 70}
\datum{64}{ 72}
\datum{64}{ 76}
\datum{64}{ 78}
\datum{64}{ 80}
\datum{64}{ 82}
\datum{64}{ 86}
\datum{64}{ 90}
\datum{64}{ 102}
\datum{64}{ 106}
\datum{64}{ 110}
\datum{64}{ 118}
\datum{64}{ 124}
\datum{64}{ 134}
\datum{64}{ 154}
\datum{64}{ 166}
\datum{66}{ 61}
\datum{66}{ 63}
\datum{66}{ 65}
\datum{66}{ 69}
\datum{66}{ 73}
\datum{66}{ 75}
\datum{66}{ 79}
\datum{66}{ 87}
\datum{66}{ 89}
\datum{66}{ 97}
\datum{66}{ 129}
\datum{66}{ 137}
\datum{66}{ 159}
\datum{68}{ 58}
\datum{68}{ 64}
\datum{68}{ 78}
\datum{68}{ 86}
\datum{68}{ 110}
\datum{68}{ 126}
\datum{68}{ 150}
\datum{70}{ 53}
\datum{70}{ 73}
\datum{70}{ 75}
\datum{70}{ 89}
\datum{70}{ 95}
\datum{70}{ 101}
\datum{70}{ 135}
\datum{72}{ 46}
\datum{72}{ 48}
\datum{72}{ 50}
\datum{72}{ 52}
\datum{72}{ 54}
\datum{72}{ 56}
\datum{72}{ 58}
\datum{72}{ 60}
\datum{72}{ 62}
\datum{72}{ 64}
\datum{72}{ 66}
\datum{72}{ 68}
\datum{72}{ 70}
\datum{72}{ 74}
\datum{72}{ 76}
\datum{72}{ 78}
\datum{72}{ 80}
\datum{72}{ 82}
\datum{72}{ 84}
\datum{72}{ 86}
\datum{72}{ 88}
\datum{72}{ 90}
\datum{72}{ 92}
\datum{72}{ 94}
\datum{72}{ 96}
\datum{72}{ 98}
\datum{72}{ 100}
\datum{72}{ 102}
\datum{72}{ 104}
\datum{72}{ 112}
\datum{72}{ 116}
\datum{72}{ 118}
\datum{72}{ 120}
\datum{72}{ 124}
\datum{72}{ 126}
\datum{72}{ 128}
\datum{72}{ 132}
\datum{72}{ 136}
\datum{72}{ 140}
\datum{72}{ 142}
\datum{72}{ 144}
\datum{72}{ 148}
\datum{72}{ 150}
\datum{72}{ 158}
\datum{72}{ 160}
\datum{72}{ 164}
\datum{72}{ 174}
\datum{72}{ 182}
\datum{72}{ 198}
\datum{72}{ 212}
\datum{74}{ 121}
\datum{74}{ 177}
\datum{76}{ 68}
\datum{76}{ 74}
\datum{76}{ 94}
\datum{76}{ 100}
\datum{76}{ 112}
\datum{76}{ 124}
\datum{76}{ 142}
\datum{78}{ 61}
\datum{78}{ 79}
\datum{78}{ 83}
\datum{78}{ 89}
\datum{78}{ 91}
\datum{78}{ 93}
\datum{78}{ 95}
\datum{78}{ 109}
\datum{78}{ 111}
\datum{78}{ 153}
\datum{78}{ 163}
\datum{78}{ 175}
\datum{80}{ 54}
\datum{80}{ 58}
\datum{80}{ 60}
\datum{80}{ 62}
\datum{80}{ 66}
\datum{80}{ 68}
\datum{80}{ 70}
\datum{80}{ 72}
\datum{80}{ 74}
\datum{80}{ 76}
\datum{80}{ 78}
\datum{80}{ 94}
\datum{80}{ 96}
\datum{80}{ 98}
\datum{80}{ 100}
\datum{80}{ 102}
\datum{80}{ 106}
\datum{80}{ 118}
\datum{80}{ 122}
\datum{80}{ 138}
\datum{80}{ 146}
\datum{84}{ 52}
\datum{84}{ 54}
\datum{84}{ 58}
\datum{84}{ 60}
\datum{84}{ 64}
\datum{84}{ 66}
\datum{84}{ 68}
\datum{84}{ 70}
\datum{84}{ 72}
\datum{84}{ 74}
\datum{84}{ 78}
\datum{84}{ 80}
\datum{84}{ 82}
\datum{84}{ 84}
\datum{84}{ 88}
\datum{84}{ 90}
\datum{84}{ 94}
\datum{84}{ 96}
\datum{84}{ 98}
\datum{84}{ 100}
\datum{84}{ 102}
\datum{84}{ 106}
\datum{84}{ 114}
\datum{84}{ 124}
\datum{84}{ 130}
\datum{84}{ 132}
\datum{84}{ 134}
\datum{84}{ 138}
\datum{84}{ 154}
\datum{84}{ 162}
\datum{84}{ 164}
\datum{84}{ 166}
\datum{84}{ 172}
\datum{84}{ 174}
\datum{84}{ 178}
\datum{84}{ 190}
\datum{84}{ 194}
\datum{84}{ 262}
\datum{84}{ 322}
\datum{86}{ 57}
\datum{86}{ 77}
\datum{86}{ 87}
\datum{86}{ 127}
\datum{86}{ 137}
\datum{88}{ 54}
\datum{88}{ 62}
\datum{88}{ 66}
\datum{88}{ 68}
\datum{88}{ 70}
\datum{88}{ 78}
\datum{88}{ 82}
\datum{88}{ 84}
\datum{88}{ 102}
\datum{88}{ 104}
\datum{88}{ 108}
\datum{88}{ 128}
\datum{88}{ 132}
\datum{88}{ 158}
\datum{90}{ 53}
\datum{90}{ 61}
\datum{90}{ 63}
\datum{90}{ 65}
\datum{90}{ 67}
\datum{90}{ 71}
\datum{90}{ 73}
\datum{90}{ 79}
\datum{90}{ 83}
\datum{90}{ 93}
\datum{90}{ 95}
\datum{90}{ 97}
\datum{90}{ 99}
\datum{90}{ 105}
\datum{90}{ 109}
\datum{90}{ 131}
\datum{90}{ 133}
\datum{90}{ 141}
\datum{90}{ 171}
\datum{92}{ 56}
\datum{92}{ 64}
\datum{92}{ 86}
\datum{92}{ 94}
\datum{92}{ 98}
\datum{92}{ 150}
\datum{94}{ 87}
\datum{96}{ 54}
\datum{96}{ 58}
\datum{96}{ 60}
\datum{96}{ 62}
\datum{96}{ 64}
\datum{96}{ 66}
\datum{96}{ 70}
\datum{96}{ 74}
\datum{96}{ 76}
\datum{96}{ 78}
\datum{96}{ 80}
\datum{96}{ 82}
\datum{96}{ 84}
\datum{96}{ 86}
\datum{96}{ 88}
\datum{96}{ 90}
\datum{96}{ 92}
\datum{96}{ 94}
\datum{96}{ 96}
\datum{96}{ 98}
\datum{96}{ 102}
\datum{96}{ 106}
\datum{96}{ 108}
\datum{96}{ 110}
\datum{96}{ 112}
\datum{96}{ 114}
\datum{96}{ 118}
\datum{96}{ 120}
\datum{96}{ 122}
\datum{96}{ 126}
\datum{96}{ 134}
\datum{96}{ 138}
\datum{96}{ 142}
\datum{96}{ 146}
\datum{96}{ 154}
\datum{96}{ 158}
\datum{96}{ 166}
\datum{96}{ 186}
\datum{96}{ 190}
\datum{96}{ 204}
\datum{96}{ 212}
\datum{96}{ 250}
\datum{96}{ 286}
\datum{96}{ 342}
\datum{96}{ 402}
\datum{98}{ 87}
\datum{100}{ 62}
\datum{100}{ 66}
\datum{100}{ 70}
\datum{100}{ 76}
\datum{100}{ 80}
\datum{100}{ 88}
\datum{100}{ 96}
\datum{100}{ 98}
\datum{100}{ 118}
\datum{100}{ 120}
\datum{100}{ 168}
\datum{102}{ 61}
\datum{102}{ 67}
\datum{102}{ 73}
\datum{102}{ 81}
\datum{102}{ 83}
\datum{102}{ 93}
\datum{102}{ 97}
\datum{102}{ 99}
\datum{102}{ 113}
\datum{104}{ 62}
\datum{104}{ 68}
\datum{104}{ 74}
\datum{104}{ 76}
\datum{104}{ 78}
\datum{104}{ 80}
\datum{104}{ 82}
\datum{104}{ 84}
\datum{104}{ 86}
\datum{104}{ 90}
\datum{104}{ 92}
\datum{104}{ 100}
\datum{104}{ 102}
\datum{104}{ 106}
\datum{104}{ 108}
\datum{104}{ 126}
\datum{104}{ 132}
\datum{104}{ 140}
\datum{104}{ 154}
\datum{106}{ 69}
\datum{106}{ 117}
\datum{106}{ 129}
\datum{106}{ 133}
\datum{108}{ 66}
\datum{108}{ 68}
\datum{108}{ 70}
\datum{108}{ 74}
\datum{108}{ 76}
\datum{108}{ 78}
\datum{108}{ 80}
\datum{108}{ 82}
\datum{108}{ 84}
\datum{108}{ 86}
\datum{108}{ 88}
\datum{108}{ 90}
\datum{108}{ 94}
\datum{108}{ 96}
\datum{108}{ 100}
\datum{108}{ 106}
\datum{108}{ 108}
\datum{108}{ 112}
\datum{108}{ 118}
\datum{108}{ 120}
\datum{108}{ 122}
\datum{108}{ 130}
\datum{108}{ 150}
\datum{108}{ 166}
\datum{108}{ 178}
\datum{108}{ 186}
\datum{108}{ 190}
\datum{108}{ 212}
\datum{110}{ 73}
\datum{110}{ 83}
\datum{110}{ 119}
\datum{110}{ 141}
\datum{110}{ 193}
\datum{112}{ 70}
\datum{112}{ 74}
\datum{112}{ 76}
\datum{112}{ 78}
\datum{112}{ 80}
\datum{112}{ 86}
\datum{112}{ 88}
\datum{112}{ 90}
\datum{112}{ 92}
\datum{112}{ 94}
\datum{112}{ 96}
\datum{112}{ 102}
\datum{112}{ 108}
\datum{112}{ 110}
\datum{112}{ 118}
\datum{112}{ 124}
\datum{112}{ 126}
\datum{112}{ 138}
\datum{112}{ 150}
\datum{112}{ 160}
\datum{114}{ 67}
\datum{114}{ 77}
\datum{114}{ 79}
\datum{114}{ 85}
\datum{114}{ 91}
\datum{114}{ 97}
\datum{114}{ 103}
\datum{114}{ 113}
\datum{114}{ 115}
\datum{114}{ 127}
\datum{114}{ 145}
\datum{114}{ 147}
\datum{114}{ 169}
\datum{114}{ 187}
\datum{114}{ 197}
\datum{116}{ 72}
\datum{116}{ 108}
\datum{116}{ 124}
\datum{120}{ 68}
\datum{120}{ 70}
\datum{120}{ 72}
\datum{120}{ 74}
\datum{120}{ 76}
\datum{120}{ 78}
\datum{120}{ 80}
\datum{120}{ 82}
\datum{120}{ 86}
\datum{120}{ 88}
\datum{120}{ 90}
\datum{120}{ 92}
\datum{120}{ 94}
\datum{120}{ 96}
\datum{120}{ 98}
\datum{120}{ 100}
\datum{120}{ 102}
\datum{120}{ 104}
\datum{120}{ 106}
\datum{120}{ 108}
\datum{120}{ 110}
\datum{120}{ 112}
\datum{120}{ 114}
\datum{120}{ 116}
\datum{120}{ 118}
\datum{120}{ 122}
\datum{120}{ 124}
\datum{120}{ 128}
\datum{120}{ 130}
\datum{120}{ 132}
\datum{120}{ 134}
\datum{120}{ 136}
\datum{120}{ 138}
\datum{120}{ 140}
\datum{120}{ 142}
\datum{120}{ 146}
\datum{120}{ 148}
\datum{120}{ 150}
\datum{120}{ 152}
\datum{120}{ 154}
\datum{120}{ 156}
\datum{120}{ 158}
\datum{120}{ 168}
\datum{120}{ 178}
\datum{120}{ 190}
\datum{120}{ 196}
\datum{120}{ 212}
\datum{120}{ 248}
\datum{120}{ 276}
\datum{120}{ 278}
\datum{124}{ 84}
\datum{124}{ 94}
\datum{124}{ 106}
\datum{124}{ 136}
\datum{124}{ 150}
\datum{124}{ 162}
\datum{126}{ 69}
\datum{126}{ 77}
\datum{126}{ 79}
\datum{126}{ 85}
\datum{126}{ 87}
\datum{126}{ 89}
\datum{126}{ 91}
\datum{126}{ 99}
\datum{126}{ 101}
\datum{126}{ 109}
\datum{126}{ 115}
\datum{126}{ 117}
\datum{126}{ 133}
\datum{126}{ 143}
\datum{126}{ 179}
\datum{128}{ 74}
\datum{128}{ 78}
\datum{128}{ 82}
\datum{128}{ 86}
\datum{128}{ 90}
\datum{128}{ 92}
\datum{128}{ 94}
\datum{128}{ 98}
\datum{128}{ 102}
\datum{128}{ 108}
\datum{128}{ 110}
\datum{128}{ 118}
\datum{128}{ 124}
\datum{128}{ 126}
\datum{128}{ 142}
\datum{130}{ 93}
\datum{130}{ 109}
\datum{132}{ 72}
\datum{132}{ 74}
\datum{132}{ 76}
\datum{132}{ 78}
\datum{132}{ 80}
\datum{132}{ 84}
\datum{132}{ 86}
\datum{132}{ 90}
\datum{132}{ 92}
\datum{132}{ 94}
\datum{132}{ 96}
\datum{132}{ 100}
\datum{132}{ 102}
\datum{132}{ 104}
\datum{132}{ 114}
\datum{132}{ 126}
\datum{132}{ 128}
\datum{132}{ 148}
\datum{132}{ 150}
\datum{132}{ 172}
\datum{132}{ 178}
\datum{132}{ 202}
\datum{132}{ 238}
\datum{132}{ 302}
\datum{134}{ 87}
\datum{136}{ 78}
\datum{136}{ 82}
\datum{136}{ 94}
\datum{136}{ 102}
\datum{136}{ 108}
\datum{136}{ 112}
\datum{136}{ 118}
\datum{136}{ 126}
\datum{136}{ 136}
\datum{136}{ 138}
\datum{136}{ 182}
\datum{138}{ 79}
\datum{138}{ 85}
\datum{138}{ 97}
\datum{138}{ 103}
\datum{138}{ 105}
\datum{138}{ 125}
\datum{138}{ 151}
\datum{140}{ 84}
\datum{140}{ 88}
\datum{140}{ 94}
\datum{140}{ 96}
\datum{140}{ 98}
\datum{140}{ 110}
\datum{140}{ 114}
\datum{140}{ 130}
\datum{140}{ 138}
\datum{140}{ 148}
\datum{140}{ 158}
\datum{142}{ 127}
\datum{142}{ 129}
\datum{144}{ 76}
\datum{144}{ 78}
\datum{144}{ 82}
\datum{144}{ 84}
\datum{144}{ 86}
\datum{144}{ 88}
\datum{144}{ 90}
\datum{144}{ 92}
\datum{144}{ 94}
\datum{144}{ 96}
\datum{144}{ 98}
\datum{144}{ 100}
\datum{144}{ 102}
\datum{144}{ 104}
\datum{144}{ 106}
\datum{144}{ 110}
\datum{144}{ 112}
\datum{144}{ 114}
\datum{144}{ 118}
\datum{144}{ 122}
\datum{144}{ 124}
\datum{144}{ 126}
\datum{144}{ 130}
\datum{144}{ 134}
\datum{144}{ 138}
\datum{144}{ 140}
\datum{144}{ 148}
\datum{144}{ 150}
\datum{144}{ 152}
\datum{144}{ 154}
\datum{144}{ 158}
\datum{144}{ 164}
\datum{144}{ 172}
\datum{144}{ 174}
\datum{144}{ 182}
\datum{144}{ 188}
\datum{144}{ 190}
\datum{144}{ 206}
\datum{144}{ 214}
\datum{146}{ 193}
\datum{148}{ 92}
\datum{148}{ 96}
\datum{148}{ 106}
\datum{148}{ 158}
\datum{150}{ 93}
\datum{150}{ 97}
\datum{150}{ 101}
\datum{150}{ 103}
\datum{150}{ 109}
\datum{150}{ 123}
\datum{150}{ 133}
\datum{150}{ 151}
\datum{150}{ 153}
\datum{150}{ 181}
\datum{152}{ 96}
\datum{152}{ 102}
\datum{152}{ 108}
\datum{152}{ 110}
\datum{152}{ 130}
\datum{152}{ 134}
\datum{152}{ 150}
\datum{154}{ 89}
\datum{154}{ 97}
\datum{156}{ 88}
\datum{156}{ 92}
\datum{156}{ 94}
\datum{156}{ 102}
\datum{156}{ 106}
\datum{156}{ 108}
\datum{156}{ 110}
\datum{156}{ 112}
\datum{156}{ 114}
\datum{156}{ 118}
\datum{156}{ 120}
\datum{156}{ 122}
\datum{156}{ 124}
\datum{156}{ 126}
\datum{156}{ 132}
\datum{156}{ 148}
\datum{156}{ 150}
\datum{156}{ 154}
\datum{156}{ 178}
\datum{156}{ 210}
\datum{156}{ 232}
\datum{156}{ 234}
\datum{156}{ 430}
\datum{160}{ 90}
\datum{160}{ 94}
\datum{160}{ 96}
\datum{160}{ 98}
\datum{160}{ 102}
\datum{160}{ 108}
\datum{160}{ 110}
\datum{160}{ 124}
\datum{160}{ 126}
\datum{160}{ 150}
\datum{160}{ 156}
\datum{160}{ 170}
\datum{160}{ 178}
\datum{160}{ 198}
\datum{160}{ 314}
\datum{162}{ 97}
\datum{162}{ 115}
\datum{162}{ 121}
\datum{162}{ 131}
\datum{162}{ 133}
\datum{162}{ 137}
\datum{162}{ 185}
\datum{164}{ 94}
\datum{166}{ 127}
\datum{168}{ 88}
\datum{168}{ 90}
\datum{168}{ 94}
\datum{168}{ 96}
\datum{168}{ 98}
\datum{168}{ 100}
\datum{168}{ 102}
\datum{168}{ 106}
\datum{168}{ 108}
\datum{168}{ 110}
\datum{168}{ 112}
\datum{168}{ 114}
\datum{168}{ 116}
\datum{168}{ 118}
\datum{168}{ 120}
\datum{168}{ 122}
\datum{168}{ 124}
\datum{168}{ 126}
\datum{168}{ 128}
\datum{168}{ 134}
\datum{168}{ 138}
\datum{168}{ 140}
\datum{168}{ 142}
\datum{168}{ 144}
\datum{168}{ 148}
\datum{168}{ 152}
\datum{168}{ 164}
\datum{168}{ 168}
\datum{168}{ 174}
\datum{168}{ 178}
\datum{168}{ 184}
\datum{168}{ 256}
\datum{170}{ 93}
\datum{170}{ 117}
\datum{172}{ 106}
\datum{172}{ 114}
\datum{172}{ 122}
\datum{172}{ 144}
\datum{174}{ 97}
\datum{174}{ 143}
\datum{174}{ 157}
\datum{174}{ 167}
\datum{176}{ 98}
\datum{176}{ 100}
\datum{176}{ 102}
\datum{176}{ 106}
\datum{176}{ 108}
\datum{176}{ 114}
\datum{176}{ 126}
\datum{176}{ 132}
\datum{176}{ 150}
\datum{176}{ 174}
\datum{180}{ 98}
\datum{180}{ 100}
\datum{180}{ 104}
\datum{180}{ 106}
\datum{180}{ 108}
\datum{180}{ 110}
\datum{180}{ 112}
\datum{180}{ 114}
\datum{180}{ 118}
\datum{180}{ 120}
\datum{180}{ 124}
\datum{180}{ 126}
\datum{180}{ 134}
\datum{180}{ 136}
\datum{180}{ 138}
\datum{180}{ 142}
\datum{180}{ 144}
\datum{180}{ 148}
\datum{180}{ 154}
\datum{180}{ 158}
\datum{180}{ 168}
\datum{180}{ 174}
\datum{180}{ 184}
\datum{180}{ 194}
\datum{180}{ 226}
\datum{180}{ 228}
\datum{180}{ 286}
\datum{180}{ 366}
\datum{184}{ 102}
\datum{184}{ 106}
\datum{184}{ 110}
\datum{184}{ 112}
\datum{184}{ 116}
\datum{184}{ 120}
\datum{184}{ 122}
\datum{184}{ 134}
\datum{184}{ 136}
\datum{184}{ 148}
\datum{184}{ 168}
\datum{184}{ 174}
\datum{184}{ 180}
\datum{186}{ 97}
\datum{186}{ 115}
\datum{186}{ 117}
\datum{186}{ 123}
\datum{186}{ 127}
\datum{186}{ 151}
\datum{186}{ 167}
\datum{190}{ 129}
\datum{192}{ 102}
\datum{192}{ 106}
\datum{192}{ 108}
\datum{192}{ 110}
\datum{192}{ 112}
\datum{192}{ 114}
\datum{192}{ 116}
\datum{192}{ 118}
\datum{192}{ 122}
\datum{192}{ 124}
\datum{192}{ 126}
\datum{192}{ 128}
\datum{192}{ 130}
\datum{192}{ 134}
\datum{192}{ 138}
\datum{192}{ 142}
\datum{192}{ 150}
\datum{192}{ 154}
\datum{192}{ 162}
\datum{192}{ 166}
\datum{192}{ 170}
\datum{192}{ 178}
\datum{192}{ 184}
\datum{192}{ 190}
\datum{192}{ 198}
\datum{192}{ 206}
\datum{192}{ 218}
\datum{192}{ 226}
\datum{192}{ 250}
\datum{194}{ 113}
\datum{194}{ 119}
\datum{196}{ 106}
\datum{196}{ 150}
\datum{196}{ 172}
\datum{198}{ 113}
\datum{198}{ 125}
\datum{198}{ 131}
\datum{198}{ 139}
\datum{200}{ 102}
\datum{200}{ 106}
\datum{200}{ 108}
\datum{200}{ 118}
\datum{200}{ 128}
\datum{200}{ 130}
\datum{200}{ 134}
\datum{200}{ 136}
\datum{200}{ 148}
\datum{200}{ 150}
\datum{200}{ 156}
\datum{200}{ 168}
\datum{202}{ 133}
\datum{204}{ 104}
\datum{204}{ 108}
\datum{204}{ 118}
\datum{204}{ 120}
\datum{204}{ 124}
\datum{204}{ 126}
\datum{204}{ 130}
\datum{204}{ 134}
\datum{204}{ 142}
\datum{204}{ 164}
\datum{204}{ 168}
\datum{204}{ 178}
\datum{204}{ 190}
\datum{204}{ 222}
\datum{208}{ 108}
\datum{208}{ 118}
\datum{208}{ 126}
\datum{208}{ 138}
\datum{208}{ 178}
\datum{210}{ 113}
\datum{210}{ 117}
\datum{210}{ 131}
\datum{210}{ 133}
\datum{210}{ 143}
\datum{210}{ 145}
\datum{210}{ 151}
\datum{210}{ 171}
\datum{210}{ 221}
\datum{210}{ 241}
\datum{212}{ 150}
\datum{212}{ 158}
\datum{216}{ 116}
\datum{216}{ 120}
\datum{216}{ 124}
\datum{216}{ 126}
\datum{216}{ 128}
\datum{216}{ 132}
\datum{216}{ 134}
\datum{216}{ 136}
\datum{216}{ 138}
\datum{216}{ 140}
\datum{216}{ 142}
\datum{216}{ 144}
\datum{216}{ 150}
\datum{216}{ 152}
\datum{216}{ 154}
\datum{216}{ 158}
\datum{216}{ 164}
\datum{216}{ 168}
\datum{216}{ 174}
\datum{216}{ 180}
\datum{216}{ 182}
\datum{216}{ 188}
\datum{216}{ 192}
\datum{216}{ 212}
\datum{216}{ 240}
\datum{216}{ 244}
\datum{216}{ 274}
\datum{216}{ 292}
\datum{220}{ 126}
\datum{220}{ 158}
\datum{220}{ 218}
\datum{222}{ 133}
\datum{222}{ 135}
\datum{224}{ 122}
\datum{224}{ 124}
\datum{224}{ 126}
\datum{224}{ 134}
\datum{224}{ 138}
\datum{224}{ 142}
\datum{224}{ 146}
\datum{224}{ 152}
\datum{224}{ 170}
\datum{224}{ 178}
\datum{228}{ 126}
\datum{228}{ 130}
\datum{228}{ 132}
\datum{228}{ 138}
\datum{228}{ 140}
\datum{228}{ 142}
\datum{228}{ 146}
\datum{228}{ 170}
\datum{228}{ 176}
\datum{228}{ 178}
\datum{228}{ 196}
\datum{228}{ 202}
\datum{228}{ 346}
\datum{230}{ 129}
\datum{232}{ 126}
\datum{232}{ 134}
\datum{232}{ 158}
\datum{234}{ 131}
\datum{234}{ 133}
\datum{234}{ 141}
\datum{234}{ 217}
\datum{236}{ 162}
\datum{236}{ 172}
\datum{236}{ 182}
\datum{240}{ 124}
\datum{240}{ 126}
\datum{240}{ 130}
\datum{240}{ 134}
\datum{240}{ 138}
\datum{240}{ 140}
\datum{240}{ 142}
\datum{240}{ 144}
\datum{240}{ 146}
\datum{240}{ 150}
\datum{240}{ 158}
\datum{240}{ 162}
\datum{240}{ 166}
\datum{240}{ 174}
\datum{240}{ 178}
\datum{240}{ 188}
\datum{240}{ 206}
\datum{240}{ 218}
\datum{240}{ 226}
\datum{240}{ 232}
\datum{240}{ 268}
\datum{240}{ 394}
\datum{242}{ 139}
\datum{242}{ 189}
\datum{244}{ 162}
\datum{246}{ 167}
\datum{246}{ 185}
\datum{246}{ 237}
\datum{248}{ 134}
\datum{248}{ 174}
\datum{252}{ 130}
\datum{252}{ 138}
\datum{252}{ 142}
\datum{252}{ 158}
\datum{252}{ 162}
\datum{252}{ 166}
\datum{252}{ 174}
\datum{252}{ 178}
\datum{252}{ 202}
\datum{252}{ 206}
\datum{252}{ 278}
\datum{256}{ 134}
\datum{256}{ 156}
\datum{256}{ 158}
\datum{256}{ 166}
\datum{256}{ 174}
\datum{256}{ 234}
\datum{258}{ 151}
\datum{258}{ 161}
\datum{258}{ 163}
\datum{260}{ 134}
\datum{260}{ 150}
\datum{264}{ 144}
\datum{264}{ 148}
\datum{264}{ 150}
\datum{264}{ 154}
\datum{264}{ 158}
\datum{264}{ 160}
\datum{264}{ 162}
\datum{264}{ 164}
\datum{264}{ 172}
\datum{264}{ 178}
\datum{264}{ 184}
\datum{264}{ 188}
\datum{264}{ 214}
\datum{264}{ 228}
\datum{264}{ 262}
\datum{266}{ 169}
\datum{270}{ 223}
\datum{272}{ 148}
\datum{272}{ 150}
\datum{272}{ 166}
\datum{272}{ 178}
\datum{276}{ 150}
\datum{276}{ 154}
\datum{276}{ 162}
\datum{276}{ 172}
\datum{276}{ 174}
\datum{276}{ 186}
\datum{276}{ 192}
\datum{276}{ 212}
\datum{276}{ 214}
\datum{276}{ 222}
\datum{276}{ 234}
\datum{276}{ 262}
\datum{276}{ 330}
\datum{280}{ 148}
\datum{280}{ 150}
\datum{280}{ 166}
\datum{284}{ 166}
\datum{286}{ 163}
\datum{286}{ 177}
\datum{288}{ 146}
\datum{288}{ 152}
\datum{288}{ 158}
\datum{288}{ 162}
\datum{288}{ 166}
\datum{288}{ 170}
\datum{288}{ 182}
\datum{288}{ 186}
\datum{288}{ 188}
\datum{288}{ 190}
\datum{288}{ 206}
\datum{288}{ 214}
\datum{288}{ 222}
\datum{288}{ 226}
\datum{288}{ 258}
\datum{288}{ 270}
\datum{288}{ 374}
\datum{292}{ 158}
\datum{294}{ 159}
\datum{294}{ 187}
\datum{294}{ 259}
\datum{296}{ 150}
\datum{296}{ 162}
\datum{296}{ 166}
\datum{296}{ 174}
\datum{300}{ 168}
\datum{300}{ 178}
\datum{300}{ 180}
\datum{300}{ 194}
\datum{300}{ 198}
\datum{300}{ 218}
\datum{300}{ 222}
\datum{300}{ 226}
\datum{300}{ 230}
\datum{304}{ 176}
\datum{306}{ 169}
\datum{306}{ 177}
\datum{306}{ 217}
\datum{312}{ 166}
\datum{312}{ 172}
\datum{312}{ 174}
\datum{312}{ 178}
\datum{312}{ 180}
\datum{312}{ 190}
\datum{312}{ 192}
\datum{312}{ 196}
\datum{312}{ 224}
\datum{312}{ 318}
\datum{316}{ 262}
\datum{318}{ 197}
\datum{320}{ 170}
\datum{320}{ 174}
\datum{320}{ 190}
\datum{320}{ 198}
\datum{320}{ 206}
\datum{320}{ 222}
\datum{322}{ 193}
\datum{324}{ 168}
\datum{324}{ 182}
\datum{324}{ 184}
\datum{324}{ 222}
\datum{324}{ 230}
\datum{324}{ 232}
\datum{324}{ 262}
\datum{330}{ 181}
\datum{330}{ 221}
\datum{330}{ 261}
\datum{336}{ 178}
\datum{336}{ 188}
\datum{336}{ 194}
\datum{336}{ 198}
\datum{336}{ 202}
\datum{336}{ 206}
\datum{336}{ 222}
\datum{336}{ 230}
\datum{336}{ 312}
\datum{336}{ 358}
\datum{340}{ 198}
\datum{342}{ 233}
\datum{344}{ 224}
\datum{348}{ 186}
\datum{348}{ 198}
\datum{348}{ 226}
\datum{348}{ 238}
\datum{348}{ 258}
\datum{352}{ 190}
\datum{354}{ 259}
\datum{356}{ 202}
\datum{356}{ 220}
\datum{356}{ 234}
\datum{360}{ 190}
\datum{360}{ 192}
\datum{360}{ 194}
\datum{360}{ 206}
\datum{360}{ 212}
\datum{360}{ 228}
\datum{360}{ 258}
\datum{360}{ 306}
\datum{364}{ 194}
\datum{364}{ 204}
\datum{368}{ 204}
\datum{372}{ 194}
\datum{372}{ 202}
\datum{372}{ 226}
\datum{372}{ 258}
\datum{372}{ 262}
\datum{372}{ 306}
\datum{372}{ 346}
\datum{376}{ 214}
\datum{380}{ 198}
\datum{384}{ 204}
\datum{384}{ 218}
\datum{384}{ 222}
\datum{384}{ 232}
\datum{384}{ 234}
\datum{384}{ 242}
\datum{384}{ 250}
\datum{384}{ 262}
\datum{390}{ 303}
\datum{396}{ 214}
\datum{396}{ 222}
\datum{396}{ 262}
\datum{396}{ 340}
\datum{408}{ 212}
\datum{408}{ 224}
\datum{408}{ 232}
\datum{408}{ 240}
\datum{408}{ 268}
\datum{416}{ 262}
\datum{420}{ 218}
\datum{420}{ 230}
\datum{420}{ 248}
\datum{420}{ 250}
\datum{420}{ 306}
\datum{420}{ 334}
\datum{426}{ 265}
\datum{432}{ 238}
\datum{432}{ 242}
\datum{432}{ 266}
\datum{432}{ 274}
\datum{432}{ 334}
\datum{444}{ 234}
\datum{444}{ 330}
\datum{450}{ 303}
\datum{450}{ 331}
\datum{456}{ 234}
\datum{456}{ 248}
\datum{456}{ 256}
\datum{456}{ 262}
\datum{456}{ 264}
\datum{456}{ 272}
\datum{456}{ 302}
\datum{468}{ 286}
\datum{468}{ 306}
\datum{476}{ 270}
\datum{480}{ 246}
\datum{480}{ 262}
\datum{480}{ 278}
\datum{480}{ 286}
\datum{480}{ 306}
\datum{480}{ 334}
\datum{492}{ 256}
\datum{504}{ 276}
\datum{504}{ 312}
\datum{510}{ 331}
\datum{512}{ 286}
\datum{516}{ 302}
\datum{516}{ 330}
\datum{528}{ 278}
\datum{528}{ 286}
\datum{528}{ 318}
\datum{528}{ 334}
\datum{540}{ 274}
\datum{540}{ 298}
\datum{540}{ 334}
\datum{552}{ 306}
\datum{564}{ 322}
\datum{564}{ 330}
\datum{564}{ 340}
\datum{576}{ 302}
\datum{576}{ 314}
\datum{588}{ 346}
\datum{612}{ 330}
\datum{612}{ 346}
\datum{624}{ 330}
\datum{624}{ 358}
\datum{636}{ 342}
\datum{648}{ 358}
\datum{660}{ 366}
\datum{672}{ 374}
\datum{720}{ 394}
\datum{732}{ 386}
\datum{744}{ 402}
\datum{804}{ 430}
\datum{840}{ 446}
\datum{900}{ 474}
\datum{960}{ 502}
\newpage
\immediate\closeout\referencewrite\referenceopenfalse
\line{\bf\hfil References\hfil}\bigskip\parindent=0pt\input referenc.texauxil
\bye